\documentclass[acmsmall,screen,nonacm]{acmart}
\usepackage{microtype}
\usepackage{amsmath,amsthm,mathtools}
\usepackage{booktabs,array,tabularx}
\usepackage{placeins}
\usepackage{xcolor}
\usepackage{longtable}
\definecolor{LinkBlue}{HTML}{174A7E}
\definecolor{BoxBg}{HTML}{F3F6FA}
\definecolor{RuleGray}{HTML}{D0D7DE}
\hypersetup{bookmarksnumbered=true}
\allowdisplaybreaks[3]
\numberwithin{equation}{section}
\theoremstyle{acmplain}
\newtheorem{theorem}{Theorem}[section]
\newtheorem{lemma}[theorem]{Lemma}
\newtheorem{proposition}[theorem]{Proposition}
\newtheorem{corollary}[theorem]{Corollary}
\theoremstyle{acmdefinition}
\newtheorem{definition}[theorem]{Definition}

\theoremstyle{remark}

\newcommand{\NN}{\mathbb{N}}
\newcommand{\RR}{\mathbb{R}}
\newcommand{\CC}{\mathbb{C}}
\newcommand{\ZZ}{\mathbb{Z}}
\newcommand{\cH}{\mathcal{H}}
\newcommand{\cK}{\mathcal{K}}
\newcommand{\cT}{\mathcal{T}}

\newcommand{\cM}{\mathcal{M}}
\newcommand{\cC}{\mathcal{C}}
\newcommand{\cR}{\mathcal{R}}

\newcommand{\id}{\mathrm{id}}
\newcommand{\dd}{\,\mathrm{d}}
\newcommand{\one}{\mathbf{1}}
\newcommand{\res}[2]{R_{#1}\!\left(#2\right)}

\newcommand{\sem}[1]{\llbracket #1 \rrbracket}
\newcommand{\ket}[1]{\lvert #1\rangle}
\newcommand{\bra}[1]{\langle #1\rvert}
\newcommand{\ketbra}[2]{\lvert #1\rangle\langle #2\rvert}
\newcommand{\ip}[2]{\langle #1,#2\rangle}
\newcommand{\norm}[1]{\lVert #1\rVert}

\newcommand{\Eff}[1]{[0,I]_{#1}}
\newcommand{\Pred}[1]{\mathsf{Pred}(#1)}

\newcommand{\Bop}[1]{\mathcal{B}(#1)}
\renewcommand{\Top}[1]{\mathcal{T}(#1)}
\DeclareMathOperator{\Tr}{tr}
\DeclareMathOperator{\Dom}{Dom}
\DeclareMathOperator{\Var}{Var}

\renewcommand{\wp}{\operatorname{wp}}
\newcommand{\wlp}{\operatorname{wlp}}
\newcommand{\E}{\mathbb{E}}
\newcommand{\Prob}{\mathbb{P}}

\hypersetup{pdftitle={Parameterized Weakest-Precondition Calculus for Fine-grained Properties of Quantum Loops}}

\title{Parameterized Weakest-Precondition Calculus for Fine-grained Properties of Quantum Loops}
\author{Tianshi Yu}
\affiliation{%
  \institution{Institute of Software, Chinese Academy of Sciences}
  \city{Beijing}
  \country{China}
}
\email{yuts@ios.ac.cn}

\author{Minbo Gao}
\affiliation{%
  \institution{Institute of Software, Chinese Academy of Sciences}
  \city{Beijing}
  \country{China}
}
\affiliation{
  \institution{University of Chinese Academy of Sciences}
  \city{Beijing}
  \country{China}
}
\email{gmb17@tsinghua.org.cn}

\author{Li Zhou}
\affiliation{%
  \institution{Institute of Software, Chinese Academy of Sciences}
  \city{Beijing}
  \country{China}
}
\email{zhouli@iscas.ac.cn}

\begin{document}
\begin{abstract}
Quantum loops may execute arbitrarily many iterations, and understanding
their runtime behavior requires knowing how often they terminate, how
long terminating runs take, and how common very long runs are.
We develop a compositional weakest-precondition framework for
analyzing these properties on possibly infinite-dimensional state spaces.
It uses parameterized predicates: families of
quantum effects indexed by a positive observation scale.
These families retain runtime information even when the expected runtime
is infinite and can be reused through sequential composition and nested loops.
We prove that weakest-precondition equations for loops have unique
solutions in the parameterized predicate domain.
From the asymptotic behavior of the derived weakest preconditions,
we then recover these runtime properties and quantify the influence
of very long terminating executions.
For the classic discrete-time quantum walk on the one-dimensional infinite line
with one absorbing boundary at the origin and the Hadamard gate as its
coin operator, we verify its termination probability $2/\pi$,
conditional mean $\pi/2$, and logarithmic growth of the conditional
capped second moment. We then reuse the walk's specification to verify
the termination probability and conditional mean runtime of an outer
loop that repeatedly invokes it.
\end{abstract}
\keywords{Quantum loops, termination and runtime analysis, weakest preconditions,
parameterized predicates, compositional program verification,
infinite-dimensional quantum systems}
\maketitle


\section{Introduction}
\label{sec:intro}

Quantum programs can contain loops that execute arbitrarily many
iterations~\cite{AmbainisEtAl2001,YingFeng2010}.
Understanding their termination and runtime behavior involves more than
deciding whether they terminate: we would like to know how often a loop
terminates, how long its terminating runs take on average, and how common
very long terminating runs are~\cite{BachEtAl2004,ChatterjeeEtAl2018}.
These questions describe different aspects of completing a computation.
Termination probability measures the likelihood of completion, conditional
expected runtime describes its average duration~\cite{KonnoEtAl2003},
and higher moments---averages of higher powers of runtime, such as its
square---give greater weight to long
executions~\cite{KuraEtAl2019,WangHoffmannReps2021}.
This information helps us both understand a program in isolation and
assess the behavior of a larger program that invokes it as a
subroutine~\cite{AvanziniEtAl2022}.
It is therefore desirable for a verification framework to establish these
properties in a form that supports compositional reasoning, whether the
relevant runtime moments are finite or infinite.

Quantum walks, a standard tool in quantum algorithm design, illustrate
this difficulty~\cite{AmbainisEtAl2001,KonnoEtAl2003,BachEtAl2004}.
Consider a particle moving along an infinite line whose positions are
labeled by integers, with a quantum coin controlling whether each move
is to the left or right.
Before each move, the program uses a quantum measurement to check
whether the particle is at the origin (position $0$).
If so, the walk stops; otherwise, the program tosses the coin
(using a Hadamard operation) and moves the particle one position
left or right according to the coin state.
The quantum walk can evolve in a superposition of positions and coin
states, allowing interference between different paths and thereby producing termination and runtime behavior different from that of a classical random walk.
Suppose the walk starts at position $1$ with its coin in the rightward state. Then the 
runtime $\tau$, measured in loop iterations, satisfies
\[
 \Prob(\tau<\infty)=\frac{2}{\pi},
 \quad
 \E[\tau]=\infty,
 \quad
 \E[\tau\mid\tau<\infty]=\frac{\pi}{2},
 \quad
 \E[\min(\tau,N)^2\mid\tau<\infty]=\log N+O(1).
\]
Therefore, the walk fails to terminate with probability greater than one third, yet takes
fewer than two iterations on average conditional on termination. 
Moreover, when each terminating run's duration is capped at
$N$, the resulting conditional second moment grows logarithmically with $N$, which implies the conditional second moment $\E[\tau^2\mid\tau<\infty]$ is infinite.

The example above exhibits two distinct sources of infinite moments. On the one hand, the unconditional mean $\E[\tau]$ is infinite because nontermination has positive probability. On the other hand,
the conditional second moment $\E[\tau^2\mid\tau<\infty]$ remains infinite even after restricting attention
to terminating runs, because long executions make a sufficiently large
contribution. Both facts matter, and the value $+\infty$ alone leaves verifying their
quantitative structure unresolved. Moreover, the walk explores an infinite-dimensional state space, so
directly representing its unbounded runtime quantities requires a theory of
unbounded quantum observables. Thus the challenge is to encode these finer aspects of unbounded behavior in a structure suitable for program reasoning.

\paragraph{Existing approaches to representing unbounded information.}
There are two approaches to representing such unbounded information \cite{CousotCousot1977,VitterFlajolet1990,ChenEtAl2022Generating}. One approach is to admit
infinite values directly into the domain of reasoning. For example, interval analyses use infinite endpoints to describe program values with no finite
upper or lower bound~\cite{CousotCousot1977}, and probabilistic runtime calculi
reason with cost functions taking values in the extended nonnegative
reals~\cite{KaminskiEtAl2018}. Another approach is to represent an infinite collection of quantitative information
through a parameterized object. A famous example is the generating
function, which encodes an infinite sequence as the coefficients of a
formal power series and allows asymptotic properties of the sequence
to be recovered from the series' analytic behavior. This approach
supports average-case analysis of sorting, searching, and hashing
algorithms~\cite{VitterFlajolet1990}, as well as reasoning about the
output distributions of probabilistic
programs~\cite{KlinkenbergEtAl2021,ChenEtAl2022Generating}.
Similarly, in finite-state Markov decision processes, rewards over an
infinite horizon are encoded in normalized discounted values, whose
limits recover optimal long-run average
rewards~\cite{GimbertZielonka2007}.

Existing works in quantum runtime analysis either do not cover infinite expected costs or follow the first approach, admitting infinite values directly into their reasoning domains.
Expected runtime calculi provide compositional methods with finite-dimensional
quantum registers or finite-dimensional computational
techniques~\cite{AvanziniEtAl2022,LiuEtAl2025}; these techniques do not cover the
coherent walk on the infinite line. Yu \cite{Yu2023} reduces flowchart
programs to a single loop while preserving runtime moments and gives algorithms
for termination and exact expected runtime; the computation of this work assumes a finite-dimensional state space and almost-sure termination, conditions under which all runtime moments are finite. 
Unbounded pre-expectations~\cite{GehnenEtAl2026} and closed positive quadratic
forms~\cite{YuEtAl2026} follow the direct approach, extending
expectation-transformer reasoning to infinite-dimensional systems through
unbounded predicates. These calculi can establish that an expected cost is
infinite, but that conclusion alone leaves the finer properties above
unresolved. Thus we want to seek a representation that retains this finer information and supports quantum program composition.

\paragraph{Design choice.}
We take the second approach: encoding the unbounded information into a parameterized object. Its central design problem is to choose
how execution times are weighted at each parameter value. The weights are
designed to meet three requirements. First, they are \emph{bounded}, so that the
resulting observations remain finite even when runtime moments are infinite.
Second, they support \emph{composition}, so that verified specifications
for individual commands can be combined according to the program structure.
For quantum programs, this involves preserving the dependence between execution
time and the state on which the next command acts. Third, they retain
enough \emph{runtime information} to recover termination probabilities,
conditional moments, and the divergence rates of infinite moments.
Meeting these requirements together allows the parameterized representation
to support both quantitative analysis and compositional verification.

To meet the above requirements, we use an auxiliary \emph{observation scale} $s>0$
and assign a finite runtime $n$ the weight
\[
 r_s(n)\triangleq \frac{s}{s+n},
 \qquad r_s(\infty)\triangleq 0.
\]
At each scale, these weights give bounded observations of execution time.
Changing the scale changes how strongly long executions contribute to the
observation. The weights also allow time consumed by a command to be accounted
for by shifting the scale, which supports sequential calls and nested loops.
Their large-scale behavior retains the information used to recover the
runtime properties above. Thus the scale provides a common means of organizing
both program composition and quantitative analysis.

\paragraph{Contributions.}
Building on the above choice of weights, we develop a weakest-precondition
calculus for reasoning about fine-grained properties of quantum loops. The concrete contributions are as follows:
\begin{itemize}
    \item \textbf{Runtime-graded semantics and parameterized predicates.} To express specifications in this calculus, we introduce
    \emph{parameterized predicates}: families of bounded quantum effects indexed
    by the observation scale. 
    We support them with a runtime-graded semantics that records, for each finite
    runtime $n$, the output state of executions that terminate after exactly $n$
    loop iterations. We also show how to translate a
     valid cost Hoare triple in \cite{YuEtAl2026} into a valid triple over
     parameterized predicates.

 \item \textbf{A compositional weakest-precondition calculus.}
    Our weakest-precondition calculus propagates these parameterized predicates
    backward through the program syntax using the runtime weights and scale shifts. Moreover, for loops, we prove that the
 weakest-precondition equation has a unique solution in the predicate domain
 and show how to certify an approximation by checking it against the
 weakest-precondition equation.

 \item \textbf{Rules for extracting runtime properties.}
 Since our parameterized weight has a clear probabilistic interpretation,
  the asymptotic behavior of a precondition yields termination probabilities and conditional moments.
 We provide rules for recovering these properties from a runtime family verified on the program's full state space. 
 When a conditional moment is infinite, the same family
 gives its rate of divergence, which is measured by how quickly the
 corresponding capped moment grows as the cap on runtime increases.
\end{itemize}

\paragraph{Applications.}
Using this calculus, we verify the termination and runtime properties of the quantum walk mentioned above. To the best of our knowledge, this is the first direct verification of all these properties for the quantum walk on the one-dimensional infinite line using a weakest-precondition calculus.

For the one-dimensional absorbing walk with a Hadamard coin, we verify the termination probability and conditional
 mean runtime from the program and show that the conditional second moment
 diverges. When the durations of terminating runs are capped, we establish
 that the resulting second moment grows logarithmically with the cap. We also extend the analysis to biased walks with arbitrary initial coin states.

 Furthermore, to demonstrate the composability of our framework, we verify the termination and runtime properties of a nested loop: an outer loop that repeatedly invokes the Hadamard walk as a subprogram. At each loop test, a fair coin measurement decides whether to
 exit or to reinitialize the state and run the walk again. By reusing the
 verified runtime family of the inner walk, we obtain the outer loop's
 termination probability and its expected runtime conditional on termination.

\paragraph{Organization.}
Section~\ref{sec:motivating-example} presents the motivating example.
Sections~\ref{sec:main-semantics}--\ref{sec:main-extraction} develop the
runtime-graded semantics, weakest-precondition calculus, and rules for
extracting runtime properties.
Section~\ref{sec:main-case-studies} presents the case studies, followed by
related work in Section~\ref{sec:related-work} and the conclusion in
Section~\ref{sec:conclusion}.
The supplemental text provides full proofs and detailed calculations.
\section{Motivating Example}
\label{sec:motivating-example}

We use the one-dimensional absorbing quantum walk~\cite{KonnoEtAl2003,BachEtAl2004} to illustrate what runtime information a verification framework should retain and what it takes to reuse that information in a larger program. Quantum walks are important algorithmic primitives and universal for quantum computation~\cite{Childs2009,ChildsGossetWebb2013}. As a classic model, the absorbing walk is a concrete and useful setting in which termination probability, finite moments, and infinite moments all matter, both for a single invocation and for a program that calls it repeatedly.

\subsection{Finite and infinite runtime properties}
\label{main:mot:runtime-properties}

\begin{figure}[!htb]
\centering
\begin{minipage}[c]{.45\linewidth}
\small
\textit{Initialization}\\[3pt]
$\mathtt{coin}:=\ket{R};\quad \mathtt{position}:=\ket{1};$\\[9pt]
\textit{The loop $\mathsf{Walk}$}\\[3pt]
$\mathtt{while}\ M[\mathtt{position}]=1\ \mathtt{do}$\\
\hspace*{1em}$\mathtt{coin}:=C_{1/2}[\mathtt{coin}];$\\
\hspace*{1em}$(\mathtt{coin},\mathtt{position}):=$\\
\hspace*{2em}$V[\mathtt{coin},\mathtt{position}]$\\
$\mathtt{od}$
\end{minipage}\hfill
\begin{minipage}[c]{.53\linewidth}
\centering\small
\setlength{\unitlength}{1pt}%
\begin{picture}(202,138)
\put(101,129){\makebox(0,0){$\ket{R,1}$}}
\put(101,120){\vector(0,-1){17}}
\put(108,112){\makebox(0,0)[l]{\footnotesize coin and shift}}
\put(101,93){\makebox(0,0){$(\ket{L,0}-\ket{R,2})/\sqrt{2}$}}
\put(101,73){\makebox(0,0){\footnotesize next guard}}
\put(77,71){\vector(-2,-1){37}}
\put(125,71){\vector(2,-1){37}}
\put(38,42){\makebox(0,0){$\ket{L,0}$}}
\put(164,42){\makebox(0,0){$\ket{R,2}$}}
\put(38,25){\makebox(0,0){\footnotesize terminate $(1/2)$}}
\put(164,25){\makebox(0,0){\footnotesize continue $(1/2)$}}
\put(38,9){\makebox(0,0){\footnotesize $\tau=1$}}
\put(164,9){\makebox(0,0){\footnotesize $\tau=1+t$}}
\end{picture}
\end{minipage}
\caption{The program and its first iteration. The guard distinguishes
the origin (outcome zero) from all nonzero sites together (outcome one).
Here $\tau$ is the total runtime and $t$ is the remaining runtime from
$\ket{R,2}$.}
\Description{The program and its first iteration.}
\label{fig:mot-first-step}
\end{figure}
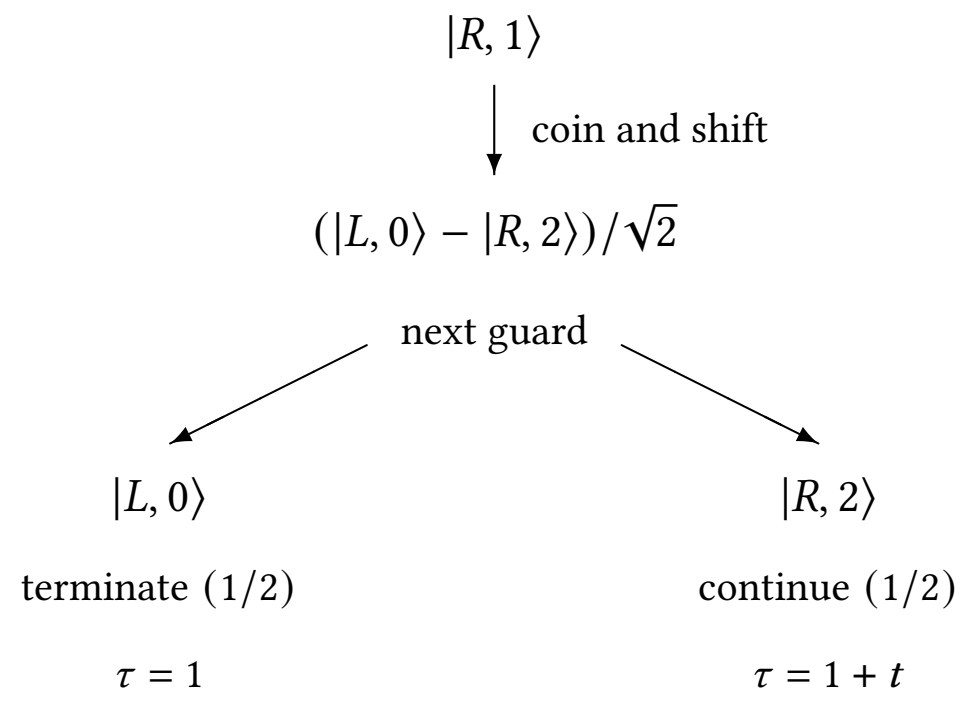

Figure~\ref{fig:mot-first-step} presents the absorbing Hadamard walk as a
quantum loop over a coin register and an integer position register. The
coin states $\ket{L}$ and $\ket{R}$ control the direction of movement along
the integer lattice. At each loop test, the guard $M$ measures whether
the particle is at the origin, grouping all nonzero positions into a
single continuation outcome. If the position is zero, the loop
terminates; otherwise, the body applies a Hadamard gate $C_{1/2}$ to the
coin and shifts the position left or right accordingly. The guard
preserves superpositions among nonzero positions. We measure runtime by
the number of entries into the loop body.

Starting from $\ket{R,1}$, the first iteration produces the superposition
shown in Figure~\ref{fig:mot-first-step}. At the next guard measurement,
half of the probability mass terminates at the origin with total runtime
$\tau=1$. The other half continues from $\ket{R,2}$ with one iteration
already spent.

To characterize the whole execution, we want to answer three questions:
\begin{enumerate}
 \item \emph{How often does the walk terminate?}
 Its termination probability on this input is $2/\pi$.
 \item \emph{How long do terminating executions take on average?}
 Their expected runtime is $\pi/2$, fewer than two iterations.
 \item \emph{How strongly do long terminating executions affect the runtime?}
 The conditional second moment is infinite. To describe its divergence,
 we ask how the conditional second moment grows when each terminating
 run's duration is capped at $N$ before squaring.
\end{enumerate}
The probability, conditional mean, and infinite conditional second moment
are known properties of this walk~\cite[Section~4]{KonnoEtAl2003}. The
capped second moment grows logarithmically with $N$; establishing this
rate is a further verification target, treated in
Section~E.2 of the supplemental text.

The expected-cost calculus of \cite{YuEtAl2026} (or equivalently \cite{GehnenEtAl2026}) allows us to
prove that the unconditional expected runtime is infinite
(Theorem~A.2 of the supplemental text). This follows from the positive
probability of nontermination. The infinite conditional second moment
has a different source: long executions still make an unbounded
contribution after nonterminating runs have been excluded. Thus a
verification framework should establish both the finite quantities and
the quantitative behavior of the infinite ones.

These questions place two requirements on the parameterized approach
outlined in the introduction. Each observation should remain bounded
when a runtime moment is infinite. At the same time, the observations
taken across parameter values must retain enough information to recover
the termination probability, conditional moments, and the growth of
capped moments. For this walk, all three questions should be answered
from the same verified representation.

\subsection{Reusing a verified loop in a caller}

Figure~\ref{fig:mot-caller} places $\mathsf{Walk}$ inside an outer loop
that draws a fresh fair bit at each guard. Outcome zero exits the outer
loop; outcome one initializes the registers to $\ket{R,1}$ and executes
$\mathsf{Walk}$. When the inner walk terminates, execution continues at
the outer guard. Each continuing pass contributes one unit of runtime
for entering the outer body plus the runtime $t_i$ of that call to
$\mathsf{Walk}$.

\begin{figure}[!htb]
\centering
\begin{minipage}[c]{.36\linewidth}
\small
\textit{The caller}\\[6pt]
$\mathtt{while}\ \mathtt{randomBit()}=1\ \mathtt{do}$\\[3pt]
\hspace*{1em}$\mathtt{coin}:=\ket{R};$\\[3pt]
\hspace*{1em}$\mathtt{position}:=\ket{1};$\\[3pt]
\hspace*{1em}$\mathsf{Walk}$\\[3pt]
$\mathtt{od}$
\end{minipage}\hfill
\begin{minipage}[c]{.61\linewidth}
\small
\renewcommand{\arraystretch}{1.25}
\begin{tabularx}{\linewidth}{@{}Xr@{}}
\toprule
Example execution & Total runtime\\
\midrule
Exit immediately & $0$\\
One call finishes, then exit & $1+t_1$\\
Two calls finish, then exit & $2+t_1+t_2$\\
A call never terminates & $\infty$\\
\bottomrule
\end{tabularx}
\end{minipage}
\caption{An outer loop calling $\mathsf{Walk}$. 
Each $t_i$ is the runtime
of a terminating call. Total runtime counts entries into both the outer
and inner loop bodies.}
\Description{An outer loop calling $\mathsf{Walk}$.}
\label{fig:mot-caller}
\end{figure}

We now ask for the outer loop's termination probability and conditional
expected runtime. As the table in Figure~\ref{fig:mot-caller} shows, the
caller can terminate before making a call or after several calls have
finished. If any call fails to terminate, the caller also fails to
terminate. A compositional analysis should obtain the caller's
properties by reusing a verified specification of the initialized walk.

In backward reasoning, we must account for the execution that follows
the call. Suppose the first call takes either one or three iterations.
Including the step spent entering the outer body, these executions reach
the outer guard after two or four iterations. If the remaining execution
takes $t$ further iterations, their total runtimes are $2+t$ and $4+t$.
Although both executions continue with the same code, an observation of
their total runtime must account for the different amounts of time
already spent.

This gives the third design requirement: the representation must support
composition. Propagating a postcondition backward through a call must
account for its possible runtimes and yield a precondition of the same
form, allowing the same rules to be reused through further commands and
enclosing loops. In a general quantum program, the remaining execution
also depends on the quantum state produced by the call. Reasoning about
the continuation must therefore retain the dependence between runtime
and output state. We seek parameterized predicates that carry this
information through program composition and support the quantitative
questions raised by the walk.

\section{Runtime Semantics and Parameterized Predicates}
\label{sec:main-semantics}

To reason compositionally about both runtime and quantum state
transformations, we refine the standard denotational semantics $\sem{S}$
of a quantum program $S$ into runtime-graded completely positive maps
$\sem{S}_{\mathrm{rt}}=(\sem{S}_n)_{n\ge0}$ and test them against
scale-indexed families of bounded effects $F=(F_s)_{s>0}$.

\subsection{Programs and Their Runtime-Graded Semantics}
\label{main:sem:programs}

Fix a finite set $\mathrm{Var}$ of quantum variables. Each variable $q$
has a separable Hilbert space $\cH_q$, possibly infinite-dimensional, and
a distinguished unit vector $\ket0_q$. The joint space and its partial
density operators are
\[
 \cH\triangleq \bigotimes_{q\in\mathrm{Var}}\cH_q,
 \qquad
 \mathcal D(\cH)\triangleq \{\rho\in\Top{\cH}:\rho\ge0,\ \Tr(\rho)\le1\}.
\]
Here $\Top{\cH}$ is the space of trace-class operators.
Following standard quantum while-language conventions~\cite{Ying11,YuEtAl2026},
programs are generated by
\[
\begin{aligned}
 S::={}&\mathbf{skip}\mid\mathbf{abort}\mid q:=\ket0
 \mid\bar q:=U[\bar q]\mid S_1;S_2\\
 &\mid\mathbf{if}\ (\square_{i\in J}\ M[\bar q]=i\to S_i)\ \mathbf{fi}
 \mid\mathbf{while}\ M[\bar q]=1\ \mathbf{do}\ B\ \mathbf{od}.
\end{aligned}
\]
Here $U$ is unitary, and operators on a subsystem are extended by the
identity on the remaining variables. Initialization resets $q$ to
$\ket0_q$. A conditional measurement has a finite
outcome set $J$ and satisfies $\sum_{i\in J}M_i^\dagger M_i=I$.
A loop uses a binary measurement: outcome zero exits, and outcome one
enters the body. Write
\[
 \cM_i(\rho)\triangleq M_i\rho M_i^\dagger,
 \quad \cR_q(\rho)\triangleq \ketbra00_q\otimes\Tr_q(\rho),
 \quad \mathcal U_{\bar q}(\rho)\triangleq U\rho U^\dagger.
\]

\begin{definition}[Runtime-graded denotational semantics]
\label{main:sem:graded-definition}
The runtime-graded denotational semantics of a program $S$ is the sequence
\[
 \sem{S}_{\mathrm{rt}}\triangleq (\sem{S}_n)_{n\ge0},
 \qquad \sem{S}_n:\Top{\cH}\to\Top{\cH},
\]
where the grade-$n$ map $\sem{S}_n$ produces the subnormalized output state of
executions terminating after exactly $n$ total loop-body entries
(accumulating iterations across nested loops, while loop exits and loop-free
commands contribute zero). For $n\ge0$, the grades are defined by
\begin{equation}
\label{main:sem:grade-rules}
\begin{aligned}
 &\sem{\mathbf{skip}}_n=\one_{\{n=0\}}\id,
 \qquad \sem{\mathbf{abort}}_n=0,\\[3pt]
 &\sem{q:=\ket0}_n=\one_{\{n=0\}}\cR_q,
 \qquad \sem{\bar q:=U[\bar q]}_n=\one_{\{n=0\}}\mathcal U_{\bar q},\\[3pt]
 &\sem{S_1;S_2}_n=\sum_{a+b=n}\sem{S_2}_b\circ\sem{S_1}_a,\\[3pt]
 &\sem{\mathbf{if}\ (\square_i\ M[\bar q]=i\to S_i)\ \mathbf{fi}}_n
 =\sum_i\sem{S_i}_n\circ\cM_i,\\[3pt]
 &\sem{\mathbf{while}\ M[\bar q]=1\ \mathbf{do}\ B\ \mathbf{od}}_n
 =\sum_{j=0}^{n}
 \sum_{\substack{k_1,\ldots,k_j\in\NN_+\\k_1+\cdots+k_j=n}}
 \cM_0\circ\cC_{k_j}\circ\cdots\circ\cC_{k_1}.
\end{aligned}
\end{equation}
Here $\cC_k\triangleq \sem{B}_{k-1}\circ\cM_1$ for $k\ge1$:
entering the body contributes one iteration, and executing $B$ contributes
the remaining $k-1$. The $j=0$ term is $\one_{\{n=0\}}\cM_0$.
\end{definition}

\begin{proposition}[Well-definedness and compatibility]
\label{main:sem:well-defined}
For every program $S$, the grades $(\sem{S}_n)_{n\ge0}$ are well-defined
completely positive, trace-nonincreasing maps, and
$\sum_{n\ge0}\sem{S}_n(\rho)=\sem{S}(\rho)$ in trace norm for every
$\rho\in\mathcal D(\cH)$.
\end{proposition}
The proof in Section~B of the supplemental text proceeds by structural
induction, using the trace-nonincreasingness of finite partial sums to
justify trace-norm convergence and series regrouping.

\subsection{From Costs to Bounded Runtime Tests}
\label{main:sem:runtime-tests}

From the grades of $S$, define the exact-runtime termination effects,
the total termination effect, and the divergence effect:
\begin{equation}
\label{main:sem:effects}
 E_n^S\triangleq \sem{S}_n^\dagger(I),
 \qquad Q_S\triangleq \sum_{n\ge0}E_n^S=\sem{S}^\dagger(I),
 \qquad D_S\triangleq I-Q_S.
\end{equation}
Here the adjoint is defined by trace duality. For a normalized input $\rho$, let $\tau_S$ denote the runtime of $S$,
with $\tau_S=\infty$ on nontermination. Then
$\Tr(E_n^S\rho)=\Prob_\rho(\tau_S=n)$ and
$\Tr(D_S\rho)=\Prob_\rho(\tau_S=\infty)$. Because runtime moments and expected costs can diverge to $+\infty$ on
infinite-dimensional spaces, they cannot in general be represented by
bounded operators. Following \cite{YuEtAl2026}, we represent the unnormalized $k$-th
runtime moment ($k > 0$) on terminating runs and the total expected
runtime as closed positive quadratic forms (which extend positive
operators by allowing the value $+\infty$):
\begin{align}
 M_k^S&\triangleq \sum_{n\ge0}n^kE_n^S,\label{main:sem:moments}\\
 C_S&\triangleq \sup_{N\ge1}\left(\sum_{n\ge0}\min(n,N)E_n^S+ND_S\right)
      =M_1^S+\infty D_S.\label{main:sem:cost}
\end{align}
Positive series and suprema are taken in the order of quadratic forms,
allowing the value $+\infty$. For a closed positive form $P$,
we write $\Tr(P\rho)$ for its extended expectation, obtained by increasing
bounded approximations to $P$.
In particular, $\Tr(C_S\rho)=\E_\rho[\tau_S]$ by monotone convergence of
the truncated expectations $\E_\rho[\min(\tau_S,N)]$.

\paragraph{Motivating the predicate domain and calculus.}
We evaluate individual execution costs via a bounded test before taking
expectations, using the scalar normalized resolvent at scale $s>0$:
\[
 r_s(t)\triangleq \frac{s}{s+t}\quad(t\ge0),\qquad r_s(\infty)\triangleq 0.
\]
Because $r_s$ is bounded and vanishes at $\infty$, its expectation remains
finite and retains information about terminating executions even when
$\E_\rho[\tau_S]=\infty$. Its large-scale behavior determines the
runtime properties studied in Section~\ref{sec:main-extraction}.

Lifting $r_s$ to a bounded positive terminal-cost operator $G$ yields the
normalized resolvent $R_s(G) \triangleq s(sI + G)^{-1}$, which is a quantum effect
(an operator between $0$ and $I$). Because $R_s$ reverses the operator
order ($0 \le G \le P$ implies $R_s(P) \le R_s(G)$), it turns an upper
bound on terminal cost into a lower bound on the bounded runtime test.
Since runtime $n$ adds $nI$ to the terminal cost,
\begin{equation}
\label{main:sem:cost-shift}
 \res{s}{nI+G}=\frac{s}{s+n}\res{s+n}{G},
 \qquad
 \sum_{n\ge0}\sem{S}_n^\dagger(\res{s}{nI+G})
 =\sum_{n\ge0}\frac{s}{s+n}\sem{S}_n^\dagger(\res{s+n}{G}).
\end{equation}
The terminal test is therefore evaluated at scale $s+n$ and weighted by
$s/(s+n)$. However, because a program may terminate after different runtimes $n$
across branches, averaging $\frac{s}{s+n} R_{s+n}(G)$ over $n$ produces
a convex combination of shifted resolvents that cannot be written as
$R_s(P)$ for any single cost operator $P$ in general. To keep the assertion domain
closed under branching and sequential pullbacks, we therefore generalize
from resolvents of single costs to arbitrary scale-indexed effect
families $F = (F_s)_{s>0}$, on which accumulating $n$ steps of runtime
acts by $F_s \mapsto \frac{s}{s+n} F_{s+n}$.

This transformation respects sequential execution because
\begin{equation}
\label{main:sem:scale-identity}
 \frac{s}{s+c+t}=\frac{s}{s+c}\frac{s+c}{s+c+t}
 \qquad(s>0,\ c,t\ge0).
\end{equation}
This scale-shift law also admits an operational interpretation via a
random deadline. Let $T_s \ge 0$ be a deadline independent of the execution,
with density $s/(s+t)^2$ on $[0,\infty)$, so that its median is $s$ and the
probability of surviving at least $t$ steps is 
$$\mathbb{P}(T_s \ge t) = \frac{s}{s+t}.$$
Then conditioning on survival past $c$ steps simply shifts the scale
of the remaining deadline to $s+c$:
\[
  \mathbb{P}(T_s - c \ge t \mid T_s \ge c) = \frac{s+c}{s+c+t}.
\]
Thus, after $c$ iterations, the remaining duration follows the same law
at scale $s+c$.

\paragraph{Running example: runtime semantics of Hadamard walk.}
For $\mathsf{Walk}$ from Section~\ref{sec:motivating-example}, write
\[
 \Pi_0\triangleq I_2\otimes\ketbra00,
 \qquad U\triangleq V(C_{1/2}\otimes I),
 \qquad W\triangleq U(I-\Pi_0).
\]
Definition~\ref{main:sem:graded-definition} gives every component of
$\sem{\mathsf{Walk}}_{\mathrm{rt}}$:
\[
 \sem{\mathsf{Walk}}_n(\rho)
 =\Pi_0W^n\rho(W^\dagger)^n\Pi_0
 \qquad(n\ge0).
\]
The corresponding runtime observables are
\[
\begin{aligned}
 E_n^{\mathsf{Walk}}
   &=(W^\dagger)^n\Pi_0W^n,\\
 Q_{\mathsf{Walk}}
   &=\sum_{n\ge0}(W^\dagger)^n\Pi_0W^n,\\
 D_{\mathsf{Walk}}
   &=I-Q_{\mathsf{Walk}}
     =\lim\limits_{N\to\infty}(W^\dagger)^NW^N,\\
 M_k^{\mathsf{Walk}}
   &=\sum_{n\ge1}n^k(W^\dagger)^n\Pi_0W^n
     \qquad(k>0),\\
 C_{\mathsf{Walk}}
   &=\sum_{n\ge1}(W^\dagger)^nW^n.
\end{aligned}
\]
Here the limit for $D_{\mathsf{Walk}}$ is taken in the strong operator
topology. The expressions for $D_{\mathsf{Walk}}$ and $C_{\mathsf{Walk}}$
follow from $W^\dagger W=I-\Pi_0$: since
$(W^\dagger)^nW^n=\sum_{m\ge n}E_m^{\mathsf{Walk}}+D_{\mathsf{Walk}}$
is the effect of entering the loop body at least $n$ times, summing
these tail effects over $n\ge1$ yields $C_{\mathsf{Walk}}$.

\subsection{Parameterized Predicates and Semantic Validity}
\label{main:sem:validity}

We now formalize the space of scale-indexed effect families and define
the semantic validity of Hoare triples.

\begin{definition}[Parameterized predicates and semantic validity]
\label{main:sem:validity-definition}
Write $\Eff{\cH}\triangleq \{E\in\Bop{\cH}:0\le E\le I\}$ for the effects.
The predicate space is
\[
 \Pred{\cH}\triangleq \prod_{s>0}\Eff{\cH},
 \qquad A\le F\iff A_s\le F_s\text{ for all }s>0,
\]
which forms an $\omega$-CPO under the pointwise L\"owner order.
Suprema of increasing sequences are taken pointwise in the strong
operator topology; $\mathbf0$ and $\mathbf I$ denote the constant zero
and identity families. For $A,F\in\Pred{\cH}$, the Hoare triple
$\{A\}\,S\,\{F\}$ is valid for total correctness, written
$\models_{\mathrm{tot}}\{A\}\,S\,\{F\}$, if for every $s>0$ and
$\rho\in\mathcal D(\cH)$,
\begin{equation}
\label{main:sem:total-validity}
 \Tr(A_s\rho)\le\sum_{n\ge0}\frac{s}{s+n}
     \Tr\bigl(F_{s+n}\sem{S}_n(\rho)\bigr).
\end{equation}
It is valid for partial correctness, written
$\models_{\mathrm{par}}\{A\}\,S\,\{F\}$, if for every such $s,\rho$,
\begin{equation}
\label{main:sem:partial-validity}
 \Tr(A_s\rho)\le\sum_{n\ge0}\frac{s}{s+n}
     \Tr\bigl(F_{s+n}\sem{S}_n(\rho)\bigr)
     +\Tr(\rho)-\Tr\bigl(\sem{S}(\rho)\bigr).
\end{equation}
\end{definition}

Both notions evaluate the same discounted postcondition on terminating
executions. Total correctness assigns nonterminating runs weight zero,
whereas partial correctness credits them with weight one (contributing the
divergence probability $\Tr(D_S\rho)=\Tr(\rho)-\Tr(\sem{S}(\rho))$).
Because the precondition is a quantitative lower bound, $\models_{\mathrm{tot}}$ can also describe programs that terminate with probability less than one.

\paragraph{Reusing total-cost proofs.}
When a subprogram already has a verified expected-cost bound in the
calculus of \cite{YuEtAl2026}, we do not need to re-verify it from scratch: the
normalized resolvent $R(P) \triangleq (R_s(P))_{s>0} \in \Pred{\cH}$
translates any valid total-cost contract directly into a parameterized
specification (proved via the Choi-Davis-Jensen inequality in Section~G of the supplemental text).

\begin{theorem}[Translation of total-cost contracts]
\label{main:sem:cost-translation}
For closed positive quadratic forms $P,G$, if
$\Tr(P\rho)\ge\Tr(G\sem{S}(\rho))+\Tr(C_S\rho)$ for every
$\rho\in\mathcal D(\cH)$, then
$\models_{\mathrm{tot}}\{R(P)\}\,S\,\{R(G)\}$.
\end{theorem}
Thus a total-cost Hoare triple from~\cite{YuEtAl2026} supplies a valid
specification in the parameterized predicate domain.
\section{A Compositional Resolvent Calculus and Loop Solutions}
\label{sec:main-calculus}

In this section, we develop a compositional calculus for the weakest
preconditions $\wp(S,F)$ and $\wlp(S,F)$.
The structural rules propagate postcondition families through the program
and reduce loops to fixed-point equations on $\Pred{\cH}$.
We then prove uniqueness of the loop solution and give certificates for
exact and approximate specifications. The proofs appear in
Section~C of the supplemental text.

\subsection{Weakest Preconditions and Structural Rules}
\label{main:calc:rules}

\begin{definition}[Weakest preconditions]
\label{main:calc:wp-definition}
For a program $S$ and postcondition $F\in\Pred{\cH}$, the weakest
precondition $\wp(S,F)$ and weakest liberal precondition $\wlp(S,F)$
are the largest valid preconditions for total and partial correctness,
respectively. Equivalently, they satisfy
\[
\begin{aligned}
 \models_{\mathrm{tot}}\{A\}\,S\,\{F\}
 &\iff A\le\wp(S,F),\\
 \models_{\mathrm{par}}\{A\}\,S\,\{F\}
 &\iff A\le\wlp(S,F).
\end{aligned}
\]
\end{definition}

\begin{theorem}[Weakest preconditions and structural rules]
\label{main:calc:semantics}
For every program $S$ and postcondition $F\in\Pred{\cH}$,
the weakest precondition exists and is given at each scale $s>0$ by
\begin{equation}
 \wp(S,F)_s
 =\sum_{n\ge0}\frac{s}{s+n}\sem{S}_n^\dagger(F_{s+n}).
 \label{main:calc:grade-wp}
\end{equation}
The transformer $\wp(S,\cdot)$ is monotone, preserves increasing
sequential suprema, and satisfies the structural rules in
Figure~\ref{main:calc:fig-rules}.
\end{theorem}

\begin{figure}[!htb]
\centering
\small
\renewcommand{\arraystretch}{1.3}
\begin{tabularx}{\linewidth}{@{}lX@{}}
\toprule
Command $S$ & $\wp(S,F)_s$\\
\midrule
$\mathbf{skip}$ & $F_s$\\
$\mathbf{abort}$ & $0$\\
$q:=\ket0$ & $\displaystyle\sum_j\ketbra{j}{0}_qF_s\ketbra{0}{j}_q$\\
$\bar q:=U[\bar q]$ & $U^\dagger F_sU$\\
$S_1;S_2$ & $\wp(S_1,\wp(S_2,F))_s$\\
$\mathbf{if}\;(\square_i\,M[\bar q]=i\to S_i)\;\mathbf{fi}$
 & $\displaystyle\sum_i M_i^\dagger\wp(S_i,F)_sM_i$\\
$\mathbf{while}\ M[\bar q]=1\ \mathbf{do}\ B\ \mathbf{od}$
 & $(\operatorname{lfp}\cT_F)_s$\\
\bottomrule
\end{tabularx}
\par\smallskip
\begin{minipage}{\linewidth}
For $L=\mathbf{while}\ M[\bar q]=1\ \mathbf{do}\ B\ \mathbf{od}$,
$\operatorname{lfp}\cT_F$ denotes the least fixed point of
\begin{equation}
 (\cT_FX)_s
 =M_0^\dagger F_sM_0+
   \frac{s}{s+1}M_1^\dagger\wp(B,X)_{s+1}M_1.
 \label{main:calc:loop-functional}
\end{equation}
\end{minipage}
\caption{Weakest-precondition rules for parameterized predicates.
The initialization rule uses an orthonormal basis $\{\ket j_q\}_j$.
The loop's continuation branch charges one iteration and applies the
body's transformer at scale $s+1$.}
\Description{Weakest-precondition rules for parameterized predicates.}
\label{main:calc:fig-rules}
\end{figure}

\begin{proposition}[Weakest liberal preconditions]
\label{main:calc:liberal}
For every program $S$ and postcondition $F$,
\begin{equation}
 \wlp(S,F)=\wp(S,F)+\mathbf D_S,
 \qquad \mathbf D_S\triangleq (D_S)_{s>0}.
 \label{main:calc:liberal-equation}
\end{equation}
\end{proposition}

This follows directly from the nontermination term in partial validity.

Choosing the terminal postcondition $\mathbf I$ defines the
\emph{runtime family} and its liberal counterpart:
\begin{equation}
 H_s^S\triangleq \wp(S,\mathbf I)_s
       =\sum_{n\ge0}\frac{s}{s+n}E_n^S,
 \qquad
 L_s^S\triangleq \wlp(S,\mathbf I)_s=H_s^S+D_S.
 \label{main:calc:runtime-families}
\end{equation}
For a normalized input $\rho$, $\Tr(\rho H_s^S)$ is the probability
of termination before the independent deadline $T_s$ from
Section~\ref{main:sem:runtime-tests}.

\subsection{Certifying Loop Specifications}
\label{main:calc:certificates}

We verify a loop specification by substituting it into the loop equation.
The check uses the body's transformer $\wp(B,\cdot)$, allowing a verified
subprogram to serve as the loop body.

\begin{theorem}[Exact loop specifications and bounds]
\label{main:calc:uniqueness}
For every loop $L$ as in~\eqref{main:calc:loop-functional} and every
$F\in\Pred{\cH}$, the equation $X=\cT_FX$ has exactly one solution
in $\Pred{\cH}$, namely $\wp(L,F)$.
Moreover, for all $X,Y\in\Pred{\cH}$,
\begin{equation}
 X\le\cT_FX\ \Longrightarrow\ X\le\wp(L,F),
 \qquad
 \cT_FY\le Y\ \Longrightarrow\ \wp(L,F)\le Y.
 \label{main:calc:two-sided}
\end{equation}
\end{theorem}

The uniqueness conclusion is notable. In general, loop equations
may admit multiple fixed points, so finding a solution alone does
not identify the program semantics: one must establish that it is
the least fixed point. Here, any solution in $\Pred{\cH}$ is already
the weakest precondition, with no separate minimality argument needed.

For example, consider
$\mathbf{while}\ \mathbf{true}\ \mathbf{do}\ \mathbf{skip}\ \mathbf{od}$,
with guard operators $M_0=0$ and $M_1=I$.
Its ordinary weakest-precondition equation is simply
$X=X$, which admits every effect as a solution. In our calculus,
a fixed point $X\in\Pred{\cH}$ must instead satisfy
\[
 X_s=\frac{s}{s+1}X_{s+1}
     =\frac{s}{s+N}X_{s+N}
 \qquad(s>0,\ N\ge1).
\]
Since $0\le X_{s+N}\le I$, we obtain
$0\le X_s\le \frac{s}{s+N}I$ for every $N$.
Letting $N\to\infty$ shows that the unique solution is $\mathbf0$,
representing the absence of terminating executions.

For an asymptotic specification, the residual $A-\cT_FA$ measures how
closely a candidate $A$ satisfies the equation for one loop step.
The next result turns its decay rate into an approximation guarantee
for the whole loop.

\begin{theorem}[Certification of an asymptotic loop solution]
\label{main:calc:residual}
Let $L$ be a loop as in \eqref{main:calc:loop-functional}, $F \in \Pred{\cH}$, and
$A = (A_s)_{s>0}$ a uniformly bounded self-adjoint family ($\sup_{s>0}\|A_s\| < \infty$,
allowing truncated asymptotic expansions that need not lie in $[0,I]_{\cH}$).
Applying $\mathcal{T}_F$ to $A$ via \eqref{main:calc:loop-functional} and writing $d_s \triangleq A_s - (\mathcal{T}_F A)_s$
for the one-step residual, we have for every $r > 0$ as $s \to \infty$:
\begin{equation}
 \norm{d_s}=O(s^{-r-1})
 \quad\Longrightarrow\quad
 \norm{A_s-\wp(L,F)_s}=O(s^{-r}).
 \label{main:calc:residual-bound}
\end{equation}
If there is $0\le\kappa<1$ such that
\[
 \frac{u}{u+1}M_1^\dagger\wp(B,\mathbf I)_{u+1}M_1
 \le\kappa I\qquad(u>0),
\]
then
\begin{equation}
 \norm{d_s}=O(s^{-r})
 \quad\Longrightarrow\quad
 \norm{A_s-\wp(L,F)_s}=O(s^{-r}).
 \label{main:calc:contractive-residual}
\end{equation}
Both implications also hold with $o$ in place of $O$.
\end{theorem}

\paragraph{Running example: backward derivation of $\wp$ and the runtime family.}
Retain $\Pi_0=I_2\otimes\ketbra{0}{0}$, $U=V(C_{1/2}\otimes I)$, and
$W=U(I-\Pi_0)$ from Section~\ref{main:sem:runtime-tests}, and set
$\Pi_{\ne0}\triangleq I-\Pi_0$. With coin basis $\ket L,\ket R$, let
$B_{\mathsf{Walk}}$ denote the loop body of $\mathsf{Walk}$
(Figure~\ref{fig:mot-first-step}), which applies the Hadamard coin and
conditional shift
\[
 C_{1/2}=\frac1{\sqrt2}
 \begin{pmatrix}1&1\\1&-1\end{pmatrix},\qquad
 V\ket{L,x}=\ket{L,x-1},\quad
 V\ket{R,x}=\ket{R,x+1}.
\]
We apply the structural rules of Theorem~\ref{main:calc:semantics}
(Figure~\ref{main:calc:fig-rules}) to derive $\wp(\mathsf{Walk},F)$ and
$H^{\mathsf{Walk}}$ backward from the program syntax, reducing the
program verification task to a concrete operator series.

First, for any postcondition family $X\in\Pred{\cH}$ at the end of
$B_{\mathsf{Walk}}$, the unitary and sequencing rules propagate $X_s$
backward through the two body commands (read from bottom to top):
\[
\begin{array}{l}
 \bigl\{(C_{1/2}^\dagger\otimes I)V^\dagger X_sV(C_{1/2}\otimes I)\bigr\}\\
 \quad\mathtt{coin}:=C_{1/2}[\mathtt{coin}];\\
 \bigl\{V^\dagger X_sV\bigr\}\\
 \quad(\mathtt{coin},\mathtt{position})
          :=V[\mathtt{coin},\mathtt{position}]\\
 \bigl\{X_s\bigr\}.
\end{array}
\]
The shift command pulls $X_s$ back to the intermediate assertion
$V^\dagger X_sV$, which the sequencing rule passes as the postcondition to
the coin update, yielding the body transformer
$\wp(B_{\mathsf{Walk}},X)_s=U^\dagger X_sU$.

Now let $Y\in\Pred{\cH}$ be a candidate precondition family at the loop
guard $M=\{M_0,M_1\}$, where $M_0=\Pi_0$ and $M_1=\Pi_{\ne0}$. 
Substituting the body transformer
$\wp(B_{\mathsf{Walk}},Y)_{s+1}=U^\dagger Y_{s+1}U$
into the loop functional~\eqref{main:calc:loop-functional} gives
the fixed-point equation
\begin{equation}
 Y_s=\Pi_0F_s\Pi_0+
     \frac{s}{s+1}\Pi_{\ne0}U^\dagger Y_{s+1}U\Pi_{\ne0}
    =\Pi_0F_s\Pi_0+\frac{s}{s+1}W^\dagger Y_{s+1}W.
 \label{main:calc:walk-equation}
\end{equation}
By Theorem~\ref{main:calc:uniqueness}, \eqref{main:calc:walk-equation} has
a unique solution in $\Pred{\cH}$, namely $Y=\wp(\mathsf{Walk},F)$.

Because the walk starts at $x=1$ and terminates as soon as it reaches
$x=0$, no continuing execution ever visits $x\le0$. All surviving states
therefore remain supported on the positive half-line
$\cH_+\triangleq \CC^2\otimes\ell^2(\NN_+)$, and one unitary step $U$ maps $\cH_+$
into $\operatorname{ran}\Pi_0\oplus\cH_+$. This invariance allows us
to compress the full-space equation~\eqref{main:calc:walk-equation} to
$\cH_+$. Let $\Pi_+$ project onto $\cH_+$ and split $U|_{\cH_+}$ into
the one-step absorbing and continuing operators
\[
 B\triangleq \Pi_0U|_{\cH_+},\qquad
 K\triangleq \Pi_+U|_{\cH_+}.
\]
On the basis $\{\ket{L,x},\ket{R,x}:x\in\NN_+\}$ of $\cH_+$, the
absorbing operator acts by
\[
\begin{aligned}
 B\ket{L,1} &= \tfrac1{\sqrt2}\ket{L,0},
 \qquad B\ket{R,1} = \tfrac1{\sqrt2}\ket{L,0},\\
 B\ket{c,x} &= 0 \qquad(c\in\{L,R\},\,x\ge2).
\end{aligned}
\]
The continuing operator acts by
\[
\begin{aligned}
 K\ket{L,1} &= \tfrac1{\sqrt2}\ket{R,2}, &
 K\ket{L,x} &= \tfrac1{\sqrt2}
   \bigl(\ket{L,x-1}+\ket{R,x+1}\bigr)
   \quad(x\ge2),\\
 K\ket{R,1} &= -\tfrac1{\sqrt2}\ket{R,2}, &
 K\ket{R,x} &= \tfrac1{\sqrt2}
   \bigl(\ket{L,x-1}-\ket{R,x+1}\bigr)
   \quad(x\ge2).
\end{aligned}
\]
In particular, every terminating run from $\cH_+$ exits at the origin in
state $\ket{L,0}$. Compressing~\eqref{main:calc:walk-equation} to $\cH_+$
and unfolding the resulting recurrence yields the restricted
precondition family $Y_s^+$
(Proposition~E.1 of the supplemental text):
\begin{equation}
\begin{aligned}
 Y_s^+ &\triangleq \Pi_+Y_s|_{\cH_+}
          =\Pi_+\wp(\mathsf{Walk},F)_s|_{\cH_+}\\
       &\phantom{:}= \sum_{n\ge1}\frac{s}{s+n}
          \bra{L,0}F_{s+n}\ket{L,0}
          (K^\dagger)^{n-1}B^\dagger BK^{n-1}.
\end{aligned}
 \label{main:calc:walk-general}
\end{equation}
Here the $n$-th term evaluates $F_{s+n}$ on the returned state $\ket{L,0}$
and weights it by $s/(s+n)$ and the $n$-step termination effect
$(K^\dagger)^{n-1}B^\dagger BK^{n-1}$.

Finally, taking $F=\mathbf I$ in~\eqref{main:calc:walk-general} as
in~\eqref{main:calc:runtime-families} gives the restricted runtime
resolvent family $H_s$ on $\cH_+$ and its expectation $h_s$ on the initial
state $\rho_0=\ketbra{R,1}{R,1}$:
\begin{equation}
\begin{aligned}
 H_s &\triangleq \Pi_+H_s^{\mathsf{Walk}}|_{\cH_+}
       = \sum_{n\ge1}\frac{s}{s+n}
          (K^\dagger)^{n-1}B^\dagger BK^{n-1},\\
 h_s &\triangleq \Tr(\rho_0H_s)
       = \sum_{n\ge1}\frac{s}{s+n}
          \norm{BK^{n-1}\ket{R,1}}^2.
\end{aligned}
 \label{main:calc:walk-resolvent}
\end{equation}
Section~E.2 of the supplemental text derives the asymptotic expansion of
$h_s$, from which Section~\ref{sec:main-extraction} extracts the walk's
runtime properties.

\section{Extracting Termination Probabilities, Moments, and Divergence Rates}
\label{sec:main-extraction}

Once the calculus has verified the runtime family $H^S=\wp(S,\mathbf I)$, this
parameterized predicate serves two roles. Within a larger program $T;S$, it can
be reused compositionally as the postcondition for the preceding command $T$
via the sequencing rule
\[
 H^{T;S}=\wp(T,H^S).
\]
For the top-level program of interest, we use $H^S$ to
answer three operational runtime questions: how likely the program is to
terminate, which runtime moments are finite, and how quickly an infinite moment
diverges.

To see how $H^S$ encodes these observables, recall
from~\eqref{main:calc:runtime-families} that
\[
 H_s^S=\sum_{n\ge0}\frac{s}{s+n}E_n^S.
\]
In $H_s^S$, the discount factor $s/(s+n)$ weights each terminating run of
duration $n$ against the scale $s$ while assigning nonterminating runs weight
zero. As $s\to\infty$, each weight $s/(s+n)$ increases to $1$, so the limit of
$H_s^S$ recovers the termination effect $Q_S$, while its large-$s$ behavior
encodes the moments and tail behavior of terminating runs.
The moment forms $M_k^S=\sum_{n\ge0}n^kE_n^S$ count only terminating runs.
To measure the divergence of an infinite moment, we cap the duration counted
on each terminating run at $N$ and let $N$ grow, defining the
bounded capped-moment operators
\begin{equation}
 \label{main:extract-capped}
 \widetilde M_k^S(N)\triangleq \sum_{n\ge0}\min(n,N)^kE_n^S
 \qquad(k>0,\ N>0).
\end{equation}

For a normalized input state $\rho$, taking expectations of these operators
gives the termination probability and capped moments:
\[
 q\triangleq \Tr(\rho Q_S)=\Prob_\rho(\tau_S<\infty),\qquad
 c_k(N)\triangleq \Tr\bigl(\rho\widetilde M_k^S(N)\bigr).
\]
For $q>0$, dividing $\Tr(\rho M_k^S)$ or $c_k(N)$ by $q$ gives the
corresponding moment conditional on termination. For unconditional properties,
nonterminating runs contribute $(1-q)N^k$ to the capped moment, giving
$\E_\rho[\min(\tau_S,N)^k]=(1-q)N^k+c_k(N)$. In particular, $q<1$ implies an
infinite unconditional expected runtime; when $q=1$, that expectation is
$\Tr(\rho M_1^S)$.

There are two routes to extract these quantities, depending on the available
specification (with proofs in Section~D of the supplemental text):
Table~\ref{main:extract-exact-table} recovers the termination effect $Q_S$ and
moment forms $M_k^S$, and bounds the capped-moment operators
$\widetilde M_k^S(N)$ directly from the exact runtime family $H^S$, whereas
Table~\ref{main:extract-asymptotic-table} reads out the scalar quantities $q$,
$\Tr(\rho M_k^S)$, and $c_k(N)$ from a certified large-$s$ expansion of the
expectation $\Tr(\rho H_s^S)$ on an input state.

\subsection{From the Exact Runtime Family}
\label{main:extract-exact}

When an exact expression for the operator family $H_s^S$ is available, it
supports all three checks in Table~\ref{main:extract-exact-table}. While the
$s\to\infty$ limit in the first row directly recovers $Q_S$, extracting higher
moments requires isolating powers of $n$ from the rational weights $s/(s+n)$.
To do so, the second and third rows combine $Q_S$ with $H^S$ at the rescaled
parameters $s,s/2,\ldots,s/k$ to form a $k$-th order finite-difference operator
$B_k^S(s)$, which recovers $M_k^S$ as $s\to\infty$ and bounds the
capped-moment operator $\widetilde M_k^S(N)$ up to constant factors when
evaluated at $s=N$.

\begingroup
\small
\setlength{\abovedisplayskip}{5pt}
\setlength{\belowdisplayskip}{5pt}
\setlength{\abovedisplayshortskip}{3pt}
\setlength{\belowdisplayshortskip}{3pt}
\setlength{\parskip}{2pt}
\renewcommand{\arraystretch}{1.15}
\setlength{\LTleft}{0pt plus 1fill}
\setlength{\LTright}{0pt plus 1fill}
\setlength{\LTcapwidth}{\linewidth}
\begin{longtable}{@{}>{\raggedright\arraybackslash}p{.21\linewidth}p{\dimexpr.79\linewidth-2\tabcolsep\relax}@{}}
\caption{Three runtime properties from an exact verified family.
Here $\rho$ is a normalized input, $q=\Tr(\rho Q_S)$, and $k\ge1$ is an
integer. Conditional quantities require $q>0$.}
\label{main:extract-exact-table}\\
\toprule
Verification target & Rule from the exact family $H^S$ \\
\midrule
\endfirsthead
\toprule
Verification target & Rule from the exact family $H^S$ \\
\midrule
\endhead
\midrule
\endfoot
\bottomrule
\endlastfoot
\textbf{Termination probability}\par
Theorem~D.1\par (supplemental text)
& Compute
\[
 Q_S=\lim_{s\to\infty}H_s^S,\qquad D_S=I-Q_S.
\]
The limit is taken in the strong operator topology. \\
\addlinespace[4pt]
\textbf{Runtime moments}\par
Theorem~D.3\par (supplemental text)
& For an integer $k\ge1$, form
\[
 B_k^S(s)\triangleq \frac{s^k}{k!}
       \left[Q_S+\sum_{j=1}^k(-1)^j\binom{k}{j}H_{s/j}^S\right].
\]
As $s\to\infty$, the quadratic forms of $B_k^S(s)$ increase monotonically
to $M_k^S$. The conditional moments are
\[
 \E_\rho[\tau_S^k\mid\tau_S<\infty]
 =\frac{\Tr(\rho M_k^S)}q
 =\frac1q\lim_{s\to\infty}\Tr\bigl(\rho B_k^S(s)\bigr).
\] \\
\addlinespace[4pt]
\textbf{Divergence rates}\par
Theorem~D.5\par (supplemental text)
& Evaluate the same expression at the runtime cap $N$:
\[
 B_k^S(N)\le\widetilde M_k^S(N)\le(k+1)!\,B_k^S(N).
\]
Taking expectations and dividing by $q$ gives the growth order of the
conditional capped moment, up to constant factors:
\[
 \E_\rho[\min(\tau_S,N)^k\mid\tau_S<\infty]=\frac{c_k(N)}q.
\] \\
\end{longtable}
\endgroup

The mechanism unifying the second and third rows of
Table~\ref{main:extract-exact-table} comes from expanding $H_{s/j}^S$ in the
definition of $B_k^S(s)$: partial-fraction cancellation turns this alternating
linear combination into a positive series with a smooth cutoff weight:
\[
 B_k^S(s)=\sum_{n\ge0}n^k\left(\prod_{j=1}^k\frac{s}{s+jn}\right)E_n^S.
\]
For short runs ($n\ll s$), the product weight is close to $1$, so the runtime
weight behaves like $n^k$; for long runs ($n\gg s$), the $k$ denominators damp
$n^k$ and cap the weight at $\approx s^k/k!$. Consequently, $B_k^S(s)$ acts as
a smooth counterpart to the hard cutoff $\min(n,s)^k$: as $s\to\infty$, it
increases monotonically to the $k$-th moment form $M_k^S$, while setting
$s=N$ directly approximates the capped-moment operator $\widetilde M_k^S(N)$.
For any input $\rho$, the moment
$\Tr(\rho M_k^S)=\lim_{s\to\infty}\Tr(\rho B_k^S(s))$ is finite if and only
if $\Tr(\rho B_k^S(s))$ remains bounded.

\subsection{From an Asymptotic Expansion}
\label{main:extract-asymptotic}

In practice, extracting properties on a fixed normalized input state $\rho$
does not require closed-form operator limits: it suffices to expand the scalar
expectation $h_s\triangleq \Tr(\rho H_s^S)$ as $s\to\infty$. Moreover, when a candidate
loop specification with a known large-$s$ expansion is certified in operator
norm via Section~\ref{main:calc:certificates}, it transfers the same error
bound to $h_s$ on every input. Table~\ref{main:extract-asymptotic-table} shows
how successive terms in the expansion of $h_s$ recover the three runtime
properties, with the target property dictating the remainder accuracy required
from the certificate: the leading constant gives $q$, the coefficients of
$(-s)^{-j}$ give the finite moments, and logarithmic or fractional-power
remainders determine the divergence rate of the next moment.

\begingroup
\small
\setlength{\abovedisplayskip}{5pt}
\setlength{\belowdisplayskip}{5pt}
\setlength{\abovedisplayshortskip}{3pt}
\setlength{\belowdisplayshortskip}{3pt}
\setlength{\parskip}{2pt}
\renewcommand{\arraystretch}{1.15}
\setlength{\LTleft}{0pt plus 1fill}
\setlength{\LTright}{0pt plus 1fill}
\setlength{\LTcapwidth}{\linewidth}
\begin{longtable}{@{}>{\raggedright\arraybackslash}p{.21\linewidth}p{\dimexpr.79\linewidth-2\tabcolsep\relax}@{}}
\caption{Three runtime properties from a verified asymptotic expansion.
Here $k\ge1$ is an integer, and all expansions are as $s\to\infty$.
Conditional quantities require $q>0$.}
\label{main:extract-asymptotic-table}\\
\toprule
Verification target & Rule from a verified expansion of $h_s$ \\
\midrule
\endfirsthead
\toprule
Verification target & Rule from a verified expansion of $h_s$ \\
\midrule
\endhead
\midrule
\endfoot
\bottomrule
\endlastfoot
\textbf{Termination probability}\par
Theorem~D.2\par (supplemental text)
& If $h_s=a_0+o(1)$, then $q=a_0$ and the divergence probability
is $1-a_0$. \\
\addlinespace[4pt]
\textbf{Runtime moments}\par
Theorem~D.4\par (supplemental text)
& If, for real coefficients $a_0,\ldots,a_k$,
\[
 h_s=\sum_{j=0}^k\frac{(-1)^j a_j}{s^j}+o(s^{-k}),
\]
then $a_0=q$ and $a_j=\Tr(\rho M_j^S)$ for $1\le j\le k$.
Such an expansion exists exactly when the $k$-th moment is finite.
The conditional moments are
\[
 \E_\rho[\tau_S^j\mid\tau_S<\infty]=\frac{a_j}{a_0}
 \qquad(1\le j\le k).
\] \\
\addlinespace[4pt]
\textbf{Divergence rates}\par
Theorem~D.6\par (supplemental text)
& Suppose, for real coefficients $a_0,\ldots,a_{k-1}$,
\[
 h_s=\sum_{j=0}^{k-1}\frac{(-1)^j a_j}{s^j}
       +\frac{(-1)^k f(s)}{s^k},\qquad f(s)=o(s).
\]
Then $a_0=q$ and $a_j=\Tr(\rho M_j^S)<\infty$ for $1\le j<k$.
The remainder determines the next moment:
\\*
& 
\textit{Bounded:} If $f(s)=O(1)$, then
$\Tr(\rho M_k^S)=\lim_{s\to\infty}f(s)<\infty$.
The conditional $k$-th moment is $q^{-1}\lim_{s\to\infty}f(s)$.
\\
\addlinespace[2pt]
& 
\textit{Logarithmic:} If $f(s)=\lambda\log s+O(1)$ with $\lambda>0$,
then $c_k(N)=\lambda\log N+O(1)$; the critical moment order is
$\alpha=k$.
\[
 \E_\rho[\min(\tau_S,N)^k\mid\tau_S<\infty]
 =\frac{\lambda}{q}\log N+O(1).
\]
\\
\addlinespace[2pt]
& 
\textit{Fractional power:} If $f(s)=\Theta(s^\beta)$ with $0<\beta<1$,
then $c_k(N)=\Theta(N^\beta)$; the critical moment order is
$\alpha=k-\beta$.
\[
 \E_\rho[\min(\tau_S,N)^k\mid\tau_S<\infty]=\Theta(N^\beta).
\] \\
\end{longtable}
\endgroup

The mechanism behind Table~\ref{main:extract-asymptotic-table} comes from the
finite geometric expansion of the discount weight: for every integer $k\ge1$
and all $n\ge0$,
\[
 \frac{s}{s+n}
 =\sum_{j=0}^{k-1}\frac{(-1)^j n^j}{s^j}
  +\frac{(-1)^k}{s^k}\,\frac{sn^k}{s+n}.
\]
Summing this identity against $\Tr(\rho E_n^S)$ accounts for all three rows in
turn. Whenever the first $k-1$ moments are finite, $a_0=q$ and
$a_j=\Tr(\rho M_j^S)$ appear as the polynomial coefficients of $(-s)^{-j}$ up
to order $s^{-(k-1)}$, leaving the order-$s^{-k}$ remainder factor in the third
row:
\[
 f(s)=\sum_{n\ge0}\frac{sn^k}{s+n}\Tr(\rho E_n^S).
\]
Here the weight $sn^k/(s+n)$ behaves like $n^k$ for $n\le s$ and is damped to
$\le sn^{k-1}$ for $n>s$. As $s\to\infty$, $f(s)$ increases monotonically to
$\Tr(\rho M_k^S)$, so the $k$-th moment is finite if and only if $f(s)$
remains bounded. When the $k$-th moment is infinite, $f(s)$ diverges as
$s\to\infty$; in the logarithmic and fractional-power regimes of
Table~\ref{main:extract-asymptotic-table}, its growth rate at scale $s=N$
directly determines the divergence rate of the capped moment $c_k(N)$.

In the logarithmic and fractional-power cases, the parameter $\alpha$ in the
third row is called the critical moment order because it governs the power-law
tail of terminating runs: $\Prob_\rho(N<\tau_S<\infty)=\Theta(N^{-\alpha})$.
Consequently, $\alpha$ marks the exact threshold for moment finiteness: for
every real moment order $k>0$, runtime
moments $\Tr(\rho M_k^S)$ are finite if and only if $0<k<\alpha$, whereas the
capped moment $c_k(N)$ grows as $\Theta(\log N)$ at the boundary $k=\alpha$ and
as $\Theta(N^{k-\alpha})$ for $k>\alpha$.

\paragraph{Running example: extracting the Hadamard walk's runtime properties.}
For the Hadamard walk on input $\rho=\ketbra{R,1}{R,1}$,
Section~\ref{sec:main-calculus} derived the runtime family $H^{\mathsf{Walk}}$
by the program rules. Starting from the explicit actions of $B$ and $K$ on
the basis states given there, Section~E.2 of the supplemental text evaluates
the trace in~\eqref{main:calc:walk-resolvent} using elementary linear algebra,
combinatorial identities, and calculus, obtaining
\begin{equation}
 \label{main:extract-hadamard-expansion}
 h_s=\Tr(\rho H_s^{\mathsf{Walk}})
     =\frac2\pi-\frac1s+\frac{2\log s}{\pi s^2}+O(s^{-2}).
\end{equation}
Combined with the compositional calculus, these explicit calculations suggest
a route to automated runtime reasoning using computer algebra systems such as
Mathematica or Maple.
Applying the three rows of Table~\ref{main:extract-asymptotic-table} recovers
the runtime properties from Section~\ref{main:mot:runtime-properties}:
\begin{enumerate}
\item \textbf{Termination probability.}
The constant term gives $q=2/\pi$ and divergence probability $1-2/\pi$.
\item \textbf{Conditional mean runtime.}
The coefficient of $-s^{-1}$ gives $\Tr(\rho M_1^{\mathsf{Walk}})=1$,
so $\E_\rho[\tau\mid\tau<\infty]=1/q=\pi/2$.
Since $q<1$, the unconditional expected runtime is infinite.
\item \textbf{Divergence rate.}
At order $k=2$, the remainder is $f(s)=(2/\pi)\log s+O(1)$. Hence
\[
 c_2(N)=\frac2\pi\log N+O(1),\qquad
 \E_\rho[\min(\tau,N)^2\mid\tau<\infty]=\log N+O(1).
\]
The critical moment order is $\alpha=2$: conditional moments are finite
exactly for $0<k<2$, and the terminating tail is
$\Prob_\rho(N<\tau<\infty)=\Theta(N^{-2})$.
\end{enumerate}

\section{Case Studies}
\label{sec:main-case-studies}

Up to this point, our running example has focused on a single program instance: the Hadamard walk started in a fixed state. In practice, however, analyzing one fixed configuration of a quantum loop is rarely an end in itself. Once we can establish a loop's runtime behavior, we want to use that analysis in two further ways: to understand how the behavior changes when we adjust the program's parameters or inputs, and to reason about larger programs that invoke the loop as a subroutine. This section demonstrates both uses and shows how they fit together, using the scalar evaluations in Section~E of the supplemental text as mathematical oracles.

First, Section~\ref{main:case:biased} examines how the absorbing walk behaves when we replace the symmetric Hadamard coin with a biased coin and allow an arbitrary initial coin state. Moving from a single instance to this parameterized family shows which aspects of the runtime behavior are structural to the loop and which depend on the configuration, clarifying how the coin bias and initial state govern the resulting runtime properties. Second, Section~\ref{main:case:nested} turns to the nested program from Figure~\ref{fig:mot-caller}, where an outer loop repeatedly resets the registers and invokes the walk. Here the verified specification of the inner walk becomes a building block for analyzing its caller, allowing us to determine the outer loop's termination probability and conditional expected runtime without re-examining the inner walk's lattice operators. Finally, these two studies connect directly: when the inner Hadamard walk is replaced by the biased walk, the results of Section~\ref{main:case:biased} plug straight into the outer proof of Section~\ref{main:case:nested} to establish the runtime properties of the modified caller.

\subsection{Changing the Coin}
\label{main:case:biased}

Let $L_p$ be the program obtained from $\mathsf{Walk}$ by replacing its
Hadamard coin with
\[
 C_p=\begin{pmatrix}\sqrt p&\sqrt{1-p}\\
                     \sqrt{1-p}&-\sqrt p\end{pmatrix},
 \qquad 0<p<1.
\]
The guard and conditional shift remain the same. We verify the termination
probability, conditional mean, and growth of the capped second moment for
an arbitrary normalized coin state $\rho$ at position $1$.

\paragraph{Program reduction.}
The backward derivation in Section~\ref{sec:main-calculus} applies with
$U_p=V(C_p\otimes I)$, $B_p=\Pi_0U_p|_{\cH_+}$, and
$K_p=\Pi_+U_p|_{\cH_+}$. In particular, the unitary and sequencing rules
give the body transformer $X_s\mapsto U_p^\dagger X_sU_p$, and the loop
rule gives the same recurrence with these operators. Writing $H_{p,s}$
for the compression of $\wp(L_p,\mathbf I)_s$ to $\cH_+$, we obtain
the scalar calculation target
\begin{equation}
 \label{main:case:biased-target}
 h_\rho(s)\triangleq \Tr\bigl((\rho\otimes\ketbra11)H_{p,s}\bigr)
 =\sum_{n\ge1}\frac{s}{s+n}
   \Tr\!\left((\rho\otimes\ketbra11)
       (K_p^\dagger)^{n-1}B_p^\dagger B_pK_p^{n-1}\right).
\end{equation}
\paragraph{Mathematical oracle.}
At this point, program reasoning hands off a pure scalar analysis problem:
given the explicit operators $B_p, K_p$ and initial coin state $\rho$,
evaluate the series \eqref{main:case:biased-target} as $s \to \infty$. Section~E.3 of the supplemental text discharges this
obligation using Legendre polynomial identities (Theorem~E.5 of the supplemental text), yielding
\begin{equation}
 \label{main:case:biased-oracle}
 h_\rho(s)=q_\rho-\frac1s+
             \frac{\lambda_\rho\log s}{s^2}+O(s^{-2}),
\end{equation}
where, with $v_-=\sqrt{1-p}\ket L-\sqrt p\ket R$,
\[
\begin{aligned}
 \beta_\rho&\triangleq \bra{v_-}\rho\ket{v_-},
 &q_\rho&\triangleq 1-(1-\kappa_p)\beta_\rho,
 &\lambda_\rho&\triangleq \frac{4\beta_\rho}{\pi\delta},\\
 \chi&\triangleq 1-2p,
 &\delta&\triangleq 2\sqrt{p(1-p)},
 &\kappa_p&\triangleq
 \frac{\frac4\pi(\delta+\chi\arcsin\chi)-1-\chi^2}{\delta^2}.
\end{aligned}
\]
Here $0<\kappa_p<1$. 

\paragraph{Verified runtime properties.}
Applying Table~\ref{main:extract-asymptotic-table} to
\eqref{main:case:biased-oracle}, with $\tau$ the runtime on
$\rho\otimes\ketbra11$, yields
\begin{equation}
 \label{main:case:biased-results}
\begin{aligned}
 \Prob(\tau<\infty)&=q_\rho,
 \qquad \E[\tau\mid\tau<\infty]=\frac1{q_\rho},\\
 \E[\min(\tau,N)^2\mid\tau<\infty]
   &=\frac{\lambda_\rho}{q_\rho}\log N+O(1)
   \qquad(\beta_\rho>0).
\end{aligned}
\end{equation}
For $\beta_\rho>0$, the unconditional mean is infinite and the conditional
moments are finite exactly for orders $0<r<2$. If $\beta_\rho=0$, the
coin sends the input directly toward the origin and $\tau=1$ almost
surely. For the original input $\rho=\ketbra RR$, we have
$\beta_\rho=p$; setting $p=1/2$ recovers $q_\rho=2/\pi$ and
$\lambda_\rho=2/\pi$.

\subsection{Reusing the Walk Inside an Outer Loop}
\label{main:case:nested}

We now verify the caller from Figure~\ref{fig:mot-caller}. Each continuing
outer iteration resets the registers and invokes $\mathsf{Walk}$; the next
outer test is reached only if that invocation terminates. The verification
targets are the caller's termination probability and conditional expected
runtime, counting entries into both loop bodies. We derive both quantities modularly from the inner walk's verified specification \eqref{main:extract-hadamard-expansion}, without unfolding $\mathsf{Walk}$ or re-analyzing its lattice operators (full details in Section~F of the supplemental text).

\begin{figure}[!htb]
\centering
\begin{minipage}[t]{.44\linewidth}
\small
\textit{The initialized subprogram $S$}\\[4pt]
$\mathtt{coin}:=\ket R;$\\[2pt]
$\mathtt{position}:=\ket1;$\\[2pt]
$\mathsf{Walk}$
\end{minipage}\hfill
\begin{minipage}[t]{.53\linewidth}
\small
\textit{The outer loop $L$}\\[4pt]
$\mathbf{while}\ M^{\mathrm{out}}[\mathtt{coin,position}]=1$
\\[2pt]
\hspace*{1em}$\mathbf{do}\ S\ \mathbf{od}$\\[4pt]
$M^{\mathrm{out}}_0=M^{\mathrm{out}}_1=I/\sqrt2$
\end{minipage}
\caption{The nested program. The measurement $M^{\mathrm{out}}$ implements
$\mathtt{randomBit()}$ in Figure~\ref{fig:mot-caller}: each test exits
or continues with probability $1/2$, leaving the registers unchanged.
A terminating call of duration $n$ contributes $1+n$
to the total runtime, including entry into the outer body.}
\Description{The nested program.}
\label{main:case:nested-program}
\end{figure}

\paragraph{1. Reuse the initialized walk's specification.}
Let $\mathsf{Init}$ denote the two reset commands in
Figure~\ref{main:case:nested-program} and put
$\sigma=\ketbra{R,1}{R,1}$. Each reset is implemented by initialization
followed by a fixed unitary. Together they prepare $\sigma$ from every
normalized input, so its backward action on any parameterized predicate
$G$ is $\wp(\mathsf{Init},G)_s=\Tr(\sigma G_s)I$.
The sequencing rule therefore gives, for every postcondition $F$,
\begin{equation}
 \label{main:case:reset-wp}
 \wp(S,F)_s
 =\wp\bigl(\mathsf{Init},\wp(\mathsf{Walk},F)\bigr)_s
 =\Tr\!\left(\sigma\wp(\mathsf{Walk},F)_s\right)I.
\end{equation}
In particular, taking $F=\mathbf I$ reuses the Hadamard runtime family:
\begin{equation}
 \label{main:case:inner-summary}
 H_s^S=h_sI,
 \qquad h_s=q-\frac1s+o(s^{-1}),
 \qquad q=\frac2\pi.
\end{equation}
The scalar expansion is the mathematical oracle result
\eqref{main:extract-hadamard-expansion}, used here only through first order.
Equation~\eqref{main:case:reset-wp} turns that input-specific calculation
into a specification for $S$ on its full state space.

\paragraph{2. Derive the outer loop equation.}
Let $X$ describe the remaining execution from the outer guard. The exit
branch contributes $I/2$. The continuing branch charges one iteration,
propagates $X$ backward through $S$ at scale $s+1$, and has weight $1/2$.
Substitution into the loop rule gives
\begin{equation}
 \label{main:case:outer-functional}
 (\cT_{\mathbf I}X)_s
 =\underbrace{\frac12I}_{\text{exit}}
  +\underbrace{\frac{s}{2(s+1)}\wp(S,X)_{s+1}}_{\text{continue}},
 \qquad H^L=\operatorname{lfp}\cT_{\mathbf I}.
\end{equation}
The subprogram's transformer accounts for all inner iterations; the
factor $s/(s+1)$ accounts for the additional outer iteration.

\paragraph{3. Set the approximation goal.}
Computing an exact solution of~\eqref{main:case:outer-functional} directly is not easy. However, for termination
probability and conditional mean, Table~\ref{main:extract-asymptotic-table}
only needs the constant and first-order coefficients of
$\Tr(\rho H_s^L)$. We therefore seek a parameterized predicate $A$ that approximates
$H^L$ in operator norm with error $o(s^{-1})$.

Theorem~\ref{main:calc:residual} reduces this approximation goal to a check
of one outer iteration. Its same-order error bound applies because the
fair outer guard supplies a uniform contraction: since $H_{u+1}^S\le I$,
\[
 \frac{u}{u+1}(M^{\mathrm{out}}_1)^\dagger
      H_{u+1}^S M^{\mathrm{out}}_1
 =\frac{u}{2(u+1)}H_{u+1}^S\le\frac12I
 \qquad(u>0).
\]
Thus, with $\kappa=1/2$, it suffices to find $A$ satisfying
\begin{equation}
 \label{main:case:residual-target}
 \norm{A_s-(\cT_{\mathbf I}A)_s}=o(s^{-1}).
\end{equation}
The theorem then controls the accumulated error over the whole outer loop.

\paragraph{4. Construct and certify the approximation.}
To meet this goal, we look for a parameterized predicate with adjustable
constant and first-order coefficients whose precondition through $S$ can
be computed from the verified inner family $H^S$. We use the rational template
\begin{equation}
 \label{main:case:candidate}
 A_s\triangleq a\frac{s}{s+b}I
     =\left(a-\frac{ab}{s}+O(s^{-2})\right)I,
 \qquad 0<a\le1,\quad b>0.
\end{equation}
It provides these two coefficients and stays in the predicate domain at
every scale. Its backward propagation reuses $H^S$ at a shifted scale:
by the graded weakest-precondition rule,
\begin{equation}
 \label{main:case:call-reuse}
\begin{aligned}
 \wp(S,A)_s
 &=\sum_{n\ge0}\frac{s}{s+n}
     \sem{S}_n^\dagger\!\left(a\frac{s+n}{s+n+b}I\right)\\
 &=a\frac{s}{s+b}H_{s+b}^S
  =a\frac{s}{s+b}h_{s+b}I.
\end{aligned}
\end{equation}
Each inner duration $n$ selects the continuation at scale $s+n$; the
rational template collects all these shifted continuations into one
evaluation of $H^S$ at $s+b$. The identity in terms of $H^S$ applies to
any subprogram with a known runtime family.

Substituting this call precondition at scale $s+1$ into
\eqref{main:case:outer-functional} gives
\begin{equation}
 \label{main:case:candidate-image}
 (\cT_{\mathbf I}A)_s
 =\left[\frac12+\frac a2\frac{s}{s+1+b}h_{s+1+b}\right]I.
\end{equation}
The program rules have now reduced the residual check to a scalar
calculation using the inner expansion~\eqref{main:case:inner-summary}.
Matching the constant and $s^{-1}$ coefficients of
\eqref{main:case:candidate-image} with those of $A_s$ gives
\begin{equation}
 \label{main:case:matching}
 a=\frac1{2-q},\quad b=\frac{1+q}{2-q}.
\end{equation}
These values satisfy $0<a<1$ and $b>0$, and make the residual meet
\eqref{main:case:residual-target}. Theorem~\ref{main:calc:residual}
therefore certifies
\begin{equation}
 \label{main:case:outer-certificate}
 \norm{H_s^L-A_s}=o(s^{-1}),
 \qquad
 H_s^L=\left(a-\frac{ab}{s}\right)I+o(s^{-1}).
\end{equation}
\paragraph{5. Extract the caller's runtime properties.}
For every normalized input state $\rho$, taking the expectation of
\eqref{main:case:outer-certificate} gives
$\Tr(\rho H_s^L)=a-ab/s+o(s^{-1})$.
The first two rows of Table~\ref{main:extract-asymptotic-table} give the
termination probability $a$ and conditional mean $ab/a=b$:
\begin{equation}
 \label{main:case:outer-results}
\begin{aligned}
 \Prob_\rho(\tau_L<\infty)
   &=a=\frac1{2-q}=\frac{\pi}{2(\pi-1)},\\
 \E_\rho[\tau_L\mid\tau_L<\infty]
   &=b=\frac{1+q}{2-q}=\frac{\pi+2}{2(\pi-1)}.
\end{aligned}
\end{equation}
The caller may terminate before invoking $S$, which increases its
termination probability from $q$ to $a$. It still has positive divergence
probability, so its unconditional expected runtime is infinite.

Because the outer proof depends on $\mathsf{Walk}$ only through the
first-order expansion \eqref{main:case:inner-summary}, replacing $\mathsf{Walk}$ with the biased
walk $L_p$ (reset to $|R,1\rangle$) simply updates the inner termination
probability in \eqref{main:case:inner-summary} to $q = 1 - p + p\kappa_p$. Substituting this $q$
into \eqref{main:case:matching} immediately yields the new caller's termination probability
$a$ and conditional mean $b$, reusing the entire outer derivation and
certificate unchanged.
\section{Related Work}
\label{sec:related-work}

\paragraph{Quantum program verification.}
Quantum weakest preconditions, Hoare logics, and operator-based
verification provide the foundations for reasoning about quantum
programs~\cite{DHondtPanangaden2006,Ying11,YingYuFengDuan2013}.
Termination analyses cover quantum loops, concurrent and nondeterministic
programs, ranking-supermartingale synthesis, and scheduler
construction~\cite{YingFeng2010,YuYing2012,LiYuYing2014,LiYing2018,XuEtAl2024}.
Related work on quantum Markov chains and decision processes studies
reachability probabilities, recursion, and long-run
behavior~\cite{FengYuYing2013,FengYuYing2013Recursive,YingFengYuYing2013,GuanFengYing2018,YingYing2018},
while quantum automata and temporal logics support verification of
linear-time properties~\cite{YingLiYuFeng2014,Yu2024}. Our calculus
connects termination probability and runtime information in specifications
that compose through sequential calls and nested loops.

\paragraph{Quantum runtime analysis.}
Avanzini et~al.~\cite{AvanziniEtAl2022} and Liu
et~al.~\cite{LiuEtAl2025} develop compositional expectation transformers
for quantum costs and expected runtimes. The former uses finite-dimensional
quantum registers with potentially unbounded classical variables; the
latter includes quantum integers in its state-based semantics, while its
observable-based computational methods specialize to finite dimensions.
Yu~\cite{Yu2023} reduces flowchart programs to a single loop while
preserving expected runtime, variance, and general moments, and gives
algorithms for termination and exact expected runtime.
Unbounded pre-expectations~\cite{GehnenEtAl2026} and closed positive
quadratic forms~\cite{YuEtAl2026} extend cost reasoning to
infinite-dimensional systems. Our bounded parameterized predicates
retain information when the expected cost is infinite and yield
termination probabilities, conditional moments, and capped-moment
divergence rates. The resolvent translation
(Theorem~\ref{main:sem:cost-translation}) also lets existing total-cost
proofs supply specifications in this domain.

\paragraph{Probabilistic termination and runtime distributions.}
Classical expectation transformers support compositional reasoning about
probabilities and expected runtimes~\cite{McIverMorgan2005,KaminskiEtAl2018}.
Martingale and lexicographic methods establish almost-sure termination
and support automated certificate
synthesis~\cite{ChakarovSankaranarayanan2013,McIverEtAl2018,AgrawalEtAl2018,ChatterjeeEtAl2021Lexicographic}.
Quantitative analyses derive expected-time and tail
bounds~\cite{ChatterjeeEtAl2018}, termination-probability
bounds~\cite{ChatterjeeNovotnyZikelic2017,ChatterjeeEtAl2022Quantitative},
and inductive lower bounds on expectations and
runtimes~\cite{HarkEtAl2020}. Higher-moment methods bound runtime moments,
variances, and central moments~\cite{KuraEtAl2019,WangHoffmannReps2021}.
Generating-function approaches retain distributions and support symbolic
verification, conditioning, and invariant
synthesis~\cite{KlinkenbergEtAl2021,ChenEtAl2022Generating,KlinkenbergEtAl2024,HaaseEtAl2026}.
These approaches provide precedents for combining program rules with
symbolic mathematical calculations. Our calculus propagates quantum
postconditions and elapsed time together through scale-shifted predicate
families. Its loop uniqueness and residual bounds certify asymptotic
specifications from which conditional moments and their divergence rates
can be extracted, including for programs with positive nontermination
probability.

\paragraph{Quantum walks and absorption.}
Quantum walks have been studied as dynamical processes and computational
primitives~\cite{AmbainisEtAl2001,AharonovEtAl2001,Childs2009,ChildsGossetWebb2013}.
Analyses of absorbing boundaries and measured hitting times establish
absorption probabilities, conditional timing properties, and examples
with infinite hitting
times~\cite{KonnoEtAl2003,BachEtAl2004,KroviBrun2006Hypercube,KroviBrun2006Infinite}.
These results supply the mathematical setting and verification targets
for our examples. Our program rules derive the trace series whose scalar
asymptotics establish reusable runtime specifications; the nested-loop
example then applies such a specification in a caller's proof.

\paragraph{Quantum-walk circuits and formal verification.}
Circuit constructions analyze the resources needed to implement walk
evolution on various graphs, including Fourier-transform and Szegedy
implementations~\cite{DouglasWang2009,LokeWang2012,LokeWang2017,Shakeel2020,RazzoliEtAl2024}.
Their resource measures concern a prescribed evolution, whereas our
runtime counts iterations until a measurement causes the loop to terminate.
Formal-verification work establishes relational properties of absorbing
walks on finite intervals~\cite{BartheEtAl2020} and verifies walk
algorithms in Qafny and QafnyC~\cite{LiEtAl2024Qafny,ChengEtAl2025}.
Our case studies verify absorption on the unbounded integer lattice,
including arbitrarily long terminating runs, and reuse the resulting
specifications through program composition.

\section{Conclusion}
\label{sec:conclusion}

This paper shows that bounded parameterized predicates can retain detailed information about runtime, even when runtime moments are infinite. The observation scale serves as the parameter that stores this information: finite moments appear as coefficients in an asymptotic expansion, and infinite moments reveal themselves through the rate at which capped moments grow. The weakest-precondition rules propagate these parameterized predicates through program syntax, so a verified specification of a loop can be reused when that loop is called by a larger program. We demonstrated this on the absorbing Hadamard walk and on an outer loop that repeatedly invokes it.

A natural next step is to automate the construction of these specifications. Predicate templates and symbolic calculations could generate candidate families, and the uniqueness and residual theorems provide the checks needed to certify them. 

\section*{AI Use}
During this work, the authors used GPT-6 Astra as an interactive assistant. It was used to assist with proof review and to draft and revise English prose and LaTeX presentation. No generated proof or text was adopted without author verification. The authors retain full responsibility for all content of the paper.

\section*{Data-Availability Statement}
 The appendices have been submitted as Supplementary text but are not yet publicly available to preserve anonymity during the review process. They will be released publicly after the review is completed.
\bibliographystyle{ACM-Reference-Format}
\bibliography{references}
\appendix

\section*{Guide to the Supplemental Text}

This supplemental text is designed as an extended version of the main
paper and can be read independently. It includes the full definitions,
proofs, and calculations underlying the main results.

Section~\ref{walk:sec-cost} introduces the absorbing quantum walk and
establishes its infinite expected runtime using an existing cost logic.
Section~\ref{sec:runtime-semantics} defines the runtime-graded semantics,
parameterized predicates, and semantic validity.
Section~\ref{sec:calculus} develops the compositional weakest-precondition
calculus and proves the uniqueness and residual bounds used to certify
loop specifications.
Section~\ref{moment:section} shows how to extract termination
probabilities, runtime moments, and divergence rates from these
specifications.
Section~\ref{walk:sec-certified-walks} applies the framework to Hadamard
and biased quantum walks, and Section~\ref{sec:nested} reuses the walk's
specification to verify an outer loop that repeatedly invokes it.
Finally, Section~\ref{bridge:section} shows how total-cost proofs can be
translated into proofs over parameterized predicates.

The core development in
Sections~\ref{sec:runtime-semantics}--\ref{sec:nested} has been formalized
in Lean with assistance from GPT-6 Astra.
Sections~\ref{walk:sec-cost} and~\ref{bridge:section} provide complementary
connections to the cost logic of~\cite{YuEtAl2026} and lie outside the
paper's central development. They are not included in the formalization,
as formalizing them would require a full formalization of the extensive
framework developed in that work. 

Furthermore, we provide an illustrative Mathematica script, developed with
assistance from GPT-6 Astra, for semi-automatic symbolic evaluation of the
scalar sums arising in the absorbing quantum walk example.

\section*{Conventions}
We write $\NN=\{0,1,2,\dots\}$ and $\NN_{+}=\{1,2,\dots\}$.
All Hilbert spaces are complex, and program state spaces are separable. Inner products $\ip{\cdot}{\cdot}$ are conjugate-linear in the first argument. We write $\Bop{\cH}$ for the algebra of bounded linear operators, $\Top{\cH}$ for trace-class operators, and $\Eff{\cH} = \{A \in \Bop{\cH} : 0 \le A \le I\}$ for the space of quantum effects. Quantum states (partial density operators) are positive trace-class operators $\rho \in \Top{\cH}$ with $\Tr(\rho) \le 1$; conditional probabilities and conditional moments are stated for normalized states ($\Tr(\rho) = 1$). The Heisenberg adjoint $\Phi^\dagger : \Bop{\cK} \to \Bop{\cH}$ of a bounded linear map $\Phi : \Top{\cH} \to \Top{\cK}$ is defined by trace duality $\Tr(\Phi^\dagger(A)\rho) = \Tr(A\Phi(\rho))$.

Positive quadratic forms $P : \cH \to [0,\infty]$ take the value $+\infty$ outside their finite domain $\Dom(P)$, with the convention $0 \cdot \infty = 0$. For a closed positive form $P$ and a positive trace-class operator with spectral decomposition $\rho=\sum_j\lambda_j\ketbra{u_j}{u_j}$, the extended trace is
\[
 \Tr(P\rho)=\Tr(\rho P)\triangleq \sum_j\lambda_jP[u_j]\in[0,\infty].
\]
For a bounded positive operator $D \ge 0$, $\infty D$ denotes the closed positive form that vanishes on $\ker(D)$ and equals $+\infty$ on $\cH \setminus \ker(D)$. Uniformly bounded monotone sequences of self-adjoint operators converge in the strong operator topology. Positive operator series are interpreted as monotone suprema of their partial sums in the order of quadratic forms; when the partial sums are uniformly bounded, they also converge strongly. Nonnegative numerical series and integrals are evaluated by monotone convergence, with interchanges justified by Tonelli's theorem. Runtime is measured by the total number of loop iterations: each time a $\mathbf{while}$ guard selects its continuing branch, the count increases by one, with iterations of nested loops added to the same total. Loop exits and all other commands contribute zero. Terminating runtimes lie in $\NN$, and nonterminating executions have runtime $+\infty$. In moment formulas, $n^0 = 1$ for all $n \ge 0$. Big-$O$ and little-$o$ estimates for operator families use the operator norm on the stated space: for $g(s)>0$, $A_s=O(g(s))$ means that $\norm{A_s}/g(s)$ remains bounded as $s\to\infty$, and $A_s=o(g(s))$ means that this ratio tends to zero. For scalar functions, the norm is replaced by absolute value. Implicit constants in quantum-walk estimates may depend on the fixed coin parameter $p \in (0,1)$.

\section{A Coherent Quantum Walk and Its Infinite Expected Cost}
\label{walk:sec-cost}

The one-dimensional discrete-time Hadamard walk with an absorbing boundary at $x=0$ is a standard benchmark in quantum computing~\cite{KonnoEtAl2003,BachEtAl2004}. Its coherent evolution over an infinite lattice exhibits a rich runtime profile: starting at $x=1$ with coin state $\ket{R}$, it terminates with probability $2/\pi \approx 0.637$ and has infinite unconditional expected runtime, yet its conditional expected runtime given termination is only $\pi/2$~\cite[Section~4]{KonnoEtAl2003}. Its absorption-time tail also gives logarithmic growth of the capped conditional second moment. These properties provide a concrete target for compositional runtime verification of programs that are not almost surely terminating (AST).

Existing quantum runtime analyses provide compositional methods for expected costs, but are restricted to finite-dimensional state spaces~\cite{AvanziniEtAl2022,LiuEtAl2025} and therefore do not cover this coherent infinite-lattice walk. The closed-positive-quadratic-form calculus of~\cite{YuEtAl2026} extends expected-cost reasoning to infinite-dimensional systems and allows us to prove that this walk has infinite expected runtime (Theorem~\ref{walk:thm-infinite-cost}). However, because the expected-cost predicate collapses to $+\infty$, it leaves the termination probability, the conditional mean, and the divergence rate of higher moments unresolved. Recovering these quantities compositionally requires moving beyond a single unbounded cost to a parameterized family of bounded runtime predicates, which we introduce at the end of this section and develop systematically in the following sections. 

\subsection{Registers, Commands, and the Runtime Convention}

We begin by formulating the absorbing Hadamard walk as a \textbf{qwhile} program over a coin register and an infinite-lattice position register. Restricting each loop iteration to the positive half-line then allows us to express the walk's runtime distribution directly in terms of its continuing and absorbing operators.

The walk operates on a coin register with state space $\CC^2 = \operatorname{span}\{\ket{L},\ket{R}\}$ and a position register with state space $\ell^2(\ZZ) = \overline{\operatorname{span}}\{\ket{x} : x \in \ZZ\}$; we write $\ket{c,x} \triangleq \ket{c}\otimes\ket{x}$ for product basis states. As a quantum loop program, the walk is written as
\[
\begin{aligned}
&\mathtt{coin := \ket{R};\ position := \ket{1};}\\
&\mathtt{while}\ M[\mathtt{position}] = 1\ \mathtt{do}\\
&\quad\mathtt{coin}:=C_{1/2}[\mathtt{coin}];\,
 (\mathtt{coin,position}):=V[\mathtt{coin,position}]\\
&\mathtt{od}
\end{aligned}
\]
Here the loop guard $M = \{M_0, M_1\}$ tests whether the particle has reached the absorbing boundary at $x=0$, with termination projector $M_0 = \Pi_0 \triangleq I_2 \otimes \ketbra{0}{0}$ and continuation projector $M_1 = \Pi_{\ne 0} \triangleq I - \Pi_0$. Each iteration of the loop body applies the Hadamard coin $C_{1/2}$ followed by the conditional shift $V$:
\begin{equation}
 C_{1/2} = \frac{1}{\sqrt{2}}
 \begin{pmatrix} 1 & 1 \\ 1 & -1 \end{pmatrix},
 \qquad
 V\ket{L,x} = \ket{L,x-1},
 \qquad
 V\ket{R,x} = \ket{R,x+1},
 \label{walk:eq-hadamard-operations}
\end{equation}
so a single unitary step on the full line is $U = V(C_{1/2} \otimes I)$.

Because the walk starts at $x=1$ and moves by $\pm 1$ per step, any state surviving the guard measurement remains supported on the positive half-line $\cH_+ \triangleq \CC^2 \otimes \ell^2(\NN_{+})$. Moreover, the boundary $x=0$ can only be entered from $\cH_+$ by stepping left from $x=1$, which always lands in the one-dimensional subspace $\CC\ket{L,0}$. Let $\Pi_+$ denote the orthogonal projection onto $\cH_+$. Restricted to $\cH_+$, one iteration of the loop is completely characterized by the \emph{continuing operator} $K$ and the \emph{absorbing operator} $B$:
\begin{equation}
 K = \Pi_+ U\big|_{\cH_+} : \cH_+ \longrightarrow \cH_+,
 \qquad
 B = \Pi_0 U\big|_{\cH_+} : \cH_+ \longrightarrow \CC\ket{L,0}.
 \label{walk:eq-K-B}
\end{equation}
Since $U$ is unitary and maps $\cH_+$ into $\cH_+ \oplus \CC\ket{L,0}$, these operators satisfy the conservation identity $K^\dagger K + B^\dagger B = I_{\cH_+}$.

Under our runtime convention, each execution of the loop body incurs one unit of time, so the runtime $\tau$ counts the number of iterations until absorption at $x=0$. For any initial unit vector $u \in \cH_+$, the unnormalized surviving state after $n$ iterations is $K^n u$, yielding the runtime distribution
\begin{equation}
 \Prob_u(\tau > n) = \norm{K^n u}^2,
 \qquad
 \Prob_u(\tau = n) = \norm{B K^{n-1} u}^2 \quad (n \ge 1).
 \label{walk:eq-survival-probabilities}
\end{equation}

\subsection{Expected-Runtime Analysis}

We derive the walk's expected-runtime predicate from its commands, counting one unit per loop iteration. For a post-cost $X$, the unitary rule gives the following exact backward annotations of the loop body, read from bottom to top:
\[
\begin{array}{l}
 \bigl\{(C_{1/2}^{\dagger}\otimes I)(V^{\dagger}XV)(C_{1/2}\otimes I)\bigr\}\\[2pt]
 \quad\mathtt{coin}:=C_{1/2}[\mathtt{coin}];\\[2pt]
 \bigl\{V^{\dagger}XV\bigr\}\\[2pt]
 \quad(\mathtt{coin,position}):=V[\mathtt{coin,position}]\\[2pt]
 \{X\}.
\end{array}
\]
The shift pulls $X$ back to $V^{\dagger}XV$, which becomes the post-cost for the coin command. The sequencing rule joins the two unitary-rule instances at this intermediate predicate, giving the body precondition $U^{\dagger}XU$.

At the loop guard, the exit branch has post-cost $0$. On the continuing branch, $I$ pays for the current iteration and $U^{\dagger}XU$ pulls the continuation cost $X$ at the next guard back through the loop body. Pulling these two branch costs back through the measurement gives
\[
 X\longmapsto
 \underbrace{M_0^{\dagger}0M_0}_{\text{exit}}
 +\underbrace{M_1^{\dagger}(I+U^{\dagger}XU)M_1}_{\text{continue}}.
\]
The while rule takes the least fixed point of this transformer, obtained by iteration from $0$. Thus the expected-runtime predicate $C$ satisfies
\begin{equation}
 C=M_1^{\dagger}(I+U^{\dagger}CU)M_1,
 \label{walk:eq-cost-fixed-point}
\end{equation}
where $(U^{\dagger}CU)[w]=C[Uw]$. This adapts the positive-form construction of~\cite[Section~5.4 and Appendix~C.5]{YuEtAl2026} to our iteration cost: the bounded predicates obtained by finite unfolding from $0$ increase, so their supremum $C$ is a closed positive quadratic form. For a unit input $u\in\cH_+$, the $N$th unfolding has value $\E_u[\min(\tau,N)]$; hence $C[u]=\E_u[\tau]$.

Equation~\eqref{walk:eq-cost-fixed-point} makes $C$ vanish on the absorbing branch and gives $C[b+w]=C[w]$ for $b\in\operatorname{ran}\Pi_0$ and $w\in\cH_+$. Since $Uu=Bu+Ku$, the loop equation reduces to
\begin{equation}
 C[u]=\norm{u}^2+C[Ku]\qquad(u\in\cH_+).
 \label{walk:eq-cost-recursion}
\end{equation}
The cost rules thus reduce our verification task to ruling out a finite value at $\psi=\ket{R,1}$ for a nonnegative solution of the one-step inequality below. The following lemma resolves this operator-theoretic question using the boundary defects of $K$.

\begin{lemma}[A rank-one obstruction to finite cost]
\label{walk:lem-cost-obstruction}
For the continuing operator $K$ in~\eqref{walk:eq-K-B}, every nonnegative quadratic form $C$ on $\cH_+$ satisfying
\begin{equation}
 C[u]\ge\norm{u}^2+C[Ku]\qquad(u\in\cH_+)
 \label{walk:eq-cost-obligation}
\end{equation}
has $C[\psi]=\infty$ at $\psi=\ket{R,1}$.
\end{lemma}

\begin{proof}
Suppose toward a contradiction that $C[\psi]<\infty$. Iterating~\eqref{walk:eq-cost-obligation} gives
\[
 \sum_{n=0}^{N-1}\norm{K^n\psi}^2+C[K^N\psi]\le C[\psi],
\]
so nonnegativity of $C$ implies $\sum_{n\ge0}\norm{K^n\psi}^2<\infty$. Our strategy is to assemble the surviving states $K^n\psi$ into the cumulative occupation operator $G$: finiteness of the sum makes $G$ trace-class, whereas the boundary defects of $K$ will force $G$ to act as the identity on an infinite-dimensional backward-reachable subspace.

Write $\ket{\pm}\triangleq (\ket{L}\pm\ket{R})/\sqrt2$. From~\eqref{walk:eq-hadamard-operations}--\eqref{walk:eq-K-B}, the forward action of $K$ on $\cH_+$ is
\begin{align}
 K\ket{L,x} &= \begin{cases}
   2^{-1/2}\ket{R,2}, & x = 1,\\
   2^{-1/2}(\ket{L,x-1} + \ket{R,x+1}), & x \ge 2,
 \end{cases} \label{walk:eq-K-on-L}\\
 K\ket{R,x} &= \begin{cases}
   -2^{-1/2}\ket{R,2}, & x = 1,\\
   2^{-1/2}(\ket{L,x-1} - \ket{R,x+1}), & x \ge 2,
 \end{cases} \label{walk:eq-K-on-R}
\end{align}
and its adjoint action is
\[
 K^\dagger\ket{L,x}=\ket{+,x+1}\quad(x\ge1),
 \qquad
 K^\dagger\ket{R,x}=\ket{-,x-1}\quad(x\ge2),
 \qquad
 K^\dagger\psi=0.
\]
These formulas reveal two rank-one boundary defects at $x=1$. 
In the forward direction, only the $\ket{+,1}$ component steps left into $\ket{L,0}$, so $B=\ket{L,0}\bra{+,1}$ and the conservation identity $K^\dagger K+B^\dagger B=I_{\cH_+}$ gives $K^\dagger K=I-\ketbra{+,1}{+,1}$.
In the backward direction, no step inside $\cH_+$ can land in $\psi=\ket{R,1}$ (reaching $x=1$ from $x=2$ requires coin state $\ket{L}$), while $K$ maps $\{\ket{+,1}\}^\perp$ isometrically onto $\{\psi\}^\perp$. Thus $K$ is a partial isometry satisfying
\begin{equation}
 K^\dagger K=I-\ketbra{+,1}{+,1},
 \qquad
 KK^\dagger=I-\ketbra{\psi}{\psi}.
 \label{walk:eq-defect-identities}
\end{equation}

Now define the cumulative occupation operator
\begin{equation}
 G\triangleq \sum_{n=0}^{\infty}K^n\ketbra{\psi}{\psi}(K^\dagger)^n,
 \label{walk:eq-Gramian-telescoping}
\end{equation}
which converges in trace norm to a positive trace-class operator with $\Tr(G)=\sum_{n=0}^{\infty}\norm{K^n\psi}^2<\infty$. Shifting the summation index gives the fixed-point equation
\begin{equation}
 G=\ketbra{\psi}{\psi}+KGK^\dagger.
 \label{walk:eq-Gramian-Stein}
\end{equation}
Substituting $\ketbra{\psi}{\psi}=I-KK^\dagger$ from~\eqref{walk:eq-defect-identities} telescopes the first $N$ terms of~\eqref{walk:eq-Gramian-telescoping} to $I-K^N(K^\dagger)^N\le I$, so $0\le G\le I$. Taking the trace in~\eqref{walk:eq-Gramian-Stein} and using the first identity in~\eqref{walk:eq-defect-identities} yields
\[
 \Tr(G)=1+\Tr(GK^\dagger K)
       =1+\Tr(G)-\bra{+,1}G\ket{+,1}.
\]
Canceling the finite quantity $\Tr(G)$ forces $\bra{+,1}G\ket{+,1}=1$. Since $I-G\ge 0$, we have $\norm{(I-G)^{1/2}\ket{+,1}}^2=\bra{+,1}(I-G)\ket{+,1}=0$, so $(I-G)\ket{+,1}=0$.

Finally, subtracting~\eqref{walk:eq-Gramian-Stein} from $I=\ketbra{\psi}{\psi}+KK^\dagger$ gives the invariance relation
\[
 I-G=K(I-G)K^\dagger.
\]
For any $u\in\ker(I-G)$, this relation implies $\norm{(I-G)^{1/2}K^\dagger u}^2=\ip{u}{K(I-G)K^\dagger u}=0$, so $\ker(I-G)$ is invariant under the backward step $K^\dagger$. By induction from $\ket{+,1}\in\ker(I-G)$, the entire backward orbit satisfies
\begin{equation}
 (I-G)(K^\dagger)^n\ket{+,1}=0\qquad(n\ge0).
 \label{walk:eq-kernel-propagation}
\end{equation}
The adjoint formulas show inductively that $(K^\dagger)^n\ket{+,1}$ is supported on sites $\{1,\ldots,n+1\}$ with highest-site component $2^{-n/2}\ket{+,n+1}\ne0$. These vectors are therefore linearly independent, and~\eqref{walk:eq-kernel-propagation} implies that $G$ acts as the identity on their infinite-dimensional span, contradicting $\Tr(G)<\infty$.
\end{proof}

\begin{theorem}[Infinite expected runtime at the entrance site]
\label{walk:thm-infinite-cost}
Write $\ket{+}\triangleq (\ket{L}+\ket{R})/\sqrt2$. For every unit coin vector $v\in\CC^2$, the coherent Hadamard walk started at $v\otimes\ket{1}$ satisfies
\begin{equation}
 \E_{v\otimes\ket{1}}[\tau]=
 \begin{cases}
  1, & v\in\CC\ket{+},\\[2pt]
  \infty, & v\notin\CC\ket{+}.
 \end{cases}
 \label{walk:eq-cost-slice}
\end{equation}
In particular, $\E_\psi[\tau]=\infty$ for $\psi=\ket{R,1}$.
\end{theorem}

\begin{proof}
By~\eqref{walk:eq-cost-recursion}, the expected-runtime predicate $C$ satisfies~\eqref{walk:eq-cost-obligation}, so Lemma~\ref{walk:lem-cost-obstruction} gives $C[\psi]=\infty$. For $\ket{+,1}$, the coin operation produces $\ket{L,1}$ and the shift steps into the absorbing state $\ket{L,0}$; hence $K\ket{+,1}=0$ and~\eqref{walk:eq-cost-recursion} gives $C[\ket{+,1}]=1$. Because the finite domain $\Dom(C)$ of the positive quadratic form $C$ is a linear subspace, if it contained $v\otimes\ket{1}$ for any $v\notin\CC\ket{+}$, it would contain the entire two-dimensional entrance space $\CC^2\otimes\ket{1}$, including $\psi$. Thus $C[v\otimes\ket{1}]=\infty$ for every $v\notin\CC\ket{+}$. Since $C[u]=\E_u[\tau]$ for unit inputs $u\in\cH_+$, the two runtime cases in~\eqref{walk:eq-cost-slice} follow.
\end{proof}

To recover finer runtime information, we introduce a parameterized family of bounded runtime expectations. For the input $\psi=\ket{R,1}$, define
\begin{equation}
 h_s \triangleq \E_{\psi}\!\left[\frac{s}{s+\tau}\,;\,\tau < \infty\right] \qquad (s > 0),
 \label{walk:eq-scalar-runtime-family}
\end{equation}
which remains in $[0,1]$ for every scale $s > 0$. Its large-scale behaviour determines the termination probability, conditional runtime moments, and growth rates of capped moments through the results of Section~\ref{moment:section}. In the notation of Section~\ref{sec:runtime-test-design}, this quantity is $h_s = \E_\psi[r_s(\tau)]$, where $r_s(t)=s/(s+t)$ for finite $t$ and $r_s(\infty)=0$. Section~\ref{sec:calculus} constructs the corresponding bounded operator family compositionally; Section~\ref{walk:sec-certified-walks} evaluates its scalar readout $h_s$ on the input $\psi$.

\section{Runtime Semantics and Parameterized Predicates}
\label{sec:runtime-semantics}

To reason compositionally about both runtime and quantum state transformations, we refine the standard denotational semantics $\sem{S}$ of a quantum program $S$ into runtime-graded completely positive maps $\sem{S}_{\mathrm{rt}} = (\sem{S}_n)_{n\ge 0}$ and test them against scale-indexed families of bounded effects $F = (F_s)_{s>0}$.

\subsection{Programs and Their Runtime-Graded Semantics}

Fix a finite set $\mathrm{Var}$ of quantum variables (registers). Each variable $q \in \mathrm{Var}$ is associated with a separable Hilbert space $\cH_q$ (possibly infinite-dimensional, such as $\ell^2(\ZZ)$) and a distinguished unit vector $\ket{0}_q \in \cH_q$. The joint Hilbert space and the corresponding set of partial density operators are
\[
\cH \triangleq \bigotimes_{q\in\mathrm{Var}} \cH_q,
\qquad
\mathcal{D}(\cH) \triangleq \bigl\{ \rho \in \Top{\cH} : \rho \ge 0,\; \Tr(\rho) \le 1 \bigr\}.
\]
Following standard quantum while-language conventions~\cite{Ying11,YuEtAl2026}, programs are generated by the syntax
\[
\begin{aligned}
	S ::= {}&\mathbf{skip} \mid \mathbf{abort} \mid q := \ket{0} \mid \bar q := U[\bar q] \mid S_1;S_2\\
	&\mid \mathbf{if}\ \bigl(\square_{i\in J}\ M[\bar q] = i \to S_i\bigr)\ \mathbf{fi}
	\mid \mathbf{while}\ M[\bar q] = 1\ \mathbf{do}\ B\ \mathbf{od}.
\end{aligned}
\]
Here $\bar q$ is a tuple of distinct variables, $U$ is a unitary operator on their joint space $\cH_{\bar q}$, and any operator acting on a subsystem is implicitly extended by the identity on the remaining variables. Initialization $q := \ket{0}$ resets register $q$ to $\ket{0}_q$, and we write $q := \ket{\phi}$ as shorthand for $q := \ket{0};\, q := U_\phi[q]$ with $U_\phi\ket{0}_q = \ket{\phi}_q$. In the conditional $\mathbf{if}$, $M = \{M_i\}_{i\in J}$ is a measurement on $\cH_{\bar q}$ with finite outcome set $J$ satisfying $\sum_{i\in J} M_i^\dagger M_i = I$, selecting branch $S_i$ upon outcome $i$. In $\mathbf{while}$, $M = \{M_0,M_1\}$ is a binary measurement ($M_0^\dagger M_0 + M_1^\dagger M_1 = I$), where outcome $0$ exits the loop and outcome $1$ enters the loop body $B$.

Write $\cM_i(\rho) \triangleq M_i\rho M_i^\dagger$ for the measurement branch maps, and
\[
\cR_q(\rho) \triangleq \ketbra{0}{0}_q\otimes\Tr_q(\rho),
\qquad
\mathcal{U}_{\bar q}(\rho) \triangleq U\rho U^\dagger
\]
for the initialization and unitary-update channels (identifying $\cH$ with $\cH_q\otimes\cH_{\mathrm{rest}}$). Recall that the standard denotational semantics~\cite{Ying11} interprets each program $S$ as a completely positive, trace-nonincreasing (CPTNI) map $\sem{S} : \Top{\cH} \to \Top{\cH}$ defined structurally by
\begin{equation}
	\label{eq:erased-composition}
	\begin{aligned}
		&\sem{\mathbf{skip}} = \id, \qquad
		\sem{\mathbf{abort}} = 0, \qquad
		\sem{q:=\ket{0}} = \cR_q, \qquad
		\sem{\bar q:=U[\bar q]} = \mathcal{U}_{\bar q},\\[2pt]
		&\sem{S_1;S_2} = \sem{S_2} \circ \sem{S_1}, \qquad
		\sem{\mathbf{if}\ (\square_{i\in J}\ M[\bar q]=i \to S_i)\ \mathbf{fi}} = \sum_{i\in J} \sem{S_i} \circ \cM_i,\\
		&\sem{\mathbf{while}\ M[\bar q]=1\ \mathbf{do}\ B\ \mathbf{od}} = \sum_{j=0}^{\infty} \cM_0 \circ (\sem{B} \circ \cM_1)^j.
	\end{aligned}
\end{equation}

To track runtime alongside state transformations, we refine $\sem{S}$ by recording the total number of loop iterations. Each entry into a loop body increments the count by one, with iterations of nested loops accumulated into the same total, while loop exits and loop-free commands contribute zero.

\begin{definition}[Runtime-graded denotational semantics]
	\label{def:cost-grades}
	The \emph{runtime-graded denotational semantics} of a program $S$ is the sequence $\sem{S}_{\mathrm{rt}} \triangleq (\sem{S}_n)_{n\ge 0}$, where for each $n \in \ZZ$, the grade-$n$ map $\sem{S}_n : \Top{\cH} \to \Top{\cH}$ is the CP map producing the subnormalized output state of executions that terminate after exactly $n$ total loop-body entries (with $\sem{S}_n = 0$ for $n < 0$):
	\begin{align}
		\sem{\mathbf{skip}}_n &= \one_{\{n=0\}}\id, &
		\sem{\mathbf{abort}}_n &= 0, \label{eq:grade-base}\\
		\sem{q := \ket{0}}_n &= \one_{\{n=0\}}\cR_q, &
		\sem{\bar q := U[\bar q]}_n &= \one_{\{n=0\}}\mathcal{U}_{\bar q}, \label{eq:grade-updates}\\
		\sem{S_1;S_2}_n &= \sum_{a+b=n} \sem{S_2}_b \circ \sem{S_1}_a, \label{eq:grade-sequence}\\
		\sem{\mathbf{if}\ (\square_{i\in J}\ M[\bar q]=i \to S_i)\ \mathbf{fi}}_n
		&= \sum_{i\in J} \sem{S_i}_n \circ \cM_i. \label{eq:grade-case}
	\end{align}
	For a loop $L = \mathbf{while}\ M[\bar q]=1\ \mathbf{do}\ B\ \mathbf{od}$ and each integer $k\ge1$, let $\cC_k \triangleq \sem{B}_{k-1} \circ \cM_1$ denote the grade-$k$ map of one guarded continuation step (with $\cC_0\triangleq 0$): entering the body contributes one iteration and executing $B$ contributes the remaining $k-1$. The grade-$n$ map of $L$ sums over $0 \le j \le n$ completed outer iterations:
	\begin{equation}
		\label{eq:grade-loop}
		\sem{L}_n = \sum_{j=0}^{n}
		\sum_{\substack{k_1,\dots,k_j\in\NN_{+}\\ k_1 + \cdots + k_j = n}}
		\cM_0 \circ \cC_{k_j} \circ \cdots \circ \cC_{k_1},
	\end{equation}
	where $n \ge 0$ and the $j=0$ term is $\one_{\{n=0\}}\cM_0$.
\end{definition}

Recall that the trace class is complete in the trace norm~\cite[Theorem~VI.20]{ReedSimon1980}. If $(\Phi_n)_{n\ge0}$ is a sequence of CP maps whose finite partial sums are trace-nonincreasing, then for every positive trace-class operator $\rho$,
\[
 \sum_{n\ge0}\norm{\Phi_n(\rho)}_1
 =\sum_{n\ge0}\Tr\bigl(\Phi_n(\rho)\bigr)
 \le\Tr(\rho).
\]
Thus $\sum_{n\ge0}\Phi_n(\rho)$ converges absolutely in trace norm (allowing arbitrary regrouping), and by linearity this convergence extends to all $\rho\in\Top{\cH}$. Since complete positivity and the trace bound pass to the limit, the sum defines a CPTNI map. Forgetting the runtime index then recovers the standard denotational semantics $\sem{S}$:

\begin{proposition}[Well-definedness and compatibility with $\sem{S}$]
	\label{prop:graded-well-defined}
	For every program $S$, the runtime grades $(\sem{S}_n)_{n\ge 0}$ are well-defined CPTNI maps. Their sum is also CPTNI and coincides with the standard denotational semantics~\eqref{eq:erased-composition}: for every $\rho \in \mathcal{D}(\cH)$,
	\[
		\sum_{n\ge 0} \sem{S}_n(\rho) = \sem{S}(\rho)
	\]
	in trace norm.
\end{proposition}

\begin{proof}
	We proceed by structural induction on $S$. In each case, every grade $\sem{S}_n$ is a finite sum of compositions of CP maps; fixing $\rho\in\mathcal D(\cH)$, it suffices to show that the finite partial sums are trace-nonincreasing and converge in trace norm to $\sem{S}(\rho)$.
	\begin{itemize}
		\item \emph{Basic commands.}
		For $\mathbf{skip}$, initialization, and unitary updates, only grade zero is nonzero and equals the corresponding trace-preserving CP map, whereas all grades of $\mathbf{abort}$ are zero. The claims follow immediately.

		\item \emph{Sequential composition.}
		Suppose the claims hold for $S_1$ and $S_2$. By the induction hypothesis,
		\[
			\sum_{a,b\ge0}\Tr\!\left(\sem{S_2}_b\bigl(\sem{S_1}_a(\rho)\bigr)\right)
			\le \sum_{a\ge0}\Tr\bigl(\sem{S_1}_a(\rho)\bigr)
			\le \Tr(\rho).
		\]
		Thus absolute convergence in trace norm and trace-norm continuity give
		\[
		\begin{aligned}
			\sum_{n\ge0}\sem{S_1;S_2}_n(\rho)
			&=\sum_{a,b\ge0}\sem{S_2}_b\bigl(\sem{S_1}_a(\rho)\bigr)\\
			&=\sem{S_2}\bigl(\sem{S_1}(\rho)\bigr)
			=\sem{S_1;S_2}(\rho).
		\end{aligned}
		\]

		\item \emph{Conditionals.}
		Let $S_{\mathrm{if}}\triangleq \mathbf{if}\ (\square_{i\in J}\ M[\bar q]=i\to S_i)\ \mathbf{fi}$. Applying the induction hypothesis to each branch yields
		\[
			\sum_{n\ge0}\sem{S_{\mathrm{if}}}_n(\rho)
			=\sum_{i\in J}\sem{S_i}\bigl(\cM_i(\rho)\bigr)
			=\sem{S_{\mathrm{if}}}(\rho)
		\]
		in trace norm, with trace at most $\sum_{i\in J}\Tr(\cM_i(\rho))=\Tr(\rho)$.

		\item \emph{While loops.}
		For a loop $L$, set $\cC\triangleq \sem{B}\circ\cM_1$. By the induction hypothesis on $B$,
		\[
			\cC(\sigma)=\sum_{k\ge1}\cC_k(\sigma),
			\qquad \sigma\in\mathcal D(\cH),
		\]
		in trace norm. Since $\sem{B}$ is trace-nonincreasing and $\sum_{i=0}^{1}M_i^\dagger M_i=I$,
		\[
			\Tr\bigl(\cM_0(\sigma)\bigr)+\Tr\bigl(\cC(\sigma)\bigr)
			\le\Tr(\sigma).
		\]
		Applying this inequality to $\sigma=\cC^j(\rho)$ and summing gives
		\[
		\begin{aligned}
			\sum_{j=0}^{N-1}\Tr\bigl(\cM_0(\cC^j(\rho))\bigr)
			&\le\sum_{j=0}^{N-1}\left[\Tr\bigl(\cC^j(\rho)\bigr)-\Tr\bigl(\cC^{j+1}(\rho)\bigr)\right]\\
			&=\Tr(\rho)-\Tr\bigl(\cC^N(\rho)\bigr)
			\le\Tr(\rho).
		\end{aligned}
		\]
		Expanding each $\cC^j$ using the induction hypothesis gives,
        for every integer $N\ge1$,
        \[
         \sum_{n=0}^{N-1}\sem{L}_n(\rho)
         \le
         \sum_{j=0}^{N-1}\cM_0\bigl(\cC^j(\rho)\bigr).
        \]
        The preceding trace bound therefore applies to every partial
        sum of $\sum_{n\ge0}\sem{L}_n(\rho)$, ensuring absolute convergence
        in trace norm. We can therefore regroup terms either by total
        runtime $n$ or by outer iteration count $j$ to obtain
		\[
			\sum_{n\ge0}\sem{L}_n(\rho)
			=\sum_{j\ge0}\cM_0\bigl(\cC^j(\rho)\bigr)
			=\sem{L}(\rho)
		\]
		in trace norm, with trace at most $\Tr(\rho)$.
	\end{itemize}
	Hence each grade and the runtime-erased sum are CPTNI, as claimed.
\end{proof}
\subsection{From Costs to Bounded Runtime Tests}
\label{sec:runtime-test-design}
\label{sec:runtime-observables}

From the runtime grades of $S$, we define the exact-runtime termination effects $E_n^S$, the total termination effect $Q_S$, and the divergence effect $D_S$:
\begin{equation}
	\label{eq:termination-effects}
	E_n^S \triangleq \sem{S}_n^\dagger(I), \qquad
	Q_S \triangleq \sum_{n\ge 0} E_n^S = \sem{S}^\dagger(I), \qquad
	D_S \triangleq I - Q_S.
\end{equation}
For a normalized state $\rho$, $p_n = \Tr(E_n^S\rho)$ and $p_\infty = \Tr(D_S\rho)$ give the distribution of the runtime $\tau_S \in \NN \cup \{\infty\}$, with $\tau_S=\infty$ on nonterminating executions. The \emph{successful $k$-th runtime moment} ($k > 0$) and the \emph{total expected runtime} are the closed positive quadratic forms
\begin{align}
	M_k^S &\triangleq \sum_{n\ge 0} n^k E_n^S, \label{eq:successful-runtime-moment}\\
	C_S &\triangleq \sup_{N\ge 1}\Bigl(\sum_{n\ge 0}\min(n,N)E_n^S + N D_S\Bigr) = M_1^S + \infty D_S. \label{eq:total-runtime-cost}
\end{align}

Each capped operator in~\eqref{eq:total-runtime-cost} lies in $[0,NI]$, and these operators increase with $N$. Their supremum is therefore a closed positive form, and scalar monotone convergence on each vector gives $M_1^S+\infty D_S$. For a normalized input $\rho$, their expectations are $\E_\rho[\min(\tau_S,N)]$; monotone convergence then gives $\Tr(C_S\rho)=\E_\rho[\tau_S]$.

Building on the expected-cost analysis of Section~\ref{walk:sec-cost}, we evaluate individual execution costs via a bounded test before taking expectations, using the scalar normalized resolvent at scale $s>0$:
\[
 r_s(t)\triangleq \frac{s}{s+t}\quad(t\ge0),\qquad r_s(\infty)\triangleq 0.
\]
Because $r_s$ is bounded and vanishes at $\infty$, the expectation $\E_\rho[r_s(\tau_S)]$ remains finite and retains information about terminating executions even when $\E_\rho[\tau_S]=\infty$. As $s$ varies, this yields a family of bounded observations whose large-scale behaviour determines termination probabilities, moments, and divergence rates (Section~\ref{moment:section}).

The same construction extends to a bounded positive terminal-cost operator $G$ through its normalized resolvent $\res{s}{G}\triangleq s(sI+G)^{-1}\in\Eff{\cH}$. These resolvents reverse operator cost bounds into ordered effects: $0\le G\le P$ implies $\res{s}{P}\le\res{s}{G}$ for bounded $P$, turning an upper bound on the cost into a lower bound on its test across all input states. Since an execution of runtime $n$ adds $nI$ to the terminal cost, we have
\begin{equation}
\label{eq:design-cost-shift}
 \res{s}{nI+G}=\frac{s}{s+n}\res{s+n}{G},
\end{equation}
so at runtime $n$ the terminal test is evaluated at the shifted scale $s+n$ and weighted by $s/(s+n)$. Pulling this test back through each runtime grade and summing gives
\[
 \sum_{n\ge0}\sem{S}_n^\dagger\!\bigl(\res{s}{nI+G}\bigr)
 =\sum_{n\ge0}\frac{s}{s+n}\sem{S}_n^\dagger\!\bigl(\res{s+n}{G}\bigr).
\]

Once averaged over executions, however, such tests need not be the resolvent of any single cost. For instance, a scalar runtime taking the values $1$ and $3$ with equal probability has the $s\to\infty$ expansion
\[
 \frac12\frac{s}{s+1}+\frac12\frac{s}{s+3}
     =1-\frac2s+\frac5{s^2}+O(s^{-3}),
\]
which cannot equal $s/(s+c)$ for any fixed scalar $c\ge0$: matching the $s^{-1}$ coefficient forces $c=2$, whereas $s/(s+2)$ has coefficient $4$ rather than $5$ at order $s^{-2}$. To keep predicates closed under branch sums and program pullbacks, we therefore work with general effect families $F=(F_s)_{s>0}$, for which accumulating runtime $n$ transforms the terminal test into $\frac{s}{s+n}F_{s+n}$. Section~\ref{bridge:section} later shows how existing cost-logic proofs yield contracts in this domain.

Under sequential execution, accumulating runtime $c$ followed by $t$ applies this transformation twice, yielding $\frac{s}{s+c}\bigl(\frac{s+c}{s+c+t}F_{s+c+t}\bigr) = \frac{s}{s+c+t}F_{s+c+t}$. This consistency rests on a simple multiplicative identity that also gives the discount factor a random-deadline interpretation:

\begin{lemma}[Scale-shift identity and random deadlines]
	\label{lem:scale-identity}
	For all $s > 0$ and $c,t \ge 0$,
	\begin{equation}
		\label{eq:scale-identity}
		\frac{s}{s + c + t} = \frac{s}{s + c} \cdot \frac{s + c}{(s + c) + t}.
	\end{equation}
	Probabilistically, if $T_s \ge 0$ is a random deadline independent of the execution of $S$ with density $\frac{s}{(s+t)^2}$ on $[0,\infty)$ (so $T_s$ has median $s$), then its survival function obeys the scale-shifting memory law
	\[
	\Prob(T_s \ge t) = \frac{s}{s+t},
	\qquad
	\Prob(T_s - c \ge t \mid T_s \ge c) = \frac{s+c}{s+c+t}.
	\]
\end{lemma}

\begin{proof}
	Identity~\eqref{eq:scale-identity} is immediate by canceling $s+c$. Integrating $\int_t^\infty \frac{s}{(s+u)^2}\dd u = \frac{s}{s+t}$ gives $\Prob(T_s \ge t)$ (which equals $1/2$ at $t=s$), and dividing $\Prob(T_s \ge c+t)$ by $\Prob(T_s \ge c)$ yields the conditional survival probability.
\end{proof}

\subsection{Parameterized Predicates and Semantic Validity}
\label{sec:parameterized-validity}

We now formalize the space of scale-indexed effect families
and define the semantic validity of Hoare triples.

\begin{definition}[Parameterized predicates and semantic validity]
\label{def:parameterized-validity}
The predicate space on $\cH$ is the pointwise ordered $\omega$-CPO of scale-indexed effect families:
\[
 \Pred{\cH} \triangleq \prod_{s>0} \Eff{\cH},
 \qquad
 A \le F \iff A_s \le F_s \text{ for all } s > 0.
\]
Increasing sequential suprema are computed pointwise in the strong operator topology, and $\mathbf{0}$ and $\mathbf{I}$ denote the constant zero and identity families.

For $A,F\in\Pred{\cH}$, the Hoare triple $\{A\}\,S\,\{F\}$
is valid in the sense of total correctness, written
$\models_{\mathrm{tot}}\{A\}\,S\,\{F\}$, if, for every
$s>0$ and $\rho\in\mathcal D(\cH)$,
\begin{equation}
\label{eq:family-validity}
 \Tr(A_s\rho)\le\sum_{n\ge0}\frac{s}{s+n}\Tr\bigl(F_{s+n}\sem{S}_n(\rho)\bigr).
\end{equation}
The triple is valid in the sense of partial correctness, written
$\models_{\mathrm{par}}\{A\}\,S\,\{F\}$, if, for every
$s>0$ and $\rho\in\mathcal D(\cH)$,
\begin{equation}
\label{eq:family-validity-partial}
 \Tr(A_s\rho)\le\sum_{n\ge0}\frac{s}{s+n}\Tr\bigl(F_{s+n}\sem{S}_n(\rho)\bigr)
 +\Tr(\rho)-\Tr\bigl(\sem{S}(\rho)\bigr).
\end{equation}
\end{definition}

Both notions of validity evaluate the same discounted postcondition on terminating executions, differing only on nontermination: total correctness assigns nonterminating runs zero weight, whereas partial correctness credits the full nontermination mass $\Tr(\rho)-\Tr\bigl(\sem{S}(\rho)\bigr)$. Rather than evaluating these grade sums directly, we next characterize both notions of validity compositionally via weakest-precondition transformers.

\section{A Compositional Resolvent Calculus and Loop Solutions}
\label{sec:calculus}

In this section, we develop a compositional calculus for the weakest preconditions $\wp(S,F)$ and $\wlp(S,F)$. Section~\ref{subsec:wp-rules} establishes their explicit grade-sum representations and syntax-directed rules, which reduce while loops to fixed-point equations on $\Pred{\cH}$. Because closed-form loop solutions are rarely available, Section~\ref{subsec:loop-certification} proves uniqueness of the loop fixed point and derives residual bounds that certify asymptotic approximations, preparing for the extraction of termination probabilities and moments in Section \ref{moment:section}.

\subsection{Weakest Preconditions and Structural Rules}
\label{subsec:wp-rules}

\begin{definition}[Weakest preconditions]
\label{def:family-weakest-preconditions}
For a program $S$ and postcondition $F\in\Pred{\cH}$, the weakest precondition $\wp(S,F)$ and weakest liberal precondition $\wlp(S,F)$ are defined, when they exist, as the largest valid preconditions in $\Pred{\cH}$ for total and partial correctness, respectively. That is, they satisfy:
\begin{enumerate}
\item[(1)] \textbf{Validity:} $\models_{\mathrm{tot}}\{\wp(S,F)\}\,S\,\{F\}$ and $\models_{\mathrm{par}}\{\wlp(S,F)\}\,S\,\{F\}$.
\item[(2)] \textbf{Maximality:} For any $A\in\Pred{\cH}$, if $\models_{\mathrm{tot}}\{A\}\,S\,\{F\}$, then $A\le\wp(S,F)$; if $\models_{\mathrm{par}}\{A\}\,S\,\{F\}$, then $A\le\wlp(S,F)$.
\end{enumerate}
\end{definition}

\begin{theorem}[Weakest preconditions and structural rules]
\label{thm:semantics}
For every program $S$ and postcondition $F\in\Pred{\cH}$, the weakest precondition $\wp(S,F)$ exists and is given at each scale $s>0$ by
\begin{equation}
\label{eq:runtime-transformer}
 \wp(S,F)_s=\sum_{n\ge0}\frac{s}{s+n}\sem{S}_n^\dagger(F_{s+n}).
\end{equation}
The transformer $\wp(S,\cdot):\Pred{\cH}\to\Pred{\cH}$ is monotone, preserves increasing sequential suprema, and satisfies the structural rules
\begin{align}
 \wp(\mathbf{skip},F)_s&=F_s,\qquad \wp(\mathbf{abort},F)_s=0, \label{eq:primitive-rules}\\
 \wp(q:=\ket0,F)_s&=\sum_j\ketbra{j}{0}_q F_s\ketbra{0}{j}_q, \label{eq:initialization-rule}\\
 \wp(\bar q:=U[\bar q],F)_s&=U^\dagger F_sU, \label{eq:unitary-rule}\\
 \wp(\mathbf{if}\,(\square_{i\in J}\,M[\bar q]=i\to S_i)\,\mathbf{fi},F)_s
 &=\sum_{i\in J}M_i^\dagger\wp(S_i,F)_sM_i, \label{eq:case-rule}\\
 \wp(S_1;S_2,F)&=\wp(S_1,\wp(S_2,F)), \label{eq:transformer-sequence}
\end{align}
where $\{\ket j_q\}_j$ in~\eqref{eq:initialization-rule} is an orthonormal basis of $\cH_q$. Moreover, for a loop $L=\mathbf{while}\ M[\bar q]=1\ \mathbf{do}\ B\ \mathbf{od}$, $\wp(L,F)$ is the least fixed point in $\Pred{\cH}$ of the characteristic functional $\cT_F:\Pred{\cH}\to\Pred{\cH}$ defined by
\begin{equation}
\label{eq:general-loop-functional}
 (\cT_FX)_s=M_0^\dagger F_sM_0+\frac{s}{s+1}M_1^\dagger\wp(B,X)_{s+1}M_1.
\end{equation}
\end{theorem}

\begin{proof}
For each $s>0$ and $n\ge0$, positivity of $\sem{S}_n^\dagger$ and $0\le F_{s+n}\le I$ give $0\le\frac{s}{s+n}\sem{S}_n^\dagger(F_{s+n})\le\sem{S}_n^\dagger(I)=E_n^S$. By~\eqref{eq:termination-effects}, $\sum_{n\ge0}E_n^S=Q_S\le I$, so the series in~\eqref{eq:runtime-transformer} converges strongly to an effect in $\Eff{\cH}$ and $\wp(S,F)\in\Pred{\cH}$ is well defined. By trace duality and monotone convergence, the right-hand side of~\eqref{eq:family-validity} equals
\[
 \sum_{n\ge0}\frac{s}{s+n}\Tr\bigl(F_{s+n}\sem{S}_n(\rho)\bigr)
 =\sum_{n\ge0}\frac{s}{s+n}\Tr\bigl(\sem{S}_n^\dagger(F_{s+n})\rho\bigr)
 =\Tr\bigl(\wp(S,F)_s\rho\bigr).
\]
Hence $\models_{\mathrm{tot}}\{A\}\,S\,\{F\}$ holds if and only if $\Tr(A_s\rho)\le\Tr(\wp(S,F)_s\rho)$ for all $\rho\in\mathcal D(\cH)$ and $s>0$, which is equivalent to $A\le\wp(S,F)$. This establishes both validity and maximality. Monotonicity and preservation of increasing sequential suprema follow immediately from the normality of each adjoint map $\sem{S}_n^\dagger$.

We now verify the structural rules~\eqref{eq:primitive-rules}--\eqref{eq:general-loop-functional} using Definition~\ref{def:cost-grades}:
\begin{itemize}
\item \emph{Basic commands.} By~\eqref{eq:grade-base}--\eqref{eq:grade-updates}, all runtime grades of $\mathbf{abort}$ vanish, while $\mathbf{skip}$, initialization, and unitary updates are concentrated at grade $n=0$ with weight $\frac{s}{s+0}=1$. This gives~\eqref{eq:primitive-rules} and~\eqref{eq:unitary-rule}. For $q:=\ket0$, writing $\cR_q(\rho)=\ketbra{0}{0}_q\otimes\Tr_q(\rho)=\sum_j\ketbra{0}{j}_q\rho\ketbra{j}{0}_q$ yields $\cR_q^\dagger(F_s)=\sum_j\ketbra{j}{0}_qF_s\ketbra{0}{j}_q$, proving~\eqref{eq:initialization-rule}.

\item \emph{Conditionals.} Let $S_{\mathrm{if}}\triangleq \mathbf{if}\,(\square_{i\in J}\,M[\bar q]=i\to S_i)\,\mathbf{fi}$. By~\eqref{eq:grade-case}, $\sem{S_{\mathrm{if}}}_n^\dagger=\sum_{i\in J}\cM_i^\dagger\circ\sem{S_i}_n^\dagger$. Substituting this into~\eqref{eq:runtime-transformer} and interchanging the finite sum over $i\in J$ with the strong sum over $n\ge0$ gives
\[
 \wp(S_{\mathrm{if}},F)_s
 =\sum_{n\ge0}\frac{s}{s+n}\sum_{i\in J}M_i^\dagger\sem{S_i}_n^\dagger(F_{s+n})M_i
 =\sum_{i\in J}M_i^\dagger\wp(S_i,F)_sM_i.
\]

\item \emph{Sequential composition.} By~\eqref{eq:grade-sequence}, $\sem{S_1;S_2}_n^\dagger=\sum_{a+b=n}\sem{S_1}_a^\dagger\circ\sem{S_2}_b^\dagger$. Re-indexing the sum over $n\ge0$ as a double sum over $a,b\ge0$ with $n=a+b$ and applying the scale-shift identity~\eqref{eq:scale-identity} with $c=a$ and $t=b$, we obtain
\[
 \wp(S_1;S_2,F)_s
 =\sum_{a,b\ge0}\frac{s}{s+a}\frac{s+a}{s+a+b}\sem{S_1}_a^\dagger\bigl(\sem{S_2}_b^\dagger(F_{s+a+b})\bigr).
\]
Using the normality of $\sem{S_1}_a^\dagger$ to bring it outside the positive strong sum over $b\ge0$ yields
\[
 \wp(S_1;S_2,F)_s
 =\sum_{a\ge0}\frac{s}{s+a}\sem{S_1}_a^\dagger\bigl(\wp(S_2,F)_{s+a}\bigr)
 =\wp(S_1,\wp(S_2,F))_s.
\]

\item \emph{While loops.} Since $\wp(B,\cdot)$ is monotone and preserves increasing sequential suprema, so does $\cT_F$; by Kleene's fixed-point theorem on the $\omega$-CPO $\Pred{\cH}$, the least fixed point of $\cT_F$ is $\sup_{N\ge0}\cT_F^N\mathbf0$. Recall from~\eqref{eq:grade-loop} that $\cC_k=\sem{B}_{k-1}\circ\cM_1$ for $k\ge1$. Expanding $\wp(B,X)_{s+1}$ via~\eqref{eq:runtime-transformer}, setting $k=m+1$, and applying~\eqref{eq:scale-identity} with $c=1$ and $t=m$, we can rewrite the continuation term of~\eqref{eq:general-loop-functional} as
\[
 \frac{s}{s+1}M_1^\dagger\wp(B,X)_{s+1}M_1
 =\sum_{m\ge0}\frac{s}{s+1+m}M_1^\dagger\sem{B}_m^\dagger(X_{s+1+m})M_1
 =\sum_{k\ge1}\frac{s}{s+k}\cC_k^\dagger(X_{s+k}).
\]
An induction on $N\ge0$, again using~\eqref{eq:scale-identity} and the normality of $\cC_k^\dagger$, shows that $(\cT_F^N\mathbf0)_s$ sums all terminating unfoldings with fewer than $N$ outer iterations:
\[
 (\cT_F^N\mathbf0)_s
 =\sum_{j=0}^{N-1}\sum_{k_1,\dots,k_j\ge1}\frac{s}{s+k_1+\dots+k_j}
  (\cC_{k_1}^\dagger\circ\dots\circ\cC_{k_j}^\dagger\circ\cM_0^\dagger)(F_{s+k_1+\dots+k_j}).
\]
Taking the strong supremum as $N\to\infty$ and regrouping the positive terms by total runtime $n=k_1+\dots+k_j$ (so that $0\le j\le n$) recovers $\sum_{n\ge0}\frac{s}{s+n}\sem{L}_n^\dagger(F_{s+n})=\wp(L,F)_s$ by~\eqref{eq:grade-loop}. Thus $\wp(L,F)$ is the least fixed point of $\cT_F$.\qedhere
\end{itemize}
\end{proof}

\begin{proposition}[Weakest liberal preconditions]
\label{prop:liberal-rules}
For every program $S$, let $\mathbf D_S\triangleq (D_S)_{s>0}\in\Pred{\cH}$ denote the constant family given by the divergence effect $D_S=I-Q_S$ in~\eqref{eq:termination-effects}. Then for every $F\in\Pred{\cH}$,
\begin{equation}
\label{eq:liberal-transformer}
 \wlp(S,F)=\wp(S,F)+\mathbf D_S.
\end{equation}
The transformer $\wlp(S,\cdot):\Pred{\cH}\to\Pred{\cH}$ is monotone, preserves increasing sequential suprema, and satisfies
\begin{align}
 \wlp(\mathbf{skip},F)_s&=F_s,\qquad \wlp(\mathbf{abort},F)_s=I, \label{eq:liberal-base}\\
 \wlp(q:=\ket0,F)_s&=\sum_j\ketbra{j}{0}_qF_s\ketbra{0}{j}_q,
 \qquad \wlp(\bar q:=U[\bar q],F)_s=U^\dagger F_sU, \label{eq:liberal-updates}\\
 \wlp(S_{\mathrm{if}},F)_s&=\sum_{i\in J}M_i^\dagger\wlp(S_i,F)_sM_i, \label{eq:liberal-if}\\
 \wlp(S_1;S_2,F)_s
 &=\wlp(S_1,\wlp(S_2,F))_s
   +\sem{S_1}^\dagger(D_{S_2})-\wp(S_1,\mathbf D_{S_2})_s, \label{eq:liberal-sequence}
\end{align}
where $S_{\mathrm{if}}$ is the conditional in~\eqref{eq:case-rule}. Moreover, the correction term in~\eqref{eq:liberal-sequence} is positive:
\begin{equation}
\label{eq:liberal-sequence-correction}
 \sem{S_1}^\dagger(D_{S_2})-\wp(S_1,\mathbf D_{S_2})_s
 =\sum_{n\ge0}\frac{n}{s+n}\sem{S_1}_n^\dagger(D_{S_2})\ge0.
\end{equation}
\end{proposition}

\begin{proof}
In~\eqref{eq:family-validity-partial}, the nontermination term equals $\Tr(\rho)-\Tr(\sem{S}(\rho))=\Tr(D_S\rho)$ by~\eqref{eq:termination-effects}, so trace duality gives $\wlp(S,F)=\wp(S,F)+\mathbf D_S$. Since $0\le\wp(S,F)_s\le Q_S=I-D_S$, this sum defines a valid effect family in $\Pred{\cH}$, inheriting monotonicity and preservation of increasing suprema from $\wp(S,\cdot)$. Rules~\eqref{eq:liberal-base}--\eqref{eq:liberal-updates} follow from~\eqref{eq:primitive-rules}--\eqref{eq:unitary-rule} because $D_{\mathbf{abort}}=I$ whereas $\mathbf{skip}$, initialization, and unitary updates have $D_S=0$;~\eqref{eq:liberal-if} follows from $D_{S_{\mathrm{if}}}=\sum_{i\in J}M_i^\dagger D_{S_i}M_i$. For sequential composition, substituting $D_{S_1;S_2}=D_{S_1}+\sem{S_1}^\dagger(D_{S_2})$ and~\eqref{eq:transformer-sequence} into~\eqref{eq:liberal-transformer} at scale $s$ gives
\[
\begin{aligned}
 \wlp(S_1;S_2,F)_s
 &=\wp(S_1,\wp(S_2,F))_s+D_{S_1}
   +\sem{S_1}^\dagger(D_{S_2})\\
 &=\wlp(S_1,\wlp(S_2,F))_s
   +\sem{S_1}^\dagger(D_{S_2})
   -\wp(S_1,\mathbf D_{S_2})_s,
\end{aligned}
\]
where the second equality uses the additivity of $\wp(S_1,\cdot)$ in~\eqref{eq:runtime-transformer} together with $\wlp(S_2,F)=\wp(S_2,F)+\mathbf D_{S_2}$ and~\eqref{eq:liberal-transformer} for $S_1$. Finally, expanding $\sem{S_1}^\dagger(D_{S_2})=\sum_{n\ge0}\sem{S_1}_n^\dagger(D_{S_2})$ and subtracting $\wp(S_1,\mathbf D_{S_2})_s$ gives~\eqref{eq:liberal-sequence-correction}.
\end{proof}

Combining the initialization and unitary rules with sequential composition also gives the derived rule for the state-preparation shorthand $q:=\ket\phi$:
\[
 \wp(q:=\ket\phi,F)_s=\wlp(q:=\ket\phi,F)_s
 =\sum_j\ketbra{j}{\phi}_q F_s\ketbra{\phi}{j}_q.
\]
In particular, resetting all registers to a pure state $\sigma$ yields $\Tr(F_s\sigma)I$, as used in Section~\ref{sec:nested}.

Specializing Theorem~\ref{thm:semantics} and Proposition~\ref{prop:liberal-rules} to the terminal postcondition $F=\mathbf I$ defines the \emph{runtime resolvent family} $H^S\triangleq \wp(S,\mathbf I)\in\Pred{\cH}$ and its \emph{liberal counterpart} $L^S\triangleq \wlp(S,\mathbf I)\in\Pred{\cH}$, given at each scale $s>0$ by
\begin{align}
	H_s^S &\triangleq \wp(S,\mathbf I)_s = \sum_{n\ge 0} \frac{s}{s+n} E_n^S, \label{eq:runtime-family}\\
	L_s^S &\triangleq \wlp(S,\mathbf I)_s = H_s^S + D_S. \label{eq:liberal-runtime-family}
\end{align}
By Lemma~\ref{lem:scale-identity}, for a normalized input state $\rho$, $\Tr(H_s^S\rho) = \Prob_\rho(\tau_S \le T_s)$ is the probability that $S$ terminates before the independent random deadline $T_s$, whereas $\Tr(L_s^S\rho) = \Prob_\rho(\tau_S \le T_s \text{ or } \tau_S = \infty)$ additionally includes nonterminating runs. Later Theorem~\ref{prop:zero-order-readout} will show that $H_s^S$ increases strongly to $Q_S=I-D_S$ as $s\to\infty$. Thus $H^S$ determines $D_S$ and $L^S$; together with $\wp(S,F)$, it also determines $\wlp(S,F)$. The subsequent analysis therefore focuses on $\wp$ and $H^S$.

\subsection{Loop Unfolding and Residual Bounds}
\label{subsec:loop-certification}
\label{sec:loop-certification}

Theorem~\ref{thm:semantics} characterizes $\wp(L,F)$ as the least fixed point of $\cT_F$, but in applications one often works with finite unfoldings or asymptotic candidates rather than closed-form solutions. Both tasks require controlling how errors propagate under $\cT_F$. In this subsection, we show that $\cT_F$ contracts boundary differences at each scale, ensuring uniqueness of the fixed point, and use this decay to bound the approximation error of a candidate family by its one-step residual.

\begin{theorem}[Loop unfolding and uniqueness]
\label{thm:loop-certificates}
For every loop $L$ as in~\eqref{eq:general-loop-functional}, every $F\in\Pred{\cH}$, and all $X,Y\in\Pred{\cH}$,
\begin{equation}
\label{eq:boundary-decay}
 \norm{(\cT_F^NX)_s-(\cT_F^NY)_s}\le\frac{s}{s+N}
 \qquad(s>0,\ N\in\NN).
\end{equation}
For every $X\in\Pred{\cH}$, $(\cT_F^NX)_s$ converges in operator norm to $\wp(L,F)_s$ at each fixed $s>0$. Consequently, $X=\cT_FX$ has exactly one solution in $\Pred{\cH}$, namely $\wp(L,F)$.
\end{theorem}

\begin{proof}
For $X,Y\in\Pred{\cH}$, subtracting~\eqref{eq:general-loop-functional}
for $X$ and $Y$ cancels the exit term $M_0^\dagger F_sM_0$; expanding $\wp(B,\cdot)_{s+1}$ via~\eqref{eq:runtime-transformer} and applying~\eqref{eq:scale-identity} gives the one-step difference formula
\begin{equation}
\label{eq:one-step-difference}
\begin{aligned}
 &\qquad(\cT_FX)_s-(\cT_FY)_s\\
 &=\frac{s}{s+1}M_1^\dagger\bigl(\wp(B,X)_{s+1}-\wp(B,Y)_{s+1}\bigr)M_1\\
 &=\sum_{m\ge0}\frac{s}{s+1+m}M_1^\dagger\sem{B}_m^\dagger(X_{s+1+m}-Y_{s+1+m})M_1.
 \end{aligned}
\end{equation}
We prove~\eqref{eq:boundary-decay} by induction on $N$. For $N=0$, $0\le X_s,Y_s\le I$ implies $-I\le X_s-Y_s\le I$, so $\norm{X_s-Y_s}\le 1=\frac{s}{s+0}$. Assume~\eqref{eq:boundary-decay} holds for $N$, and set $\Delta_u^{(N)}\triangleq (\cT_F^NX)_u-(\cT_F^NY)_u$. At each shifted scale $u=s+1+m$, the induction hypothesis reads $-\frac{s+1+m}{s+1+m+N}I\le\Delta_{s+1+m}^{(N)}\le\frac{s+1+m}{s+1+m+N}I$. Since each map $M_1^\dagger\sem{B}_m^\dagger(\cdot)M_1$ is positive and
\[
\begin{aligned}
   & \frac{s}{s+1+m}\cdot\frac{s+1+m}{s+1+m+N}
 =\frac{s}{s+1+m+N}\le\frac{s}{s+N+1},\\[10pt]
 &\sum_{m\ge0}M_1^\dagger\sem{B}_m^\dagger(I)M_1
 =M_1^\dagger Q_BM_1\le M_1^\dagger M_1\le I,
\end{aligned}
\]
applying~\eqref{eq:one-step-difference} to $\cT_F^NX$ and $\cT_F^NY$ yields $-\frac{s}{s+N+1}I\le\Delta_s^{(N+1)}\le\frac{s}{s+N+1}I$, which completes the induction.

By Theorem~\ref{thm:semantics}, $\wp(L,F)$ is a fixed point of $\cT_F$. Taking $Y=\wp(L,F)$ in~\eqref{eq:boundary-decay} gives $\norm{(\cT_F^NX)_s-\wp(L,F)_s}\le\frac{s}{s+N}\to0$ as $N\to\infty$. In particular, if $X\in\Pred{\cH}$ satisfies $X=\cT_FX$, then $X_s=(\cT_F^NX)_s\to\wp(L,F)_s$, so $X=\wp(L,F)$.
\end{proof}

To certify an approximate solution, we substitute a candidate family $A$ into the loop equation and estimate its one-step residual $A-\cT_FA$. Because truncated asymptotic expansions need not lie in $\Eff{\cH}$, we evaluate~\eqref{eq:runtime-transformer} and~\eqref{eq:general-loop-functional} on uniformly bounded self-adjoint families by the same formulas. Note that if a candidate is uniformly bounded for $s\ge s_0>0$, extending it boundedly to $(0,s_0)$ does not affect its residual at scales $s\ge s_0$, since~\eqref{eq:general-loop-functional} only queries larger scales.

\begin{theorem}[Certification of an asymptotic loop solution]
\label{thm:residual}
For a loop $L$ as in~\eqref{eq:general-loop-functional}, a postcondition $F\in\Pred{\cH}$, and a self-adjoint family $A$ with $\sup_{s>0}\norm{A_s}<\infty$, the approximation error satisfies
\begin{equation}
\label{eq:variable-residual-bound}
 \norm{A_s-\wp(L,F)_s}
 \le\sum_{j\ge0}\sup_{u\ge s+j}\frac{s}{u}
       \norm{A_u-(\cT_FA)_u}\qquad(s>0).
\end{equation}
Consequently, a residual $\norm{A_s-(\cT_FA)_s}=O(s^{-r-1})$ gives an error $\norm{A_s-\wp(L,F)_s}=O(s^{-r})$ for every $r>0$. If, in addition, there is $0\le\kappa<1$ such that
\[
 \frac{u}{u+1}M_1^\dagger\wp(B,\mathbf I)_{u+1}M_1
 \le\kappa I\qquad(u>0),
\]
then
\begin{equation}
\label{eq:tail-contraction-error}
 \sup_{u\ge s}\norm{A_u-\wp(L,F)_u}
 \le\frac{1}{1-\kappa}\sup_{u\ge s}\norm{A_u-(\cT_FA)_u}.
\end{equation}
In this case a residual $O(s^{-r})$ gives an error of the same order. Both rate statements also hold with $o$ in place of $O$.
\end{theorem}

\begin{proof}
The proof combines a bound on the propagation of differences with
a telescoping sum of one-step residuals.
Firstly, for any uniformly bounded self-adjoint families $X,Y$, we prove
by induction on $j\ge0$ that
\begin{equation}
\label{eq:weighted-iteration-difference}
 \norm{(\cT_F^jX)_s-(\cT_F^jY)_s}
 \le\sup_{u\ge s+j}\frac{s}{u}\norm{X_u-Y_u}\qquad(s>0).
\end{equation}
The case $j=0$ holds trivially since the supremum includes $u=s$,
where $\frac{s}{u}=1$. Assuming~\eqref{eq:weighted-iteration-difference} for $j$, its bound at the shifted scale $s+1+m$ gives
\[
\begin{aligned}
 &\qquad   \frac{s}{s+1+m}\norm{(\cT_F^jX)_{s+1+m}-(\cT_F^jY)_{s+1+m}}\\
 &\le\frac{s}{s+1+m}\sup_{u\ge s+1+m+j}\frac{s+1+m}{u}\norm{X_u-Y_u}\\
 &\le\sup_{u\ge s+j+1}\frac{s}{u}\norm{X_u-Y_u}.
\end{aligned}
\]
Substituting this operator-norm bound into~\eqref{eq:one-step-difference} and using $\sum_{m\ge0}M_1^\dagger\sem{B}_m^\dagger(I)M_1\le I$ proves~\eqref{eq:weighted-iteration-difference} for $j+1$.

Taking $X=A$ and $Y=\wp(L,F)$ in~\eqref{eq:weighted-iteration-difference} gives $\norm{(\cT_F^NA)_s-\wp(L,F)_s}\le\frac{s}{s+N}\sup_{u\ge s+N}\norm{A_u-\wp(L,F)_u}\to 0$ as $N\to\infty$. Applying~\eqref{eq:weighted-iteration-difference} with $X=A$ and $Y=\cT_FA$ to each term of the telescoping sum
\[
 A_s-(\cT_F^NA)_s
 =\sum_{j=0}^{N-1}\bigl((\cT_F^jA)_s-(\cT_F^j(\cT_FA))_s\bigr)
\]
and letting $N\to\infty$ establishes~\eqref{eq:variable-residual-bound}. If $\norm{A_u-(\cT_FA)_u}\le Cu^{-r-1}$ for large $u$, then $\sup_{u\ge s+j}\frac{s}{u}\norm{A_u-(\cT_FA)_u}\le Cs(s+j)^{-r-2}$, and summing over $j\ge0$ yields $C\sum_{j\ge0}s(s+j)^{-r-2}=O(s^{-r})$.

Now assume the contraction condition with $\kappa<1$. Fix $s>0$ and set $e_s\triangleq \sup_{v\ge s}\norm{A_v-\wp(L,F)_v}<\infty$. For every $u\ge s$ and $m\ge0$, we have $u+1+m\ge s$ and hence $-e_sI\le A_{u+1+m}-\wp(L,F)_{u+1+m}\le e_sI$. Applying the positive map $\wp(B,\cdot)_{u+1}$ in~\eqref{eq:runtime-transformer} gives
\[
 -e_s\wp(B,\mathbf I)_{u+1}
 \le\wp(B,A)_{u+1}-\wp(B,\wp(L,F))_{u+1}
 \le e_s\wp(B,\mathbf I)_{u+1}.
\]
By~\eqref{eq:one-step-difference} and $\cT_F(\wp(L,F))=\wp(L,F)$, conjugating by $M_1$ and multiplying by $\frac{u}{u+1}$ yields
\[
\begin{aligned}
 &\qquad  \norm{(\cT_FA)_u-\wp(L,F)_u}\\
 &=\norm{\frac{u}{u+1}M_1^\dagger\bigl(\wp(B,A)_{u+1}-\wp(B,\wp(L,F))_{u+1}\bigr)M_1}\\
 &\le e_s\norm{\frac{u}{u+1}M_1^\dagger\wp(B,\mathbf I)_{u+1}M_1}\\
 &\le\kappa e_s. 
\end{aligned}
\]
Taking the supremum over $u\ge s$ and using the triangle inequality $A_u-\wp(L,F)_u=\bigl(A_u-(\cT_FA)_u\bigr)+\bigl((\cT_FA)_u-\wp(L,F)_u\bigr)$ gives $e_s\le\sup_{u\ge s}\norm{A_u-(\cT_FA)_u}+\kappa e_s$, which rearranges to~\eqref{eq:tail-contraction-error}. Finally, the little-$o$ statements follow by applying the same bounds with arbitrarily small tail constants as $s\to\infty$.
\end{proof}

\section{Extracting Termination Probabilities, Moments, and Divergence Rates}
\label{moment:section}

The compositional calculus and loop rules of Section~\ref{sec:calculus} do not compute the runtime observables of Section~\ref{sec:runtime-observables} directly; instead, for a program $S$, they characterize the runtime operator family $H_s^S = \wp(S,\mathbf I)_s$ either exactly or through operator-norm approximations (Theorem~\ref{thm:residual}). Because the operator norm directly bounds the error in expectation for any normalized input state $\rho$, an operator-norm approximation whose error decays sufficiently fast as $s \to \infty$ yields the asymptotic expansion of $\Tr(\rho H_s^S)$ to the desired order. This section establishes how to recover the operational runtime properties of $S$ from the large-$s$ behavior of $H_s^S$. Specifically, we recover termination probabilities in Section~\ref{subsec:probability-readout}, finite runtime moments on terminating runs in Section~\ref{subsec:moment-readout}, and divergence rates of truncated moments in Section~\ref{subsec:divergence-readout}. In each subsection, we show how to obtain the target quantities both directly from the operator family $H_s^S$ and from the asymptotic expansion of $\Tr(\rho H_s^S)$ as $s \to \infty$.

\subsection{Termination Probabilities}
\label{subsec:probability-readout}

As the parameter $s$ tends to infinity, the rational weights $\frac{s}{s+n}$ approach $1$ for every runtime $n$. Consequently, the large-$s$ limit of $H_s^S$ recovers the total weight of all terminating executions, which appears as the leading constant term in the asymptotic expansion of $\Tr(\rho H_s^S)$.

\begin{theorem}[Termination probability from the exact family]
\label{prop:zero-order-readout}
For every program $S$, as $s \to \infty$, the family $H_s^S$ increases monotonically with respect to the L\"owner order and converges in the strong operator topology to $Q_S$:
\begin{equation}
\label{moment:termination-limit}
 \lim_{s\to\infty} H_s^S = Q_S,
 \qquad D_S = I - \lim_{s\to\infty} H_s^S.
\end{equation}
Consequently, for any normalized input state $\rho$, the termination probability is $\Tr(\rho Q_S) = \lim_{s\to\infty}\Tr(\rho H_s^S)$, and the divergence probability is $\Tr(\rho D_S) = 1 - \lim_{s\to\infty}\Tr(\rho H_s^S)$.
\end{theorem}

\begin{proof}
For each $n \ge 0$, the weight $\frac{s}{s+n}$ increases monotonically to $1$ as $s \to \infty$. Since $H_s^S = \sum_{n=0}^\infty \frac{s}{s+n} E_n^S$ and $\sum_{n=0}^\infty E_n^S = Q_S \le I$, monotone convergence for bounded positive operators yields the strong operator limits in~\eqref{moment:termination-limit}, and the trace identities follow immediately.
\end{proof}

\begin{theorem}[Termination probability from the leading asymptotic constant]
\label{thm:asymptotic-probability}
If a normalized input state $\rho$ satisfies $\Tr(\rho H_s^S) = a_0 + o(1)$ as $s \to \infty$, then the termination and divergence probabilities of $S$ on $\rho$ are $\Tr(\rho Q_S) = a_0$ and $\Tr(\rho D_S) = 1 - a_0$, respectively.
\end{theorem}

\subsection{Runtime Moments on Terminating Runs}
\label{subsec:moment-readout}

Beyond termination probabilities, the exact family $H_s^S$ determines every successful runtime moment $M_k^S$ of integer order through a positive $k$-th order finite difference across rescaled parameters $s/j$, whereas in the asymptotic expansion of $\Tr(\rho H_s^S)$ as $s \to \infty$, the finite moments $\Tr(\rho M_j^S)$ appear as the coefficients of $(-s)^{-j}$.

\begin{theorem}[Moments from finite differences of the exact family]
\label{thm:universal-difference-readout}
For every integer $k \ge 1$ and parameter $s > 0$, the finite-difference operator
\begin{equation}
\label{eq:Bk-definition}
 B_k^S(s) \triangleq \frac{s^k}{k!}\left[Q_S + \sum_{j=1}^k (-1)^j \binom{k}{j} H_{s/j}^S\right]
\end{equation}
is a bounded positive operator admitting the series representation
\begin{equation}
\label{eq:Bk-positive-series}
 B_k^S(s) = \sum_{n=0}^\infty n^k \left(\prod_{j=1}^k \frac{s}{s+jn}\right) E_n^S.
\end{equation}
As $s \to \infty$, the quadratic forms of $B_k^S(s)$ increase monotonically to the quadratic form of $M_k^S$. Consequently, for every normalized input state $\rho$,
\begin{equation}
\label{eq:Bk-monotone-limit}
 \Tr(\rho M_k^S) = \lim_{s\to\infty} \Tr(\rho B_k^S(s)) \in [0,\infty],
\end{equation}
and $\Tr(\rho M_k^S)$ is finite if and only if $\Tr(\rho B_k^S(s))$ remains bounded as $s \to \infty$.
\end{theorem}

\begin{proof}
Expanding the rational function $\prod_{j=1}^k (u+j)^{-1}$ into partial fractions in $u$ and substituting $u = 1/x$ yields the identity
\[
 \frac{1}{k!}\left[1 + \sum_{j=1}^k (-1)^j \binom{k}{j} \frac{1}{1+jx}\right]
 = \frac{x^k}{\prod_{j=1}^k (1+jx)} \qquad (x \ge 0).
\]
Setting $x = n/s$, multiplying by $s^k E_n^S$, and summing over $n \ge 0$ gives~\eqref{eq:Bk-positive-series}. For each $n \ge 0$, the weight $n^k \prod_{j=1}^k \frac{s}{s+jn}$ is bounded above by $\frac{s^k}{k!}$ and increases monotonically to $n^k$ as $s \to \infty$. Monotone convergence therefore yields the form limit $M_k^S$ and the trace identity~\eqref{eq:Bk-monotone-limit}.
\end{proof}

\begin{theorem}[Moments from asymptotic coefficients]
\label{moment:extraction}
Let $M_0^S \triangleq Q_S$. For any integer $k \ge 1$ and normalized input state $\rho$, the moment $\Tr(\rho M_k^S)$ is finite if and only if there exist real coefficients $a_0, \ldots, a_k$ such that, as $s \to \infty$,
\begin{equation}
\label{moment:finite-expansion}
 \Tr(\rho H_s^S) = \sum_{j=0}^k \frac{(-1)^j a_j}{s^j} + o(s^{-k}).
\end{equation}
Whenever this expansion holds, the coefficients are given by $a_j = \Tr(\rho M_j^S)$ for all $0 \le j \le k$.
\end{theorem}

\begin{proof}
Write $p_n \triangleq \Tr(\rho E_n^S)$. Applying the finite geometric series identity
\[
 \frac{1}{1+x} = \sum_{j=0}^{m-1} (-x)^j + \frac{(-x)^m}{1+x}
\]
at $x = n/s$ and summing against $p_n$ shows that for any integer $m \ge 1$ with $\Tr(\rho M_{m-1}^S) < \infty$,
\begin{equation}
\label{moment:remainder-expansion}
 \Tr(\rho H_s^S)
 = \sum_{j=0}^{m-1} \frac{(-1)^j \Tr(\rho M_j^S)}{s^j}
 + \frac{(-1)^m}{s^m} \sum_{n=0}^\infty \frac{s n^m}{s+n} p_n.
\end{equation}
The remainder sum $\sum_{n=0}^\infty \frac{s n^m}{s+n} p_n$ is bounded by $s\Tr(\rho M_{m-1}^S) < \infty$ and increases monotonically to $\Tr(\rho M_m^S)$ as $s \to \infty$. If $\Tr(\rho M_k^S) < \infty$, taking $m = k$ in~\eqref{moment:remainder-expansion} and applying dominated convergence to the $k$-th term yields~\eqref{moment:finite-expansion} with $a_j = \Tr(\rho M_j^S)$.

Conversely, suppose~\eqref{moment:finite-expansion} holds. Theorem~\ref{thm:asymptotic-probability} gives $a_0 = \Tr(\rho Q_S) = \Tr(\rho M_0^S)$. Proceeding by induction on $m \in \{1,\ldots,k\}$, once $a_j = \Tr(\rho M_j^S) < \infty$ has been established for all $j < m$, comparing~\eqref{moment:finite-expansion} with~\eqref{moment:remainder-expansion} shows that
\[
 \sum_{n=0}^\infty \frac{s n^m}{s+n} p_n = a_m + o(1) \qquad (s \to \infty).
\]
By monotone convergence as $s \to \infty$, the left-hand side converges to $\Tr(\rho M_m^S)$, whence $\Tr(\rho M_m^S) = a_m < \infty$.
\end{proof}

Because $M_r^S = \sum_{n=0}^\infty n^r E_n^S$ sums only over finite runtimes, its expectation $\Tr(\rho M_r^S) = \sum_{n=0}^\infty n^r \Tr(\rho E_n^S) = \E_\rho[\tau_S^r \mathbf{1}_{\{\tau_S < \infty\}}]$ is the $r$-th runtime moment restricted to terminating runs. When the termination probability $\Tr(\rho Q_S)$ is strictly positive, conditioning on the event $\{\tau_S < \infty\}$ normalizes the runtime distribution to $\Prob_\rho(\tau_S = n \mid \tau_S < \infty) = \frac{\Tr(\rho E_n^S)}{\Tr(\rho Q_S)}$, so the conditional $r$-th moment given termination is
\[
 \E_\rho[\tau_S^r \mid \tau_S < \infty] = \frac{\Tr(\rho M_r^S)}{\Tr(\rho Q_S)}.
\]
By contrast, the total runtime form $C_S = M_1^S + \infty D_S$ assigns infinite cost to nonterminating runs; it is recovered as the $s \to \infty$ monotone form limit of the bounded positive operators $s(I - H_s^S)$, yielding the unconditional expected runtime
\begin{equation}
\label{eq:total-cost-deficit-limit}
 s(I - H_s^S) = s D_S + B_1^S(s), \qquad
 \E_\rho[\tau_S] = \Tr(\rho C_S) = \lim_{s\to\infty} s\bigl(1 - \Tr(\rho H_s^S)\bigr).
\end{equation}

\subsection{Divergence Rates of Truncated Moments}
\label{subsec:divergence-readout}

When $\Tr(\rho M_r^S) = \infty$ for a given input state $\rho$, taking the limit $s \to \infty$ only indicates divergence. To quantify the rate of divergence as the truncation threshold $N > 0$ grows, we consider the bounded truncated (or capped) moment operators
\begin{equation}
\label{moment:capped-moments}
 \widetilde M_r^S(N) \triangleq \sum_{n=0}^\infty \min(n,N)^r E_n^S
 \qquad (r > 0,\ N > 0).
\end{equation}
As $N \to \infty$, the expectation $\Tr(\rho\widetilde M_r^S(N))$ either converges to $\Tr(\rho M_r^S)$ when the $r$-th moment on $\rho$ is finite, or diverges at a rate governed by the tail of the terminating runtime distribution when the moment is infinite.

For the exact family $H_s^S$, evaluating the finite-difference operator $B_k^S(s)$ directly at $s = N$ already captures the asymptotic growth order of $\widetilde M_k^S(N)$ up to constant factors.

\begin{theorem}[Divergence rates from the exact family]
\label{thm:exact-divergence-readout}
For every integer $k \ge 1$ and truncation threshold $N > 0$, the finite-difference operator $B_k^S(N)$ in~\eqref{eq:Bk-definition} satisfies the two-sided operator inequality
\begin{equation}
\label{eq:Bk-universal-sandwich}
 \frac{1}{(k+1)!}\widetilde M_k^S(N)
 \le B_k^S(N) \le \widetilde M_k^S(N).
\end{equation}
Consequently, $\Tr(\rho\widetilde M_k^S(N))$ and $\Tr(\rho B_k^S(N))$ have the same asymptotic order of growth as $N \to \infty$ up to constant factors.
\end{theorem}

\begin{proof}
In the series representation~\eqref{eq:Bk-positive-series} at $s = N$, the $n$-th weight is $w_n(N) \triangleq n^k \prod_{j=1}^k \frac{N}{N+jn}$. For $n \le N$, each factor $\frac{N}{N+jn}$ lies in $[\frac{1}{j+1}, 1]$, so $\frac{n^k}{(k+1)!} \le w_n(N) \le n^k$. For $n > N$, rewriting $w_n(N) = N^k \prod_{j=1}^k \frac{n}{N+jn}$ shows that each factor lies in $[\frac{1}{j+1}, \frac{1}{j}]$, whence $\frac{N^k}{(k+1)!} \le w_n(N) \le N^k$. Combining both ranges gives $\frac{1}{(k+1)!}\min(n,N)^k \le w_n(N) \le \min(n,N)^k$ for all $n \ge 0$, and summing against $E_n^S \ge 0$ yields~\eqref{eq:Bk-universal-sandwich}.
\end{proof}

When an asymptotic expansion of $\Tr(\rho H_s^S)$ is available and the first $k-1$ moments on $\rho$ are finite, subtracting the polynomial terms up to $s^{-(k-1)}$ leaves a $k$-th order remainder $(-1)^k f(s)/s^k$. The next theorem shows how the large-$s$ behavior of $f(s)$ governs the truncated moments: a bounded remainder yields a finite $k$-th moment with an explicit truncation bound, whereas logarithmic or fractional-power growth determines the tail exponent $\alpha$ on terminating runs and the resulting divergence and convergence rates across all orders $r > 0$.

\begin{theorem}[Divergence and convergence rates from the asymptotic remainder]
\label{thm:moment-readout}
For an integer $k \ge 1$, a normalized input state $\rho$, and real
coefficients $a_0,\ldots,a_{k-1}$, define the remainder $f(s)$ for every
$s>0$ by
\begin{equation}
\label{eq:remainder-classification}
 \Tr(\rho H_s^S)
 = \sum_{j=0}^{k-1}\frac{(-1)^j a_j}{s^j}
   + \frac{(-1)^k f(s)}{s^k}.
\end{equation}
Suppose that $f(s)=o(s)$ as $s\to\infty$.
Then $a_j = \Tr(\rho M_j^S) < \infty$ for all $0 \le j < k$ (with $M_0^S \triangleq Q_S$), and the remainder $f(s)$ determines the asymptotic behavior of the truncated moments as follows:
\begin{enumerate}
\item \textbf{Bounded case:} If $f(s)$ is bounded, then $\Tr(\rho M_k^S) = \lim_{s\to\infty} f(s) =: a_k < \infty$, and the truncation error at threshold $N$ satisfies
\begin{equation}
\label{eq:finite-moment-convergence}
 0 \le \Tr(\rho M_k^S) - \Tr(\rho\widetilde M_k^S(N))
 \le 2\bigl(a_k - f(N)\bigr).
\end{equation}
\item \textbf{Logarithmic case:} If $f(s) = \lambda\log s + O(1)$ with $\lambda > 0$, then
\begin{equation}
\label{moment:critical-conclusion}
 \Tr(\rho\widetilde M_k^S(N)) = \lambda\log N + O(1),
\end{equation}
and the tail exponent is $\alpha = k$.
\item \textbf{Fractional-power case:} If $f(s) = \Theta(s^\beta)$ with $0 < \beta < 1$, then
$\Tr(\rho\widetilde M_k^S(N)) = \Theta(N^\beta)$, and the tail exponent is $\alpha = k - \beta$.
\end{enumerate}
Moreover, in both the logarithmic and fractional-power cases, the tail probability on terminating runs decays as $\sum_{n>N} \Tr(\rho E_n^S) = \Theta(N^{-\alpha})$. Consequently, $\alpha$ is the critical moment threshold: $\Tr(\rho M_r^S)$ is finite if and only if $0 < r < \alpha$, and for all $r > 0$,
\begin{equation}
\label{eq:classified-moment-rates}
\begin{aligned}
 \Tr(\rho M_r^S) - \Tr(\rho\widetilde M_r^S(N))
 &= \Theta(N^{r-\alpha}) && (0 < r < \alpha),\\
 \Tr(\rho\widetilde M_\alpha^S(N)) &= \Theta(\log N),\\
 \Tr(\rho\widetilde M_r^S(N)) &= \Theta(N^{r-\alpha}) && (r > \alpha).
\end{aligned}
\end{equation}
\end{theorem}

\begin{proof}
Write $p_n \triangleq \Tr(\rho E_n^S)$. Since $f(s)/s^k = o(s^{-(k-1)})$, Theorems~\ref{thm:asymptotic-probability} and~\ref{moment:extraction} give $a_j = \Tr(\rho M_j^S) < \infty$ for $0 \le j < k$. Comparing~\eqref{eq:remainder-classification} with the geometric remainder identity~\eqref{moment:remainder-expansion} at $m = k$ shows that
\[
 f(s) = \sum_{n=0}^\infty \frac{s n^k}{s+n} p_n, \qquad
 f(2s) - f(s) = \sum_{n=0}^\infty \frac{s n^{k+1}}{(s+n)(2s+n)} p_n.
\]
As $s \to \infty$, $f(s)$ increases monotonically to $\Tr(\rho M_k^S)$. If $f$ is bounded, this limit $a_k = \Tr(\rho M_k^S)$ is finite; since $n^k - N^k \le n^k \le \frac{2 n^{k+1}}{N+n}$ for $n > N$, comparing $\Tr(\rho M_k^S) - \Tr(\rho\widetilde M_k^S(N)) = \sum_{n>N} (n^k - N^k)p_n$ with $a_k - f(N) = \sum_{n=0}^\infty \frac{n^{k+1}}{N+n}p_n$ yields~\eqref{eq:finite-moment-convergence}.

Next, we relate $f(s)$ to the truncated moment $\Tr(\rho\widetilde M_k^S(s)) = \sum_{n=0}^\infty \min(n,s)^k p_n$ and bound the tail $\sum_{n>N} p_n$ from above.
In the logarithmic case, comparing the rational weight $\frac{sn^k}{s+n}$ with the truncated weight $\min(n,s)^k$ separately on $n \le s$ and $n > s$ gives the pointwise bounds
\[
 \left|\frac{s n^k}{s+n} - \min(n,s)^k\right|
 \le \frac{6 s n^{k+1}}{(s+n)(2s+n)} \quad (n \ge 0), \qquad
 \frac{s n^{k+1}}{(s+n)(2s+n)} \ge \frac{s^k}{6} \quad (n > s).
\]
Summing against $p_n$ bounds both the difference $|\Tr(\rho\widetilde M_k^S(s)) - f(s)|$ and the tail by $f(2s) - f(s)$:
\[
 \left|\Tr(\rho\widetilde M_k^S(s)) - f(s)\right|
 \le 6\bigl(f(2s) - f(s)\bigr), \qquad
 s^k \sum_{n>s} p_n \le 6\bigl(f(2s) - f(s)\bigr).
\]
Because $f(s) = \lambda\log s + O(1)$ implies $f(2s) - f(s) = O(1)$, setting $s = N$ proves~\eqref{moment:critical-conclusion} and $\sum_{n>N} p_n = O(N^{-k})$.
In the fractional-power case, $\min(n,s)^k \le \frac{2sn^k}{s+n}$ immediately gives $\Tr(\rho\widetilde M_k^S(s)) \le 2f(s)$. For a lower bound, splitting the sum for $f(s)$ at $n = bs$ with $b \ge 1$ and noting that $\frac{sn^k}{s+n} \le \frac{2}{b}\frac{(bs)n^k}{bs+n}$ for $n > bs$ yields
\[
 \Tr(\rho\widetilde M_k^S(bs)) \ge f(s) - \frac{2}{b} f(bs).
\]
Since $f(s) = \Theta(s^\beta)$ with $\beta < 1$, choosing $b$ sufficiently large ensures $\frac{2}{b}f(bs) \le \frac{1}{2}f(s)$ for large $s$, so $\Tr(\rho\widetilde M_k^S(N)) = \Theta(N^\beta)$. Moreover, $N^k \sum_{n>N} p_n \le \Tr(\rho\widetilde M_k^S(N))$ gives $\sum_{n>N} p_n = O(N^{\beta-k})$.

To obtain matching tail lower bounds, observe that raising the truncation threshold from $N$ to $bN$ ($b > 1$) only increases terms with $n > N$, each by at most $(b^k - 1)N^k$:
\[
 \Tr(\rho\widetilde M_k^S(bN)) - \Tr(\rho\widetilde M_k^S(N))
 \le (b^k - 1) N^k \sum_{n>N} p_n \qquad (b > 1).
\]
In the logarithmic case,~\eqref{moment:critical-conclusion} makes the left-hand side $\lambda\log b + O(1)$, which is strictly positive for a sufficiently large fixed $b$, yielding $\sum_{n>N} p_n = \Omega(N^{-k})$. In the fractional-power case, $\Tr(\rho\widetilde M_k^S(N)) = \Theta(N^\beta)$ makes the left-hand side $\Omega(N^\beta)$ for a sufficiently large fixed $b$, yielding $\sum_{n>N} p_n = \Omega(N^{\beta-k})$. Thus $\sum_{n>N} p_n = \Theta(N^{-\alpha})$ in both cases.

Finally, for any $r > 0$, substituting $\min(n,N)^r = r\int_0^N x^{r-1}\mathbf{1}_{\{n>x\}}\dd x$ and applying Fubini's theorem expresses the truncated moments in terms of the tail sum:
\[
 \Tr(\rho\widetilde M_r^S(N))
 = r \int_0^N x^{r-1} \sum_{n>x} p_n \dd x, \qquad
 \Tr(\rho M_r^S) - \Tr(\rho\widetilde M_r^S(N))
 = r \int_N^\infty x^{r-1} \sum_{n>x} p_n \dd x,
\]
where the second identity holds whenever $\Tr(\rho M_r^S) < \infty$. Substituting $\sum_{n>x} p_n = \Theta(x^{-\alpha})$ into both integrals immediately gives the finiteness range $0 < r < \alpha$ and all three asymptotic rates in~\eqref{eq:classified-moment-rates}.
\end{proof}

As in Section~\ref{subsec:moment-readout}, $\Tr(\rho\widetilde M_r^S(N)) = \E_\rho[\min(\tau_S,N)^r \mathbf{1}_{\{\tau_S < \infty\}}]$ sums only over terminating executions. When $\Tr(\rho Q_S) > 0$, dividing by the termination probability gives the conditional truncated moment given termination, whereas including the divergence probability $\Tr(\rho D_S) = 1 - \Tr(\rho Q_S)$ (on which $\min(\tau_S,N)^r = N^r$) gives the unconditional truncated moment:
\begin{equation}
\label{moment:conditioning}
\begin{aligned}
 \E_\rho[\min(\tau_S,N)^r \mid \tau_S < \infty]
 &= \frac{\Tr(\rho\widetilde M_r^S(N))}{\Tr(\rho Q_S)},\\
 \E_\rho[\min(\tau_S,N)^r]
 &= \Tr(\rho D_S) N^r + \Tr(\rho\widetilde M_r^S(N)).
\end{aligned}
\end{equation}
\section{Certified First-Passage Analysis of Coherent Quantum Walks}
\label{walk:sec-certified-walks}

We now apply the parameterized weakest-precondition calculus and asymptotic extraction theorems of Sections~\ref{sec:runtime-semantics}--\ref{moment:section} to the absorbing quantum walk~\cite{KonnoEtAl2003,BachEtAl2004} of Section~\ref{walk:sec-cost}, recovering the quantitative runtime properties that were lost when the single expected-cost predicate collapsed to $+\infty$ (Theorem~\ref{walk:thm-infinite-cost}). We first carry out this analysis for the Hadamard walk started at $\ket{R,1}$ and then extend it to biased coins and arbitrary entrance states.

\subsection{Deriving the Runtime Family}
\label{walk:subsec-first-crossing}

Let $\mathsf{Walk}$ be the while command of Section~\ref{walk:sec-cost}, with guard projectors $M_0=\Pi_0$ and $M_1=\Pi_{\ne0}$, and retain $U$, $B$, and $K$ from~\eqref{walk:eq-hadamard-operations}--\eqref{walk:eq-K-B}. For a postcondition family $F\in\Pred{\cH}$, we derive $\wp(\mathsf{Walk},F)$ by propagating predicates backwards through the body and then combining the two guard branches.

For an intermediate family $X$, the unitary rule of Theorem~\ref{thm:semantics} gives the following exact annotations at scale $s$, read from bottom to top:
\[
\begin{array}{l}
 \bigl\{(C_{1/2}^{\dagger}\otimes I)V^{\dagger}X_sV(C_{1/2}\otimes I)\bigr\}\\[2pt]
 \quad\mathtt{coin}:=C_{1/2}[\mathtt{coin}];\\[2pt]
 \bigl\{V^{\dagger}X_sV\bigr\}\\[2pt]
 \quad(\mathtt{coin,position}):=V[\mathtt{coin,position}]\\[2pt]
 \{X_s\}.
\end{array}
\]
The sequencing rule joins these annotations, giving the body precondition $U^{\dagger}X_sU$. At the guard, the exit branch returns $F_s$, while the continuing branch contributes the factor $s/(s+1)$ and uses the body precondition at scale $s+1$. The loop rule therefore gives $Y\triangleq \wp(\mathsf{Walk},F)$ satisfying
\begin{equation}
\label{walk:eq-full-space-wp-recurrence}
 Y_s=
 \underbrace{\Pi_0F_s\Pi_0}_{\text{exit}}
 +\underbrace{\frac{s}{s+1}\Pi_{\ne0}U^\dagger Y_{s+1}U\Pi_{\ne0}}_{\text{continue}}.
\end{equation}
Theorem~\ref{thm:loop-certificates} makes this the unique solution in $\Pred{\cH}$.

Equation~\eqref{walk:eq-full-space-wp-recurrence} gives $\Pi_0Y_s\Pi_0=\Pi_0F_s\Pi_0$ and zero cross terms between the boundary and its complement. Since $Uu=Bu+Ku$ for $u\in\cH_+$, the compression $Y_s^+\triangleq \Pi_+Y_s\big|_{\cH_+}$ satisfies
\begin{equation}
\label{walk:eq-compressed-general-recurrence}
\begin{aligned}
 Y_s^+
 &=\frac{s}{s+1}\bigl(B^\dagger F_{s+1}B+K^\dagger Y_{s+1}^+K\bigr)\\
 &=\frac{s}{s+1}\Bigl(
   \Tr\!\left(\ketbra{L,0}{L,0}F_{s+1}\right)B^\dagger B
   +K^\dagger Y_{s+1}^+K\Bigr),
\end{aligned}
\end{equation}
where the second equality uses $\operatorname{ran}B=\CC\ket{L,0}$. For $F=\mathbf I$, write $H_s\triangleq \Pi_+\wp(\mathsf{Walk},\mathbf I)_s\big|_{\cH_+}$. The recurrence becomes
\begin{equation}
\label{eq:fixed-step-functional}
 H_s=\frac{s}{s+1}\bigl(B^\dagger B+K^\dagger H_{s+1}K\bigr).
\end{equation}
Unfolding these equations gives the operator series used in the scalar calculations below.

\begin{proposition}[Runtime family from loop unfolding]
\label{walk:prop-full-space-certificate}
The runtime family on $\cH_+$ is
\begin{equation}
\label{walk:eq-full-candidate}
 H_s=\sum_{n\ge1}\frac{s}{s+n}(K^\dagger)^{n-1}B^\dagger BK^{n-1}.
\end{equation}
More generally, for any postcondition $F\in\Pred{\cH}$,
\begin{equation}
\label{walk:eq-general-postcondition-series}
 \Pi_+\wp(\mathsf{Walk},F)_s\big|_{\cH_+}
 =\sum_{n\ge1}\frac{s}{s+n}
       \Tr\!\left(\ketbra{L,0}{L,0}F_{s+n}\right)
       (K^\dagger)^{n-1}B^\dagger BK^{n-1}.
\end{equation}
Both series converge in operator norm at each fixed $s>0$.
\end{proposition}

\begin{proof}
Unfolding~\eqref{walk:eq-compressed-general-recurrence} $N$ times and canceling successive scale factors gives
\[
\begin{aligned}
 Y_s^+={}&\sum_{n=1}^N\frac{s}{s+n}
       \Tr\!\left(\ketbra{L,0}{L,0}F_{s+n}\right)
       (K^\dagger)^{n-1}B^\dagger BK^{n-1}\\
 &+\frac{s}{s+N}(K^\dagger)^NY_{s+N}^+K^N.
\end{aligned}
\]
Since $0\le Y_{s+N}^+\le I$ and $K$ is a contraction, the remainder satisfies
\[
 0\le\frac{s}{s+N}(K^\dagger)^NY_{s+N}^+K^N
 \le\frac{s}{s+N}I.
\]
At each fixed $s>0$, letting $N\to\infty$ proves~\eqref{walk:eq-general-postcondition-series} in operator norm. Taking $F=\mathbf I$ gives~\eqref{walk:eq-full-candidate}. The same derivation applies to the biased walks in Section~\ref{walk:subsec-biased}.
\end{proof}

\subsection{The Hadamard Walk}
\label{walk:subsec-Hadamard}

For the input $\rho=\ketbra{R,1}{R,1}$, Proposition~\ref{walk:prop-full-space-certificate} reduces the runtime test~\eqref{walk:eq-scalar-runtime-family} to
\begin{equation}
\label{walk:eq-Hadamard-wp-target}
 h_s=\Tr(H_s\rho)
 =\sum_{n\ge1}\frac{s}{s+n}\norm{BK^{n-1}\ket{R,1}}^2.
\end{equation}
We evaluate this expression from the matrix entries of $B$ and $K$, then apply the extraction theorems of Section~\ref{moment:section} to the resulting asymptotic expansion.

\begin{lemma}[Matrix coefficients of $BK^{n-1}$]
\label{walk:lem-Hadamard-coefficients}
Let $K$ have the matrix entries
in~\eqref{walk:eq-K-on-L}--\eqref{walk:eq-K-on-R}, and let
\[
 B=\frac1{\sqrt2}\ket{L,0}\bigl(\bra{L,1}+\bra{R,1}\bigr).
\]
For $n\ge1$, write
\[
 BK^{n-1}\ket{R,1}=a_n\ket{L,0},
 \qquad
 \norm{BK^{n-1}\ket{R,1}}^2=|a_n|^2.
\]
The coefficients $a_n$ are given by
\[
 a_1=\frac1{\sqrt2},\qquad
 a_{4j-1}=-\frac1{\sqrt2}\binom{1/2}{j}\quad(j\ge1),
 \qquad a_n=0\quad\text{otherwise}.
\]
\end{lemma}

\begin{proof}
Expand the matrix product $BK^{n-1}$ in the standard basis. Its nonzero terms are indexed by integer sequences $x_0=1,x_n=0$, with $x_i\ge1$ for $0<i<n$ and $x_i-x_{i-1}\in\{-1,1\}$. Each difference determines the next basis label: $-1$ gives $L$ and $+1$ gives $R$. Thus even $n$ give zero, and $a_1=2^{-1/2}$. For $n=2m+1\ge3$, deleting the final difference $-1$ leaves a word with $m$ entries $+1$ and $m$ entries $-1$, all of whose partial sums are nonnegative.

Group these words by their number $r$ of positive runs, writing each as
\[
 (+1)^{\alpha_1}(-1)^{\beta_1}\cdots
 (+1)^{\alpha_r}(-1)^{\beta_r},
 \qquad
 \sum_{i=1}^r\alpha_i=\sum_{i=1}^r\beta_i=m,
\]
where all run lengths are positive integers. Set $p_i\triangleq \sum_{\ell=1}^i\alpha_\ell$ and $q_i\triangleq \sum_{\ell=1}^i\beta_\ell$ for $1\le i<r$. These define $(r-1)$-element subsets $P=\{p_i\}$ and $Q=\{q_i\}$ of $\{1,\ldots,m-1\}$. Nonnegativity of the word is equivalent to $p_i\ge q_i$ for every $i$, or, equivalently, $|P\cap\{1,\ldots,t\}|\le|Q\cap\{1,\ldots,t\}|$ for every $t$. There are $\binom{m-1}{r-1}^2$ unrestricted pairs. For a violating pair, exchange the two prefixes at the first $t$ where the first count exceeds the second by one. The resulting subsets have sizes $r-2$ and $r$. Conversely, exchanging prefixes at the first excess of the second count over the first recovers the violating pair. Hence the number of words with $r$ positive runs is
\[
 N(m,r)=\binom{m-1}{r-1}^{\!2}
       -\binom{m-1}{r-2}\binom{m-1}{r}
       =\frac1m\binom mr\binom m{r-1},
\]
with out-of-range binomial coefficients interpreted as zero.

Every nonzero matrix entry has modulus $2^{-1/2}$. Negative entries occur exactly when an $R$ label is followed by another $R$ label. A word with $r$ positive runs therefore contributes $m-r$ negative entries within those runs and one from the initial label $R$, giving sign $(-1)^{m-r+1}$. Consequently,
\[
\begin{aligned}
 a_{2m+1}
 &=\frac{2^{-(2m+1)/2}}m
   \sum_{r=1}^m(-1)^{m-r+1}\binom mr\binom m{r-1}\\
 &=\frac{2^{-(2m+1)/2}}m
   [z^{m+1}](1+z)^m(1-z)^m
  =\frac{2^{-(2m+1)/2}}m[z^{m+1}](1-z^2)^m.
\end{aligned}
\]
Here $[z^\ell]$ denotes coefficient extraction. This coefficient vanishes for even $m$. For $m=2j-1$, its value divided by $m$ is
\[
 \frac{(-1)^j}{2j-1}\binom{2j-1}{j}
 =\frac{(-1)^j}{j}\binom{2j-2}{j-1}.
\]
Together with
\[
 \binom{1/2}{j}
 =\frac{(-1)^{j-1}}{2^{2j-1}j}\binom{2j-2}{j-1},
\]
this proves the stated coefficients.
\end{proof}

Since $BK^{n-1}\ket{R,1}=a_n\ket{L,0}$, the scalar series is now explicit:
\begin{equation}
\label{walk:eq-Hadamard-exact-family}
 h_s=\sum_{n\ge1}\frac{s}{s+n}|a_n|^2
 =\frac12\frac{s}{s+1}
       +\frac12\sum_{j\ge1}\binom{1/2}{j}^{\!2}\frac{s}{s+4j-1}.
\end{equation}
Two finite summation identities determine its constant and $s^{-1}$ terms. The coefficient asymptotics determine the logarithmic remainder.

\begin{theorem}[Asymptotic expansion of the rational series]
\label{walk:thm-Hadamard-expansion}
The function $h_s$ in~\eqref{walk:eq-Hadamard-exact-family} satisfies, as $s\to\infty$,
\begin{equation}
\label{walk:eq-Hadamard-runtime-expansion}
 h_s=\frac2\pi-\frac1s+\frac{2\log s}{\pi s^2}+O(s^{-2}).
\end{equation}
\end{theorem}

\begin{proof}
Put $c_j=\binom{1/2}{j}^2$ and $u_m=4^{-m}\binom{2m}{m}$. For $m\ge1$, the identities $c_m=u_m^2/(2m-1)^2$ and $u_m/u_{m-1}=(2m-1)/(2m)$ give
\[
\begin{aligned}
 (4m+1)u_m^2-(4m-3)u_{m-1}^2&=c_m,\\
 m u_m^2-(m-1)u_{m-1}^2&=m c_m.
\end{aligned}
\]
Summing these differences and using $c_0=u_0=1$ yields
\[
 \sum_{j=0}^m c_j=(4m+1)u_m^2,\qquad
 \sum_{j=1}^m j c_j=m u_m^2.
\]
Since Stirling's formula gives $u_m^2\sim1/(\pi m)$, we obtain $\sum_{j\ge0}c_j=4/\pi$ and $\sum_{j\ge1}jc_j=1/\pi$. Thus
\[
\begin{aligned}
 \sum_{n\ge1}|a_n|^2
 &=\frac12+\frac12\left(\frac4\pi-1\right)=\frac2\pi,\\
 \sum_{n\ge1}n|a_n|^2
 &=\frac12+\frac12\left[\frac4\pi-\left(\frac4\pi-1\right)\right]=1.
\end{aligned}
\]

The geometric remainder identity used in~\eqref{moment:remainder-expansion} now gives
\[
\begin{aligned}
 s^2\left(h_s-\frac2\pi+\frac1s\right)
 &=\sum_{n\ge1}\frac{sn^2}{s+n}|a_n|^2\\
 &=\frac12\frac{s}{s+1}
   +\frac12\sum_{j\ge1}\frac{s(4j-1)^2}{s+4j-1}c_j.
\end{aligned}
\]
Stirling's formula also gives
\[
 c_j=\frac1{4\pi j^3}+O(j^{-4}),\qquad
 \frac12(4j-1)^2c_j=\frac2{\pi j}+O(j^{-2}).
\]
The $O(j^{-2})$ error is summable uniformly after multiplication by $s/(s+4j-1)\le1$, so
\[
 s^2\left(h_s-\frac2\pi+\frac1s\right)
 =\frac2\pi\sum_{j\ge1}\frac{s}{j(s+4j-1)}+O(1).
\]
To estimate the last sum, split it at $J=\lfloor s\rfloor$. For $j\le J$,
\[
 0\le\frac1j-\frac{s}{j(s+4j-1)}
 =\frac{4j-1}{j(s+4j-1)}\le\frac4s,
\]
so this part equals $\sum_{j=1}^Jj^{-1}+O(1)=\log s+O(1)$. For $j>J$, the bound $s/[j(s+4j-1)]\le s/(3j^2)$ makes the remaining sum $O(1)$. This proves the expansion.
\end{proof}

\paragraph{Extracting runtime properties.}
The constant term in~\eqref{walk:eq-Hadamard-runtime-expansion} gives the termination probability by Theorem~\ref{thm:asymptotic-probability}:
\[
 \Prob(\tau<\infty)=\frac2\pi.
\]
Since $h_s=2/\pi-s^{-1}+o(s^{-1})$, Theorem~\ref{moment:extraction} with $k=1$ reads the successful first moment from the coefficient of $-s^{-1}$:
\[
 \E[\tau;\tau<\infty]=1.
\]
For the second moment, subtracting these two terms leaves
\[
 s^2\left(h_s-\frac2\pi+\frac1s\right)
 =\frac2\pi\log s+O(1).
\]
The logarithmic case of Theorem~\ref{thm:moment-readout}, with $k=2$, therefore gives
\begin{equation}
\label{walk:eq-Hadamard-readouts}
 \E[\min(\tau,N)^2;\tau<\infty]=\frac2\pi\log N+O(1).
\end{equation}
It also gives $\Prob(N<\tau<\infty)=\Theta(N^{-2})$ and critical moment order $2$. By the conditioning formula~\eqref{moment:conditioning}, the conditional moments are finite exactly for $0<r<2$, and their capped versions grow as $\Theta(N^{r-2})$ for $r>2$.

Dividing the successful first moment and the capped second moment by $2/\pi$ gives the conditional mean $\pi/2$ and conditional capped second moment $\log N+O(1)$. Since the conditional capped mean is at most $\pi/2$, its squared contribution to the variance is bounded. Hence
\begin{equation}
\label{walk:eq-Hadamard-capped-variance}
 \E[\tau\mid\tau<\infty]=\frac\pi2,\qquad
 \Var(\min(\tau,N)\mid\tau<\infty)=\log N+O(1).
\end{equation}
The positive divergence probability $1-2/\pi$ gives $\E[\tau]=\infty$.

\subsection{Extension to Biased Coins}
\label{walk:subsec-biased}

Replace the Hadamard coin by
\begin{equation}
\label{walk:eq-general-coin}
 C_p=\begin{pmatrix}a&b\\b&-a\end{pmatrix},\qquad
 a=\sqrt p,\quad b=\sqrt{1-p},\quad p\in(0,1).
\end{equation}
Let $L_p$ be the resulting loop, and put
\[
 U_p\triangleq V(C_p\otimes I),\qquad
 B_p\triangleq \Pi_0U_p\big|_{\cH_+},\qquad
 K_p\triangleq \Pi_+U_p\big|_{\cH_+}.
\]
The derivation in Section~\ref{walk:subsec-first-crossing} applies with $U,B,K$ replaced by $U_p,B_p,K_p$. Thus the compression $H_{p,s}$ of $\wp(L_p,\mathbf I)_s$ to $\cH_+$ satisfies
\[
 H_{p,s}=\frac{s}{s+1}
       \bigl(B_p^\dagger B_p+K_p^\dagger H_{p,s+1}K_p\bigr),
\]
and Proposition~\ref{walk:prop-full-space-certificate} gives its operator series. For a normalized coin state $\rho$ at position $1$, the scalar calculation target is
\begin{equation}
\label{walk:eq-biased-wp-target}
\begin{aligned}
 h_\rho(s)&\triangleq \Tr\bigl((\rho\otimes\ketbra{1}{1})H_{p,s}\bigr)\\
 &=\sum_{n\ge1}\frac{s}{s+n}
       \Tr\bigl((\rho\otimes\ketbra{1}{1})
          (K_p^\dagger)^{n-1}B_p^\dagger B_pK_p^{n-1}\bigr).
\end{aligned}
\end{equation}
We evaluate this trace using the same matrix expansion as in Section~\ref{walk:subsec-Hadamard}. For the calculation, write
\[
 v_+=a\ket L+b\ket R,\quad v_-=b\ket L-a\ket R,\quad
 \chi=1-2p,\quad\delta=2\sqrt{p(1-p)},
\]
so that $v_+,v_-$ are orthonormal and $\delta^2=1-\chi^2$.

\begin{lemma}[Matrix coefficients of powers of $K_p$]
\label{walk:lem-biased-coefficients}
For every $\psi\in\CC^2$ and $n\ge1$,
\[
 \bra{L,0}B_pK_p^{n-1}\ket{\psi,1}
 =\begin{cases}
   \ip{v_+}{\psi}, & n=1,\\
   t_k\ip{v_-}{\psi}, & n=2k+1,\ k\ge1,\\
   0, & n\text{ even},
  \end{cases}
\]
where, for the Legendre polynomials $\mathsf P_k$,
\begin{equation}
\label{walk:eq-biased-coefficients}
 t_k=\frac{\delta\,\mathsf P'_k(\chi)}{k(k+1)}
    =\frac{\mathsf P_{k-1}(\chi)-\mathsf P_{k+1}(\chi)}{\delta(2k+1)}.
\end{equation}
\end{lemma}

\begin{proof}
The identities $C_pv_+=\ket L$ and $C_pv_-=\ket R$ give
\[
\begin{aligned}
 B_p\ket{v_+,1}&=\ket{L,0}, & K_p\ket{v_+,1}&=0,\\
 B_p\ket{v_-,1}&=0, & K_p\ket{v_-,1}&=\ket{R,2}.
\end{aligned}
\]
By linearity, it remains to compute the coefficients with initial vector $\ket{v_-,1}$. Parity again excludes even $n$. For $n=2k+1$ with $k\ge1$, use the words counted by $N(k,r)$ in the proof of Lemma~\ref{walk:lem-Hadamard-coefficients}. The first $+1$ has coefficient $1$. For a word with $r$ positive runs, the remaining factors contain $k-r$ transitions $R\to R$, $k+1-r$ transitions $L\to L$ (including the final $B_p$ factor), and $2r-1$ changes of label. Their entries are respectively $-a,a,b$, giving the product $(-a)^{k-r}a^{k+1-r}b^{2r-1}$. Summing over $r$ therefore gives
\[
 t_k=\frac{ab}{k}\sum_{r=1}^k\binom kr\binom k{r-1}
                  (-p)^{k-r}(1-p)^{r-1}.
\]
Rodrigues' formula identifies this finite sum with $\delta\mathsf P'_k(\chi)/(k(k+1))$. The identity
$(1-\chi^2)\mathsf P'_k(\chi)=k(k+1)(\mathsf P_{k-1}(\chi)-\mathsf P_{k+1}(\chi))/(2k+1)$
gives the second expression in~\eqref{walk:eq-biased-coefficients}.
\end{proof}

For a positive $2\times2$ matrix $\rho$ of trace one, put 
\[\beta_\rho\triangleq \bra{v_-}\rho\ket{v_-}\in[0,1].\]
The lemma gives quadratic forms proportional to $\ketbra{v_+}{v_+}$ at $n=1$ and to $\ketbra{v_-}{v_-}$ at $n=2k+1$. Taking their traces against $\rho$ evaluates~\eqref{walk:eq-biased-wp-target} as
\begin{equation}
\label{walk:eq-entrance-exact-family}
 h_\rho(s)=(1-\beta_\rho)\frac{s}{s+1}
             +\beta_\rho\sum_{k\ge1}\frac{s}{s+2k+1}t_k^2.
\end{equation}
As in Section~\ref{walk:subsec-Hadamard}, the expansion follows from two coefficient sums and a tail estimate. Here the Legendre representation makes the two sums accessible through Parseval's identity.

\begin{theorem}[Asymptotic expansion of the parameterized series]
\label{walk:thm-entrance-expansion}
For fixed $p\in(0,1)$ and every positive $2\times2$ matrix $\rho$ of trace one, the scalar series~\eqref{walk:eq-entrance-exact-family} satisfies
\begin{equation}
\label{walk:eq-biased-resolvent-expansion}
 h_\rho(s)=q_\rho-\frac1s
              +\frac{\lambda_\rho\log s}{s^2}+O(s^{-2})
 \qquad(s\to\infty),
\end{equation}
where $q_\rho\triangleq 1-(1-\kappa_p)\beta_\rho$, $\lambda_\rho\triangleq 4\beta_\rho/(\pi\delta)$, and
\begin{equation}
\label{walk:eq-kappa-closed-form}
 \kappa_p=\frac{\frac4\pi(\delta+\chi\arcsin\chi)-1-\chi^2}{\delta^2}
 \in(0,1).
\end{equation}
\end{theorem}

\begin{proof}
To compute the coefficient sums, set $u_k\triangleq \int_\chi^1\mathsf P_k(x)\dd x$. Then $u_0=1-\chi$ and $u_k=\delta t_k$ for $k\ge1$. The squared norm of $\mathsf P_k$ in $L^2(-1,1)$ is $2/(2k+1)$, so Parseval's identity for $\one_{[\chi,1]}$ gives
\[
 \sum_{k\ge0}(2k+1)u_k^2=2(1-\chi).
\]
Subtracting the $k=0$ term yields
\[
 \sum_{k\ge1}(2k+1)t_k^2
 =\frac{2(1-\chi)-(1-\chi)^2}{\delta^2}=1.
\]

For the unweighted sum, put $\phi\triangleq \arccos\chi\in(0,\pi)$, so $\sin\phi=\delta$. Integrating the Mehler--Dirichlet formula~\cite[Eq.~(4.8.6)]{Szego1975} over $x\in[\chi,1]$ and interchanging the absolutely integrable terms gives
\[
 u_k=\frac{2\sqrt2}{\pi}\int_0^\phi
       \sqrt{\cos\theta-\chi}\,
       \cos\bigl((k+\tfrac12)\theta\bigr)\dd\theta.
\]
The functions $\cos((k+\tfrac12)\theta)$ form a complete orthogonal system in $L^2(0,\pi)$ with squared norm $\pi/2$. Applying Parseval to $\sqrt{\cos\theta-\chi}$ on $[0,\phi]$, extended by zero, gives
\[
 \sum_{k\ge0}u_k^2
 =\frac4\pi\int_0^\phi(\cos\theta-\chi)\dd\theta
 =\frac4\pi(\delta-\chi\phi).
\]
Subtracting $u_0^2$ and using $\phi=\pi/2-\arcsin\chi$ yields $\sum_{k\ge1}t_k^2=\kappa_p$. Since $t_1=ab>0$ and $3\sum_{k\ge1}t_k^2\le\sum_{k\ge1}(2k+1)t_k^2=1$, we have $0<\kappa_p<1$.

The two sums determine the constant and $s^{-1}$ coefficients in~\eqref{walk:eq-entrance-exact-family}. Applying the geometric remainder identity from~\eqref{moment:remainder-expansion}, as in the proof of Theorem~\ref{walk:thm-Hadamard-expansion}, gives
\[
\begin{aligned}
 s^2\bigl(h_\rho(s)-q_\rho+s^{-1}\bigr)
 &=(1-\beta_\rho)\frac{s}{s+1}
   +\beta_\rho\sum_{k\ge1}\frac{s(2k+1)^2}{s+2k+1}t_k^2.
\end{aligned}
\]
To estimate this remainder, apply Darboux's formula~\cite[Theorem~8.21.8]{Szego1975} to $\mathsf P_{k-1}(\chi)-\mathsf P_{k+1}(\chi)$ in~\eqref{walk:eq-biased-coefficients}. Squaring the resulting leading term gives
\[
 t_k^2=\frac{1-\sin((2k+1)\phi)}{\pi\delta k^3}+O(k^{-4}).
\]
Consequently,
\[
 \sum_{k\ge1}\frac{s(2k+1)^2}{s+2k+1}t_k^2
 =\frac4{\pi\delta}\sum_{k\ge1}
    \frac{s\bigl(1-\sin((2k+1)\phi)\bigr)}{k(s+2k+1)}+O(1).
\]
The nonoscillating sum is $\log s+O(1)$ by the same split at $\lfloor s\rfloor$ used in Theorem~\ref{walk:thm-Hadamard-expansion}. For the oscillating sum, the partial sums of $\sin((2k+1)\phi)$ are bounded, while $s/[k(s+2k+1)]$ decreases to zero and has first term at most $1$. Summation by parts therefore bounds it uniformly in $s$. The remainder is thus $\lambda_\rho\log s+O(1)$, proving~\eqref{walk:eq-biased-resolvent-expansion}.
\end{proof}

\paragraph{Extracting runtime properties.}
Let $\tau$ be the runtime of $L_p$ on input $\rho\otimes\ketbra{1}{1}$. Theorem~\ref{thm:asymptotic-probability} reads the termination probability $q_\rho$ from the constant term in~\eqref{walk:eq-biased-resolvent-expansion}. Since $h_\rho(s)=q_\rho-s^{-1}+o(s^{-1})$, Theorem~\ref{moment:extraction} with $k=1$ gives $\E_\rho[\tau;\tau<\infty]=1$. As $q_\rho\ge\kappa_p>0$, conditioning gives
\begin{equation}
\label{walk:eq-successful-mean-identity}
 \Prob_\rho(\tau<\infty)=q_\rho,\qquad
 \E_\rho[\tau\mid\tau<\infty]=\frac1{q_\rho}.
\end{equation}

If $\beta_\rho>0$, then $\lambda_\rho>0$, so the logarithmic case of Theorem~\ref{thm:moment-readout} with $k=2$ applies to the remainder in~\eqref{walk:eq-biased-resolvent-expansion}. It gives
\begin{equation}
\label{walk:eq-biased-capped-second}
 \E_\rho[\min(\tau,N)^2;\tau<\infty]
 =\lambda_\rho\log N+O(1),
\end{equation}
and the conditional $r$-th moment is finite exactly for $0<r<2$. By~\eqref{moment:conditioning}, the conditional capped second moment has leading term $(\lambda_\rho/q_\rho)\log N$. As in Section~\ref{walk:subsec-Hadamard}, the conditional capped mean is bounded by $1/q_\rho$, so its square contributes only $O(1)$ to the variance. Hence
\begin{equation}
\label{walk:eq-biased-capped-variance}
 \Var_\rho(\min(\tau,N)\mid\tau<\infty)
 =\frac{4\beta_\rho}{\pi\delta q_\rho}\log N+O(1).
\end{equation}
The divergence probability $(1-\kappa_p)\beta_\rho$ is positive, giving $\E_\rho[\tau]=\infty$. If $\beta_\rho=0$, Lemma~\ref{walk:lem-biased-coefficients} gives probability one at runtime $1$, so $\tau=1$ almost surely.
\section{Compositional Verification of a Nested Loop}
\label{sec:nested}

To illustrate how our calculus composes across nested loops without re-analyzing inner loop operators, we verify the termination probability and conditional expected runtime of a probabilistic outer loop $L$ that repeatedly invokes the absorbing Hadamard walk of Section~\ref{walk:sec-cost}.

Use the coin and position registers of Section~\ref{walk:sec-cost}, with identities $I_c$, $I_x$, and $I=I_c\otimes I_x$. The inner subprogram $S$ resets the registers to $\ket{R,1}$ and runs the Hadamard walk loop $\mathsf{Walk}$ of Section~\ref{walk:subsec-first-crossing}, while the outer loop $L$ guards each invocation of $S$ with a fair coin measurement $M^{\mathrm{out}}_0=M^{\mathrm{out}}_1=I/\sqrt2$:
\[
\begin{aligned}
 S &\equiv \mathtt{coin}:=\ket R;\ \mathtt{position}:=\ket1;\ \mathsf{Walk},\\
 L &\equiv \mathbf{while}\ M^{\mathrm{out}}[\mathtt{coin,position}]=1\ \mathbf{do}\ S\ \mathbf{od}.
\end{aligned}
\]
At each outer guard, outcome $0$ exits and outcome $1$ executes $S$, each with probability $1/2$. One continuing outer test adds one unit of runtime to that of $S$; if an invocation of $S$ does not terminate, $L$ also diverges.

\paragraph{Reusing the inner runtime family.}
Set $\sigma\triangleq \ketbra{R,1}{R,1}$. By the sequencing rule~\eqref{eq:transformer-sequence} and the state-preparation rule of Section~\ref{subsec:wp-rules}, pulling any postcondition $F\in\Pred{\cH}$ back through $S$ gives
\begin{equation}
\label{eq:nested-general-inner-contract}
 \wp(S,F)_s
 =\wp\bigl(\mathtt{coin}:=\ket R;\,\mathtt{position}:=\ket1,\,
           \wp(\mathsf{Walk},F)\bigr)_s
 =\Tr\!\left(\sigma\,\wp(\mathsf{Walk},F)_s\right)I.
\end{equation}
Specializing~\eqref{eq:nested-general-inner-contract} to $F=\mathbf I$ directly reuses the scalar runtime test $h_s=\Tr\bigl(\sigma\,\wp(\mathsf{Walk},\mathbf I)_s\bigr)$ expanded in Theorem~\ref{walk:thm-Hadamard-expansion}:
\begin{equation}
\label{eq:nested-inner-data}
 H_s^S = \wp(S,\mathbf I)_s = h_s I,
 \qquad
 h_s = q - \frac1s + o(s^{-1})
 \quad (s\to\infty),
 \qquad
 q = \frac2\pi.
\end{equation}

\paragraph{Outer loop equation and rational candidate.}
Applying the loop rule~\eqref{eq:general-loop-functional} of Theorem~\ref{thm:semantics} to $L$ characterizes $H^L\triangleq \wp(L,\mathbf I)=\operatorname{lfp}(\cT_{\mathbf I})$ via
\begin{equation}
\label{eq:nested-loop-equation}
 (\cT_{\mathbf I}X)_s
 =(M^{\mathrm{out}}_0)^\dagger I M^{\mathrm{out}}_0
  +\frac{s}{s+1}(M^{\mathrm{out}}_1)^\dagger\wp(S,X)_{s+1}M^{\mathrm{out}}_1
 =\frac12I+\frac{s}{2(s+1)}\wp(S,X)_{s+1}.
\end{equation}
To capture the first-order asymptotics of $H_s^L$ while shifting cleanly under $\wp(S,\cdot)$, we test $\cT_{\mathbf I}$ on a bounded rational candidate $A\in\Pred{\cH}$ of the form
\begin{equation}
\label{eq:nested-reused-candidate}
 A_s \triangleq a\frac{s}{s+b}I
      = \left(a-\frac{ab}{s}+O(s^{-2})\right)I
 \qquad (a\in(0,1],\ b>0).
\end{equation}
By the grade representation~\eqref{eq:runtime-transformer} and the scale-shift identity~\eqref{eq:scale-identity}, $\wp(S,A)_s$ reduces to $H^S$ at the shifted scale $s+b$:
\[
\begin{aligned}
 \wp(S,A)_s
 &=\sum_{n\ge0}\frac{s}{s+n}\sem{S}_n^\dagger\!
      \left(a\frac{s+n}{s+b+n}I\right)\\
 &=a\frac{s}{s+b}
      \sum_{n\ge0}\frac{s+b}{s+b+n}E_n^S\\
 &=a\frac{s}{s+b}H_{s+b}^S
  =a\frac{s}{s+b}h_{s+b}I.
\end{aligned}
\]
Substituting $\wp(S,A)_{s+1}$ into~\eqref{eq:nested-loop-equation} and canceling $s+1$ via~\eqref{eq:scale-identity} yields
\begin{equation}
\label{eq:nested-candidate-image}
 (\cT_{\mathbf I}A)_s
 =\left[\frac12+\frac a2\frac{s}{s+1+b}h_{s+1+b}\right]I.
\end{equation}

\begin{theorem}[Termination probability and conditional mean runtime of $L$]
\label{thm:nested-expansion}
Let $q=2/\pi$, and set
\[
 a\triangleq \frac1{2-q}=\frac{\pi}{2(\pi-1)},
 \qquad
 b\triangleq \frac{1+q}{2-q}=\frac{\pi+2}{2(\pi-1)}.
\]
Then the outer runtime family $H_s^L=\wp(L,\mathbf I)_s$ satisfies
\begin{equation}
\label{eq:nested-certified-expansion}
 \norm{H_s^L-A_s} = o(s^{-1}) \qquad (s\to\infty),
\end{equation}
and for every normalized input state $\rho$,
\begin{equation}
\label{eq:nested-conditional-readouts}
\begin{aligned}
 \Prob_\rho(\tau_L<\infty) &= a = \frac1{2-q},\\
 \E_\rho[\tau_L;\tau_L<\infty] &= ab = \frac{1+q}{(2-q)^2},
 \qquad
 \E_\rho[\tau_L\mid\tau_L<\infty] = b = \frac{1+q}{2-q}.
\end{aligned}
\end{equation}
\end{theorem}

\begin{proof}
Inserting $h_{s+1+b} = q - s^{-1} + o(s^{-1})$ from~\eqref{eq:nested-inner-data} and $\frac{s}{s+1+b} = 1 - \frac{1+b}{s} + O(s^{-2})$ into~\eqref{eq:nested-candidate-image} gives
\[
 (\cT_{\mathbf I}A)_s
 =\left[\frac12+\frac{aq}{2}
   -\frac{a\bigl(q(1+b)+1\bigr)}{2s}
   +o(s^{-1})\right]I.
\]
The choices $a=(2-q)^{-1}$ and $b=(1+q)(2-q)^{-1}$ satisfy $\frac12+\frac{aq}{2}=a$ and $\frac12\bigl(q(1+b)+1\bigr)=b$, matching~\eqref{eq:nested-reused-candidate} to order $s^{-1}$ and leaving
\[
 \norm{A_s-(\cT_{\mathbf I}A)_s} = o(s^{-1}) \qquad (s\to\infty).
\]
Since $H_{u+1}^S\le I$, we have
\[
 \frac{u}{u+1}(M^{\mathrm{out}}_1)^\dagger H_{u+1}^S M^{\mathrm{out}}_1
 =\frac{u}{2(u+1)}H_{u+1}^S \le \frac12 I \qquad (u>0),
\]
so Theorem~\ref{thm:residual} with $\kappa=1/2<1$ gives~\eqref{eq:nested-certified-expansion}.

Consequently, for any normalized input $\rho$,~\eqref{eq:nested-reused-candidate} and~\eqref{eq:nested-certified-expansion} yield
\[
 \Tr(\rho H_s^L) = \Tr(\rho A_s) + o(s^{-1})
 = a - \frac{ab}{s} + o(s^{-1}) \qquad (s\to\infty).
\]
Theorems~\ref{thm:asymptotic-probability} and~\ref{moment:extraction} ($k=1$) extract $\Prob_\rho(\tau_L<\infty)=a$ and $\E_\rho[\tau_L;\tau_L<\infty]=ab$, and dividing by $a$ gives $\E_\rho[\tau_L\mid\tau_L<\infty]=b$.
\end{proof}
\section{Reusing Total-Cost Proofs}
\label{bridge:section}

The cost logic of~\cite{YuEtAl2026} proves upper bounds on expected execution costs and terminal observables using closed positive quadratic forms. Such proofs can also supply contracts for the parameterized calculus developed here. We first recall total-cost Hoare triples and their proof rules under our loop-iteration cost convention, then show how normalized resolvents translate these contracts into valid total-correctness Hoare triples on bounded effect families.

\subsection{Total-Cost Logic with Closed Positive Forms}
\label{bridge:subsec-cost-logic}

A cost predicate is a closed positive quadratic form $P:\cH\to[0,\infty]$, with finite domain $\Dom(P)=\{u:P[u]<\infty\}$. The domain may be non-dense, allowing the predicate to impose an infinite penalty outside a prescribed subspace. Cost predicates are ordered pointwise:
\[
 P\le G\quad\Longleftrightarrow\quad P[u]\le G[u]\quad(u\in\cH).
\]
Increasing sequences have pointwise suprema that are again closed positive forms. This order structure supports the loop constructions in the cost logic~\cite[Sections~4--5]{YuEtAl2026}.

To define expectations, let $\cH_P\triangleq \overline{\Dom(P)}$ and let $T_P$ be the positive self-adjoint operator on $\cH_P$ representing $P$, so that $P[u]=\norm{T_P^{1/2}u}^2$ on $\Dom(P)=\Dom(T_P^{1/2})$. Its bounded spectral truncations are
\[
 P^{[N]}\triangleq \min(T_P,NI_{\cH_P})\oplus NI_{\cH_P^\perp}
 \qquad(N\ge1),
\]
where the first term is defined by spectral calculus. They increase to $P$ as quadratic forms, and the extended trace is
\[
 \Tr(P\rho)\triangleq \sup_{N\ge1}\Tr(P^{[N]}\rho)\in[0,\infty].
\]
For a CPTNI map $\mathcal E$, its Heisenberg adjoint extends to cost forms by
\begin{equation}
\label{bridge:eq:form-pullback}
 \widehat{\mathcal E^\dagger}(G)\triangleq
 \sup_{N\ge1}\mathcal E^\dagger(G^{[N]}),
 \qquad
 \Tr\!\left(\widehat{\mathcal E^\dagger}(G)\rho\right)
 =\Tr\bigl(G\mathcal E(\rho)\bigr).
\end{equation}
The supremum is taken in the order of closed positive forms; the trace identity follows by monotone convergence~\cite[Theorem~4.1]{YuEtAl2026}. In particular, $K^\dagger GK$ denotes the form $u\mapsto G[Ku]$ for a bounded operator $K$.

We use the cost convention of Section~\ref{sec:runtime-semantics}: each entry into a loop body costs one, terminating loop-free commands and loop exits cost zero, and a nonterminating execution has infinite cost. Thus the cost predicate is $C_S=M_1^S+\infty D_S$ from~\eqref{eq:total-runtime-cost}. For cost predicates $P,G$, the validity of the total-cost Hoare triple $\{P\}\,S\,\{G\}$ is defined by
\begin{equation}
\label{bridge:eq:cost-validity}
 \models_{C,\mathrm{tot}}\{P\}\,S\,\{G\}
 \quad\Longleftrightarrow\quad
 \Tr(P\rho)\ge\Tr\bigl(G\sem{S}(\rho)\bigr)+\Tr(C_S\rho)
 \quad\text{for every state }\rho.
\end{equation}
The precondition $P$ therefore bounds the sum of the execution cost and the terminal cost $G$. Its least valid value is the cost weakest precondition
\begin{equation}
\label{bridge:eq:cost-wp}
 \wp_C(S,G)\triangleq \widehat{\sem{S}^\dagger}(G)+C_S,
 \qquad
 \models_{C,\mathrm{tot}}\{P\}\,S\,\{G\}
 \quad\Longleftrightarrow\quad P\ge\wp_C(S,G).
\end{equation}
This is the cost-transformer construction of~\cite[Proposition~5.2]{YuEtAl2026}, with its command costs replaced by the loop-iteration costs above.

The structural rules pull terminal costs backwards and add the cost of each command. For the basic commands they give
\begin{align*}
 \wp_C(\mathbf{skip},G)&=G,
 &\wp_C(\mathbf{abort},G)&=\infty I,\\
 \wp_C(q:=\ket0,G)&=\sum_j\ketbra{j}{0}_qG\ketbra{0}{j}_q,
 &\wp_C(\bar q:=U[\bar q],G)&=U^\dagger GU.
\end{align*}
Sequential composition uses the precondition of the second program as the postcondition of the first:
\[
 \wp_C(S_1;S_2,G)=\wp_C(S_1,\wp_C(S_2,G)).
\]
For a conditional $S_{\mathrm{if}}=\mathbf{if}\,(\square_{i\in J}\,M[\bar q]=i\to S_i)\,\mathbf{fi}$, the branch costs combine as
\[
 \wp_C(S_{\mathrm{if}},G)=\sum_{i\in J}M_i^\dagger\wp_C(S_i,G)M_i.
\]
All sums and pullbacks here are interpreted as quadratic forms. For a loop $L=\mathbf{while}\ M[\bar q]=1\ \mathbf{do}\ B\ \mathbf{od}$, define
\begin{equation}
\label{bridge:eq:cost-loop-functional}
 \mathcal K_G(X)\triangleq M_0^\dagger GM_0+
 M_1^\dagger\bigl(I+\wp_C(B,X)\bigr)M_1.
\end{equation}
The exit branch returns $G$, while the continuing branch pays one unit before executing the body. Starting from $X_0=0$ and setting $X_{m+1}=\mathcal K_G(X_m)$ gives
\[
 \wp_C(L,G)=\sup_{m\ge0}X_m.
\]
These rules agree with~\eqref{bridge:eq:cost-wp} by structural induction using Definition~\ref{def:cost-grades}: sequencing adds runtimes, and conditionals sum branch costs. For a loop, $X_m$ records the costs of up to $m$ outer iterations, including each reached body's full cost, and $G$ on exits after fewer than $m$ outer iterations. Monotone convergence recovers the terminal expectation plus total expected runtime; a divergent run either enters infinitely many outer iterations or incurs infinite cost inside a reached body.

Consequently, a cost predicate $P$ satisfying $P\ge\mathcal K_G(P)$ proves $\models_{C,\mathrm{tot}}\{P\}\,L\,\{G\}$. Indeed, monotonicity gives $X_m\le P$ by induction. This is the invariant-and-potential rule underlying the expected-runtime analysis in Section~\ref{walk:sec-cost}.

\subsection{Resolvent Translation of Total Correctness}
\label{bridge:subsec-total-reduction}

To turn a cost bound into a bounded guarantee, we apply the normalized resolvent to its pre- and postconditions. The proof uses two order properties: the resolvent reverses cost inequalities, and the operator Jensen inequality compares a resolvent of an expected cost with the expectation of its resolvent. We establish these properties first for bounded operators and then extend them to closed positive forms.

\begin{definition}[Normalized resolvents of cost forms]
\label{bridge:def:resolvent}
For a closed positive form $P$ and $s>0$, define
\[
 \res{s}{P}\triangleq s(sI_{\cH_P}+T_P)^{-1}\oplus0_{\cH_P^\perp},
 \qquad R(P)\triangleq (\res{s}{P})_{s>0}.
\]
Thus $\res{s}{P}$ is an effect on $\cH$, and $R(P)\in\Pred{\cH}$. For bounded $P$ it equals $(I+P/s)^{-1}$; in particular, $\res{s}{0}=I$ and $\res{s}{\infty I}=0$.
\end{definition}

The resolvent has the variational characterization
\begin{equation}
\label{bridge:eq:variational}
 \ip{u}{\res{s}{P}u}
 =\sup_{v\in\Dom(P)}
 \left\{2\operatorname{Re}\ip{u}{v}-\norm{v}^2-\frac{P[v]}s\right\}.
\end{equation}
To see this, write $u=u_0+u_\perp$ according to $\cH=\cH_P\oplus\cH_P^\perp$ and complete the square with $(I+T_P/s)^{1/2}v$. The supremum is attained at $v=(I+T_P/s)^{-1}u_0$ and equals $\ip{u_0}{(I+T_P/s)^{-1}u_0}$. If $P\ge G$, then $\Dom(P)\subseteq\Dom(G)$ and the expression being maximized for $P$ is no larger than that for $G$. Hence
\begin{equation}
\label{bridge:eq:antitonicity}
 P\ge G\quad\Longrightarrow\quad\res{s}{P}\le\res{s}{G}
 \qquad(s>0).
\end{equation}

\begin{definition}[Operator convexity]
A continuous function $f:[0,\infty)\to\RR$ is operator convex if, for all bounded positive operators $A,B$ on any Hilbert space and every $\lambda\in[0,1]$,
\[
 f\bigl(\lambda A+(1-\lambda)B\bigr)
 \le\lambda f(A)+(1-\lambda)f(B).
\]
\end{definition}

\begin{lemma}[Choi--Davis--Jensen inequality]
\label{bridge:lem:jensen}
Let $\mathcal A$ be a unital $C^*$-algebra and let $\Phi:\mathcal A\to\Bop{\cH}$ be a unital completely positive map. If $f:[0,\infty)\to\RR$ is operator convex, then for every positive $X\in\mathcal A$,
\[
 f\bigl(\Phi(X)\bigr)\le\Phi\bigl(f(X)\bigr).
\]
\end{lemma}

\begin{proof}
We give the compression argument for the operator Jensen inequality~\cite{HansenPedersen2003}. By Stinespring's theorem, $\Phi(X)=V^\dagger\pi(X)V$, where $\pi:\mathcal A\to\Bop{\cK}$ is a unital $*$-representation and $V:\cH\to\cK$ is an isometry. Put $E=VV^\dagger$ and $U=2E-I_{\cK}$. Then $U$ is a self-adjoint unitary. For $Y=\pi(X)$, averaging with its conjugate by $U$ removes the off-diagonal blocks relative to $E$:
\[
 Z\triangleq \frac12(Y+UYU^\dagger)
   =EYE+(I-E)Y(I-E).
\]
Operator convexity and unitary invariance of functional calculus give
\[
\begin{aligned}
 f(Z)&\le\frac12\bigl(f(Y)+Uf(Y)U^\dagger\bigr)\\
     &=Ef(Y)E+(I-E)f(Y)(I-E).
\end{aligned}
\]
Compressing by $V$ therefore yields $V^\dagger f(Z)V\le V^\dagger f(Y)V$.

Since $Z$ commutes with $E$ and $V^\dagger EV=I$, its restriction to $E\cK$ is unitarily identified with $V^\dagger ZV=V^\dagger YV$. More explicitly, $(V^\dagger ZV)^m=V^\dagger Z^mV$ for every nonnegative integer $m$, so polynomial approximation gives
\[
 V^\dagger f(Z)V=f(V^\dagger ZV)=f(V^\dagger YV).
\]
Combining the two identities and using $f(\pi(X))=\pi(f(X))$, we obtain
\[
 f(\Phi(X))
 =f(V^\dagger YV)
 \le V^\dagger f(Y)V
 =V^\dagger\pi(f(X))V
 =\Phi(f(X)).
\]
\end{proof}

\begin{proposition}[Operator convexity of the normalized resolvent]
\label{bridge:prop:convexity}
For every $s>0$, the function $r_s(t)=s/(s+t)$ is operator convex on $[0,\infty)$.
\end{proposition}

\begin{proof}
We prove convexity of inversion on bounded positive invertible operators. For such an operator $Z$, the block matrix
\[
 \begin{pmatrix}Z&I\\I&Z^{-1}\end{pmatrix}
\]
is positive, because its quadratic form at $(u,v)$ is $\norm{Z^{1/2}u+Z^{-1/2}v}^2$. Thus, for positive invertible $X,Y$ and $\lambda\in[0,1]$,
\[
 \begin{pmatrix}
  \lambda X+(1-\lambda)Y&I\\
  I&\lambda X^{-1}+(1-\lambda)Y^{-1}
 \end{pmatrix}\ge0.
\]
Taking the Schur complement of the positive invertible upper-left block gives
\[
 \bigl(\lambda X+(1-\lambda)Y\bigr)^{-1}
 \le\lambda X^{-1}+(1-\lambda)Y^{-1}.
\]
For bounded $A,B\ge0$, substitute $X=sI+A$ and $Y=sI+B$, then multiply by $s$. This gives
\[
 \res{s}{\lambda A+(1-\lambda)B}
 \le\lambda\res{s}{A}+(1-\lambda)\res{s}{B},
\]
as required.
\end{proof}

\begin{corollary}[Resolvent Jensen inequality for closed positive forms]
\label{bridge:cor:form-jensen}
Let $P$ be a closed positive form on $\cK$, and let $\Phi:\Bop{\cK}\to\Bop{\cH}$ be a normal unital completely positive map. Define its extension to $P$ by
\[
 \widehat\Phi(P)\triangleq \sup_{N\ge1}\Phi(P^{[N]}).
\]
Then, for every $s>0$,
\begin{equation}
\label{bridge:eq:form-jensen}
 \res{s}{\widehat\Phi(P)}\le\Phi(\res{s}{P}).
\end{equation}
\end{corollary}

\begin{proof}
Each $P^{[N]}$ is bounded and positive, so Lemma~\ref{bridge:lem:jensen} and Proposition~\ref{bridge:prop:convexity} give
\[
 \res{s}{\Phi(P^{[N]})}\le\Phi(\res{s}{P^{[N]}}).
\]
As $N\to\infty$, the quadratic forms of $\Phi(P^{[N]})$ increase monotonically to $\widehat\Phi(P)$. Monotone convergence for closed positive forms implies that the operators $\res{s}{\Phi(P^{[N]})}$ decrease monotonically and converge strongly to $\res{s}{\widehat\Phi(P)}$~\cite{Simon1978,YuEtAl2026}. Spectral calculus likewise shows that $\res{s}{P^{[N]}}$ decreases monotonically and converges strongly to $\res{s}{P}$. Normality of $\Phi$ preserves this bounded decreasing limit, as follows by applying it to the increasing sequence $I-\res{s}{P^{[N]}}$. Taking matrix elements and passing to the limit proves~\eqref{bridge:eq:form-jensen}.
\end{proof}

Total-cost validity accounts for divergence through $C_S$. The next lemma isolates the bound that makes the divergent contribution disappear after applying the resolvent.

\begin{lemma}[Divergence bound]
\label{bridge:lem:divergence}
For every program $S$ and every integer $N\ge1$,
\begin{equation}
\label{bridge:eq:capped-cost-bound}
 C_S\ge\sum_{n\ge0}\min(n,N)E_n^S+ND_S\ge ND_S.
\end{equation}
Consequently, if $\models_{C,\mathrm{tot}}\{P\}\,S\,\{G\}$ and $\Tr(P\rho)<\infty$ for a normalized state $\rho$, then $S$ has finite expected runtime and terminates almost surely on $\rho$.
\end{lemma}

\begin{proof}
The first inequality follows from~\eqref{eq:total-runtime-cost}, which defines $C_S$ as the supremum of these capped costs; the second follows from positivity of the sum. Validity of the assumed total-cost Hoare triple gives $\Tr(C_S\rho)\le\Tr(P\rho)<\infty$. Since $N\Tr(D_S\rho)\le\Tr(C_S\rho)$ for every $N$, we obtain $\Tr(D_S\rho)=0$.
\end{proof}

\begin{theorem}[Translation of total-cost contracts]
\label{bridge:main}
For closed positive cost forms $P,G$ and a program $S$,
\begin{equation}
\label{bridge:eq:total}
 \models_{C,\mathrm{tot}}\{P\}\,S\,\{G\}
 \quad\Longrightarrow\quad
 \models_{\mathrm{tot}}\{R(P)\}\,S\,\{R(G)\}.
\end{equation}
Equivalently, a valid total-cost contract gives
\[
 \res{s}{P}\le\wp(S,R(G))_s
 =\sum_{n\ge0}\frac{s}{s+n}
       \sem{S}_n^\dagger(\res{s+n}{G})
 \qquad(s>0).
\]
\end{theorem}

\begin{proof}
Fix $s>0$. We record each terminating runtime in a separate output block and add one block for divergence. Let
\[
 \cK\triangleq \left(\bigoplus_{n\in\NN}\cH\right)\oplus\CC.
\]
For a positive trace-class operator $\rho$, define
\begin{equation}
\label{bridge:eq:completed-map}
 \Psi_*(\rho)
 \triangleq \left(\bigoplus_{n\ge0}\sem{S}_n(\rho)\right)
   \oplus\Tr(D_S\rho).
\end{equation}
The direct sum converges in trace norm, and
\[
 \Tr\bigl(\Psi_*(\rho)\bigr)
 =\sum_{n\ge0}\Tr\bigl(\sem{S}_n(\rho)\bigr)
   +\Tr(D_S\rho)
 =\Tr\bigl((Q_S+D_S)\rho\bigr)
 =\Tr(\rho).
\]
Complete positivity holds blockwise, so linear extension gives a trace-preserving CP map $\Psi_*:\Top{\cH}\to\Top{\cK}$. Its Heisenberg adjoint $\Psi:\Bop{\cK}\to\Bop{\cH}$ is therefore normal and unital.

For $N\in\NN_{+}$, define the closed positive form on $\cK$
\[
 X_N\triangleq
 \left(\bigoplus_{n\ge0}
       \bigl(G+\min(n,N)I\bigr)\right)\oplus NI_{\CC}.
\]
Extended trace duality gives
\[
 \widehat\Psi(X_N)
 =\widehat{\sem{S}^\dagger}(G)
  +\sum_{n\ge0}\min(n,N)E_n^S+ND_S.
\]
Here the terminal-cost terms combine by monotone convergence:
\[
 \sum_{n\ge0}\Tr\bigl(G\sem{S}_n(\rho)\bigr)
 =\Tr\bigl(G\sem{S}(\rho)\bigr)
 \qquad(\rho\ge0).
\]
Thus total-cost validity,~\eqref{bridge:eq:cost-wp}, and Lemma~\ref{bridge:lem:divergence} imply
\[
 P\ge\widehat{\sem{S}^\dagger}(G)+C_S
   \ge\widehat\Psi(X_N).
\]
Applying antitonicity~\eqref{bridge:eq:antitonicity} and Corollary~\ref{bridge:cor:form-jensen} yields
\begin{equation}
\label{bridge:eq:finite-penalty}
\begin{aligned}
 \res{s}{P}
 &\le\res{s}{\widehat\Psi(X_N)}
 \le\Psi(\res{s}{X_N})\\
 &=\sum_{n\ge0}\sem{S}_n^\dagger
      \!\left(\res{s}{G+\min(n,N)I}\right)
   +\frac{s}{s+N}D_S.
\end{aligned}
\end{equation}

For each fixed $n$, the corresponding block of $\res{s}{X_N}$ equals $\res{s}{G+nI}$ once $N\ge n$, while its final scalar block tends to zero. Together with the bound $0\le\res{s}{X_N}\le I_{\cK}$, these blockwise limits show that, as $N\to\infty$, the operators $\res{s}{X_N}$ decrease monotonically and converge in the strong operator topology to
\[
 \left(\bigoplus_{n\ge0}\res{s}{G+nI}\right)\oplus0.
\]
Normality of $\Psi$ preserves this bounded decreasing limit. Letting $N\to\infty$ in~\eqref{bridge:eq:finite-penalty} therefore gives
\[
 \res{s}{P}
 \le\sum_{n\ge0}\sem{S}_n^\dagger(\res{s}{G+nI}).
\]
The scale-shift identity extends to closed positive forms by spectral calculus on $\cH_G$, with both sides zero on $\cH_G^\perp$:
\[
 \res{s}{G+nI}=\frac{s}{s+n}\res{s+n}{G}.
\]
Substituting this identity and invoking~\eqref{eq:runtime-transformer} gives $\res{s}{P}\le\wp(S,R(G))_s$. Since this holds for every $s>0$, Definition~\ref{def:parameterized-validity} establishes the validity of the total-correctness Hoare triple in~\eqref{bridge:eq:total}.
\end{proof}


\end{document}